\documentclass[pdftex,aps,preprintnumbers,nofootinbib,superscriptaddress,showpacs,twocolumn]{revtex4}
\usepackage{amsmath,amssymb,amsbsy,booktabs,array}
\usepackage{bm}
\usepackage[pdftex]{graphicx,color}
\usepackage[sort&compress]{natbib}
\usepackage[colorlinks=true, linkcolor=blue, filecolor=blue, urlcolor=blue, citecolor=blue, pdftex=true, plainpages=false]{hyperref}

\newcommand\riken{RIKEN-BNL Research Center, Brookhaven National
Laboratory, Upton, NY 11973, USA}
\newcommand\bnlaf{Physics Department, Brookhaven
National Laboratory, Upton, NY 11973, USA}

\newcommand\cuaff{Physics Department, Columbia University, New York, NY 10027, USA}
\newcommand\soton{School of Physics and Astronomy, University of
Southampton, Southampton SO17 1BJ, UK}

\newcommand\nagoya{Kobayashi-Maskawa Institute for the Origin of Particles and the Universe, Nagoya University, Nagoya 464-8602, Japan}

\newcommand\ST{\rule[-1em]{0pt}{2.5em}}

\def\CO{{\cal O}}
\def\CC{{\cal C}}
\def\bar{\overline}

\def\beq{\begin{equation}}
\def\eeq{\end{equation}}
\def\bea{\begin{eqnarray}}
\def\eea{\end{eqnarray}}

\def\spose#1{\hbox to 0pt{#1\hss}}
\def\ltapprox{\mathrel{\spose{\lower 3pt\hbox{$\mathchar"218$}}
\raise 2.0pt\hbox{$\mathchar"13C$}}}
\def\gtapprox{\mathrel{\spose{\lower 3pt\hbox{$\mathchar"218$}}
\raise 2.0pt\hbox{$\mathchar"13E$}}}
\def\inapprox{\mathrel{\spose{\lower 3pt\hbox{$\mathchar"218$}}
\raise 2.0pt\hbox{$\mathchar"232$}}}

\begin{document}
\bibliographystyle{apsrev}

\title{Nonperturbative tuning of an improved relativistic heavy-quark action with application to bottom spectroscopy}

\author{Y.~Aoki}\affiliation\riken\affiliation\nagoya
\author{N.~H.~Christ}\affiliation\cuaff
\author{J.~M.~Flynn}\affiliation\soton
\author{T.~Izubuchi}\affiliation\riken\affiliation\bnlaf
\author{C.~Lehner}\affiliation\riken
\author{M.~Li}\affiliation\cuaff
\author{H.~Peng}\affiliation\cuaff
\author{A.~Soni}\affiliation\bnlaf
\author{R.~S.~Van~de~Water}\affiliation\bnlaf
\author{O.~Witzel\footnote{Present address: Center for Computational Science, Boston University, Boston, MA 02215, USA}}\affiliation\bnlaf
\collaboration{RBC and UKQCD Collaborations}\noaffiliation

%
%


\date{\today}

\begin{abstract}
We calculate the masses of bottom mesons using an improved relativistic action for the $b$-quarks
and the RBC/UKQCD Iwasaki gauge configurations with 2+1 flavors of dynamical domain-wall light quarks.  We analyze configurations with two lattice spacings: $a^{-1} = 1.729$~GeV ($a \approx 0.11$~fm) and $a^{-1} = 2.281$~GeV ($a \approx 0.086$~fm).  We use an anisotropic, clover-improved Wilson action for the $b$-quark, and tune the three parameters of the action nonperturbatively such that they reproduce the experimental values of the $B_s$ and $B_s^*$ heavy-light meson states.  The masses and mass-splittings of the low-lying
bottomonium states (such as the $\eta_b$ and $\Upsilon$) can then be computed with no additional inputs, and comparison between these predictions and experiment provides a test of the validity of our method.  We obtain bottomonium masses with total uncertainties of $\sim 0.5$--$0.6\%$ and fine-structure splittings with uncertainties of $\sim 35$--$45\%$;  for all cases we find good agreement with experiment.  The parameters of the relativistic heavy-quark action tuned for $b$-quarks presented in this work can be used for precise calculations of weak matrix elements such as $B$-meson decay constants and mixing parameters with lattice discretization errors that are of the same size as in light pseudoscalar meson quantities.  This general method can also be used for charmed meson masses and matrix elements if the parameters of the heavy-quark action are appropriately tuned.
\end{abstract}

\maketitle


\section{Introduction}
\label{sec:Intro}

Precise knowledge of mass spectrum, decay, and mixing properties of hadrons
containing one or more bottom or charm quarks is essential to
advancing our understanding of the parameters of the Standard Model.
Lattice Quantum Chromodynamics (QCD) provides methods to compute these
quantities from first principles. Conventional lattice
calculations with heavy quarks are challenging, however, because it is
impractical to use a sufficiently small lattice spacing to control
the $O(ma)^n$ discretization errors directly.

One way to address this challenge is to adapt the lattice description
of heavy quarks to correctly describe heavy-quark physics in a
carefully circumscribed kinematic range.  Such approaches include
heavy-quark effective theory (HQET)~\cite{Eichten:1989zv} in which
the limit of infinite quark mass is considered and the continuum
limit of the lattice theory reproduces the continuum heavy quark
effective theory.  A second method is non-relativistic
QCD (NRQCD)~\cite{Thacker:1990bm,Lepage:1992tx} in which the mass of the heavy
quark is assumed to be much greater than the inverse lattice
spacing but the momentum-dependence of the heavy quark energy
is included in the non-relativistic limit.  Each of these approaches
has its own limitations.   Specifically, radiative corrections to the coefficients of the NRQCD Lagrangian contain power-law divergences that blow up in the limit $ma \to 0$, while HQET cannot deal with quarkonia.

The Fermilab or relativistic heavy quark (RHQ) method~\cite{ElKhadra:1996mp, Aoki:2001ra,Christ:2006us} provides a more complete framework for
heavy-quark physics.  It applies for all values of the heavy quark mass $m_{Q}$,
for both heavy-heavy and heavy-light systems, and allows a continuum limit.
The improved RHQ action accurately describes energies and amplitudes of on-shell states containing
heavy quarks whose spatial momentum $\vec p$ is small compared to the
lattice spacing.  It can be shown~\cite{Christ:2006us} that all errors
of order $|\vec p a|$, $(m_{Q} a)^n$ and $|\vec p a| (m_{Q} a)^n$ for all
non-negative integers $n$ can be removed if an anisotropic, clover-improved
Wilson action is used for the heavy quark.  This action depends
on three relevant parameters: the bare quark mass $m_0$, an anisotropy
parameter $\zeta(m_0 a)$ and the coefficient $c_P(m_0 a)$ of an isotropic
Sheikholeslami and Wohlert term.

In order to exploit this RHQ approach, values for these three parameters are
needed.  The bare charm or bottom quark mass, $m_0$, must be determined
from experiment, usually by equating the known mass of a physical state
containing one or two heavy quarks with the mass determined from a lattice
calculation with the RHQ action.  The remaining two parameters, $\zeta$ and $c_P$, may be
estimated from lattice perturbation theory or determined with a
nonperturbative technique.  We cannot use the nonperturbative method of Rome-Southampton~\cite{Martinelli:1994ty} to tune $c_P$ and $\zeta$ because the Rome-Southampton approach depends on evaluating off-shell amplitudes, whereas the 3-parameter RHQ action only controls discretization
errors for  on-shell states.  On-shell step-scaling methods can be used,
either via the Schr\"odinger functional approach or a simple comparison of
small volume spectra between calculations with varying lattice scale
but identical physical volumes~\cite{Lin:2006ur}.  Both of these
step-scaling approaches, however, involve substantial computational effort, requiring a series
of carefully matched finite volume simulations with varying lattice
spacing.

In the work reported here, we use another approach and determine
the two remaining parameters $\zeta$ and $c_P$ nonperturbatively by imposing two simple
conditions.  The first condition is the often-exploited requirement
that the energy of a specific heavy-heavy or heavy-light state
depend on that state's spatial momentum in a fashion consistent
with continuum relativity:  $E(\vec{p})^2 = \vec p^2 + M^2$.  The second constraint
is that a specific mass splitting agree with its experimental
value.  For the case of bottom, a natural choice is the $B_s^* - B_s$ mass splitting.  Thus, using the bottom system as an example
and including the bare quark mass $m_0$, we determine our three
parameters $m_0$, $\zeta(m_0 a)$ and $c_P(m_0 a)$ by requiring
that experimental values are obtained for $m_{B_s}$ and
$m_{B_s^*}$ and that $E_{B_s}$ has the proper dependence on
$\vec p_{B_s}$.

As is described below, these three conditions are straightforward
to impose and yield quite precise results for the three unknown
parameters.  This approach has the disadvantage that a possible
experimental prediction from lattice QCD, a non-trivial spin-spin
splitting, cannot be made.  With this approach, however, we can
immediately determine the masses of a large number of heavy-heavy
and heavy-light states.  These results can be viewed as tests of
QCD and can be used to explore the accuracy and limitations
of the RHQ approach.  Finally, once the RHQ
action has been determined by fixing these three parameters, it can be used to compute phenomenologically-important charm and bottom decay constants and mixing matrix elements, which are needed for determinations of CKM matrix elements and constraints on the CKM unitarity triangle

\bigskip

In this paper we present results for the bottom
system.  Our calculation is performed on the 2+1 flavor,
domain wall fermion (DWF) + Iwasaki gauge-field ensembles generated by the LHP, RBC, and UKQCD
collaborations with several values of the light dynamical quark mass at two lattice spacings, $a \approx 0.11$~fm and $a \approx 0.086$~fm~\cite{Allton:2008pn,Aoki:2010dy}.  For the heavy-light
mesons, the heavy quark will typically carry a small
spatial momentum, $|\vec p| \approx \Lambda_{\rm QCD}$.  Thus, for
these systems the expected $|\vec p a|^2$ errors are of the same size
as those encountered in calculations involving only light
quarks.  For heavy-heavy systems, however, the spatial
momenta will be larger: $|\vec p| \approx \alpha_s m_{Q}$, where $m_{Q}$
is the heavy-quark mass and $\alpha_s$ the strong interaction coupling
constant evaluated at a scale appropriate for such a bound state.
While for charmonium $\alpha_s m_{Q}$ may be on the order of
$\Lambda_{\rm QCD}$, this is not the case for bottomonium where
discretization errors are expected to be three to four times larger due to the larger $b$-quark mass ($m_b/m_c \approx 3.3$).
Thus we choose to tune the RHQ action for $b$-quarks using bottom-light states in order to minimize systematic uncertainties.  In particular, we match to the experimentally-measured masses of the bottom-strange states $B_s$ and $B_s^*$ in order to avoid the need to perform a chiral extrapolation in the valence light-quark mass.  Once we have determined the values of the parameters $\{m_0, c_P, \zeta\}$ we make predictions for the masses and mass-splittings of several low-lying bottomonium states:  $\eta_B$, $\Upsilon$, $\chi_{b0}$, $\chi_{b1}$, and $h_b$.  Our results agree with experiment within estimates of systematic uncertainties, confirming the validity of the RHQ approach and bolstering confidence in future computations of weak matrix elements with the RHQ action.

This work was begun by Li, who presented preliminary values for the RHQ parameters and bottomonium masses on the coarser $a \approx 0.11$~fm ensemble at Lattice 2008~\cite{Li:2008kb}.  We improve upon his results in several ways, most notably in determining the RHQ parameters solely from quantities in the bottom-strange system.  (Li used a hybrid of bottom-strange and bottomonium observables for the tuning.)  This reduces the systematic errors in the resulting parameters due to heavy-quark discretization effects, as discussed above.  We also significantly increased the statistics, more than quadrupling the number of configurations analyzed, and optimized the spatial smearing wavefunction used for the $b$-quarks in order to reduce excited-state contamination in the bottom-strange 2-point correlators.  More recently Peng extended this work to the finer $a \approx 0.086$~fm ensembles and presented preliminary values for the RHQ parameters and bottomonium masses at Lattice 2010~\cite{Peng:Lattice10}.  Again, we polish this result with increased statistics and improved $b$-quark smearings.

This paper is organized as follows.  In Section~\ref{sec:Framework} we first present the form of the relativistic heavy-quark action used in this work.  We then describe the numerical strategy used to determine the three parameters $m_0$,
$\zeta(m_0 a)$ and $c_P(m_0 a)$.  Next, in Section~\ref{sec:bottom_RHQ} we present the tuning of the RHQ parameters for bottom.  We give the actions and parameters used in our numerical simulations, and then discuss the fits of heavy-light meson 2-point correlators needed to extract the lattice values of the $B_s$ and $B_s^*$ meson masses.  Using this data we tune the parameters of the RHQ action such that it applies to $b$-quarks.  In Section~\ref{sec:MassPred} we use the resulting RHQ parameters to predict the masses of several bottomonium states and compare the results with experiment.  Finally, we summarize our results and discuss future plans in Section~\ref{sec:Conc}.

\section{Framework of the calculation}
\label{sec:Framework}

\subsection{Heavy-quark action}
\label{sec:HQAct}

The relativistic heavy-quark method provides a consistent framework for describing both light quarks ($am_0 \ll 1$) and heavy quarks ($am_0 \gtapprox 1$)~\cite{ElKhadra:1996mp,Christ:2006us,Aoki:2001ra}.  This approach relies upon the fact that, in the rest frame of bound states containing one or more heavy quarks, the spatial momentum carried by each heavy quark is smaller than the mass of the heavy quark:  for heavy-heavy systems $|\vec{p}| \sim \alpha_s m_0$ and for heavy-light systems $|\vec{p}| \sim \Lambda_\textrm{QCD}$.  Then one can perform the usual Symanzik expansion in powers of the spatial derivative $D_i$ (which brings down powers of $a\vec{p})$.  One must, however, include terms of all orders in the mass $m_0 a$ and the temporal derivative $D_0$.  This suggests that a suitable lattice formulation for heavy quarks should break the axis-interchange symmetry between the spatial and temporal directions.  

In this work we use the following anisotropic clover-improved Wilson action for the $b$-quarks:
\begin{widetext}
\begin{align}
S_\textrm{lat} &= a^4 \sum_{x,x'} \bar{\psi}(x') \left( m_0 + \gamma_0 D_0 + \zeta \vec{\gamma} \cdot \vec{D} - \frac{a}{2} (D^0)^2 - \frac{a}{2} \zeta (\vec{D})^2+ \sum_{\mu,\nu} \frac{ia}{4} c_P \sigma_{\mu\nu} F_{\mu\nu} \right)_{x' x} \psi(x) \,,
\label{eq:HQAct}
\intertext{where}
D_\mu\psi(x) &= \frac{1}{2a} \left[ U_\mu(x)\psi(x+\hat{\mu}) - U_\mu^\dagger(x-\hat{\mu})\psi(x-\hat{\mu}) \right] \,, \\
D^2_\mu \psi(x) &= \frac{1}{a^2} \left[  U_\mu(x) \psi(x+\hat{\mu}) + U_\mu^\dagger(x - \hat{\mu})\psi(x-\hat{\mu}) - 2 \psi(x)  \right] \,, \\
F_{\mu\nu} \psi(x)&= \frac{1}{8 a^2} \sum_{s,s'= \pm 1} s s' \left[ U_{s\mu}(x) U_{s'\nu}(x+s\hat{\mu}) U_{s\mu}^\dagger(x + s'\hat{\nu}) U_{s'\nu}^\dagger (x)- \textrm{h.c.} \right] \psi(x) \,,
\end{align}
\end{widetext}
and  $\gamma_\mu = \gamma_\mu^\dagger$ , $\{\gamma_\mu,\gamma_\nu\} = 2 \delta_{\mu\nu}$ and $\sigma_{\mu\nu} = \frac{i}{2} [ \gamma_\mu , \gamma_\nu ]$.
Christ, Li, and Lin showed in Ref.~\cite{Christ:2006us} that one can absorb all positive powers of the temporal derivative by allowing the coefficients $c_P (m_0 a)$ and $\zeta (m_0 a)$ to be functions of the bare-quark mass $m_0 a$. Thus, by suitably tuning the three coefficients in the action -- the bare-quark mass $m_0 a$, anisotropy parameter $\zeta$, and clover coefficient $c_P$ -- one can eliminate errors of $\CO(|\vec{p}|a)$, $\CO([m_0a]^n)$, and $\CO(|\vec{pa}|[m_0a]^n)$ from on-shell Green's functions.   The resulting action can be used to describe heavy quarks with $m_0a \gtapprox 1$ with discretization errors that are comparable to those for light-quark systems.  

The relativistic heavy quark formulation based on Ref.~\cite{Christ:2006us} and used in this work is one of several variations.  This general approach was first introduced by El-Khadra, Kronfeld, and Mackenzie in Ref.~\cite{ElKhadra:1996mp}, and has been used recently by the Fermilab Lattice and MILC collaborations for many phenomenolgical applications such as decay constant and form-factor computations~\cite{Bernard:2008dn,Bailey:2008wp,Evans:2009du,Simone:2010zz}.  In practice, however, Fermilab/MILC use a different approach to tune the parameters in the action, Eq.~(\ref{eq:HQAct}), than our method described below in Sec.~\ref{sec:Method}.  They fix the anisotropy parameter $\zeta$ to unity and the clover coefficient $c_P$ to its tree-level mean-field improved lattice perturbation theory value $(1/u_0^3)$, and then tune only the hopping parameter $\kappa$ (which is equivalent to the bare-quark mass) nonperturbatively~\cite{Bernard:2010fr}.  The Tsukuba group uses a slightly different formulation of the action in which they do not use field rotations to eliminate redundant operators~\cite{Aoki:2003dg}; hence their version of the action has four unknown coefficients rather than three in the RHQ or Fermilab variants.  For on-shell Green's functions the Tsukuba and RHQ/Fermilab actions are equivalent.  In practice, however, the inclusion of redundant couplings means that one cannot nonperturbatively tune all four parameters simultaneously by only adjusting the energies of heavy hadrons because one will run into flat directions in parameter space, as was shown in Ref.~\cite{Christ:2006us}.  Hence they rely upon lattice perturbation theory for quark-quark scattering
amplitudes to determine at least one of the coefficients~\cite{Aoki:2003dg}.

Because the lattice action breaks Lorentz symmetry, mesons receive corrections to their energy-momentum dispersion relation due to lattice artifacts:
\begin{align} 
	(aE)^2 = (aM_1)^2 + \left( \frac{M_1}{M_2} \right) (a\vec{p})^2 + \CO( [a\vec{p}]^4) \,.
\label{eq:DispRel}
\end{align}
The quantities $M_1$ and $M_2$ are known as the rest mass and kinetic mass, respectively,
\begin{align}
	M_1 = E(\vec{p} =0) \,, \qquad M_2 = M_1 \times \left( \frac{\partial E^2}{\partial p_i^2} \right)^{-1}_{\vec{p} = 0} \,,
\end{align}
and will generally be different for generic values of the parameters $\{m_0a, c_P, \zeta\}$.  We will exploit this fact in the RHQ tuning procedure described in the following subsection.

\subsection{Parameter tuning methodology}
\label{sec:Method}

We tune the values for the RHQ parameters $\{m_0a, c_P, \zeta\}$ to describe bottom or charm quarks by requiring that calculations of specified physical on-shell quantities with the action in Eq.~(\ref{eq:HQAct}) correctly reproduce the experimentally-measured results.  In particular, for $b$-quarks we determine the RHQ action using the bottom-strange system because both discretization errors and chiral extrapolation errors are expected to be small.  We match to the experimental values of the spin-averaged $B_s$ meson mass:
\begin{align}
\bar{M}_{B_s} = \frac{1}{4} \left( M_{B_s} + 3 M_{B_s^*} \right)
\end{align}
and hyperfine splitting:
\begin{align}
	\Delta M_{B_s} = M_{B_s^*} - M_{B_s} \,.
\end{align}
We also require that the $B_s$ meson rest and kinetic masses are equal: 
\begin{align}
\frac{M_1^{B_s}}{M_2^{B_s}} = 1 \,,
\end{align}
so that the $B_s$ meson satisfies the continuum energy-momentum dispersion relation $E^2_{B_s}(\vec{p})~=~\vec{p}^2_{B_s}~+~M^2_{B_s}$.  (Note that throughout this work we denote meson masses with a capital ``$M$" and quark masses with a lower-case ``$m$" in order to avoid confusion in situations where the context is insufficient.)  We could in principle have used other states ({\it e.g.}~scalar or vector mesons), other mass-splittings ({\it e.g.}~the spin-orbit splitting), or even other systems ({\it e.g.}~heavy-heavy mesons) to tune the parameters of the RHQ action, since the same parameters should describe $b$-quarks in all of these arenas.  Instead, however, we can make predictions for these quantities using the tuned values of $\{m_0a, c_P, \zeta\}$ and use them to test the validity of our approach.

\begin{figure}[t]
\centering
\includegraphics[scale=0.5,clip]{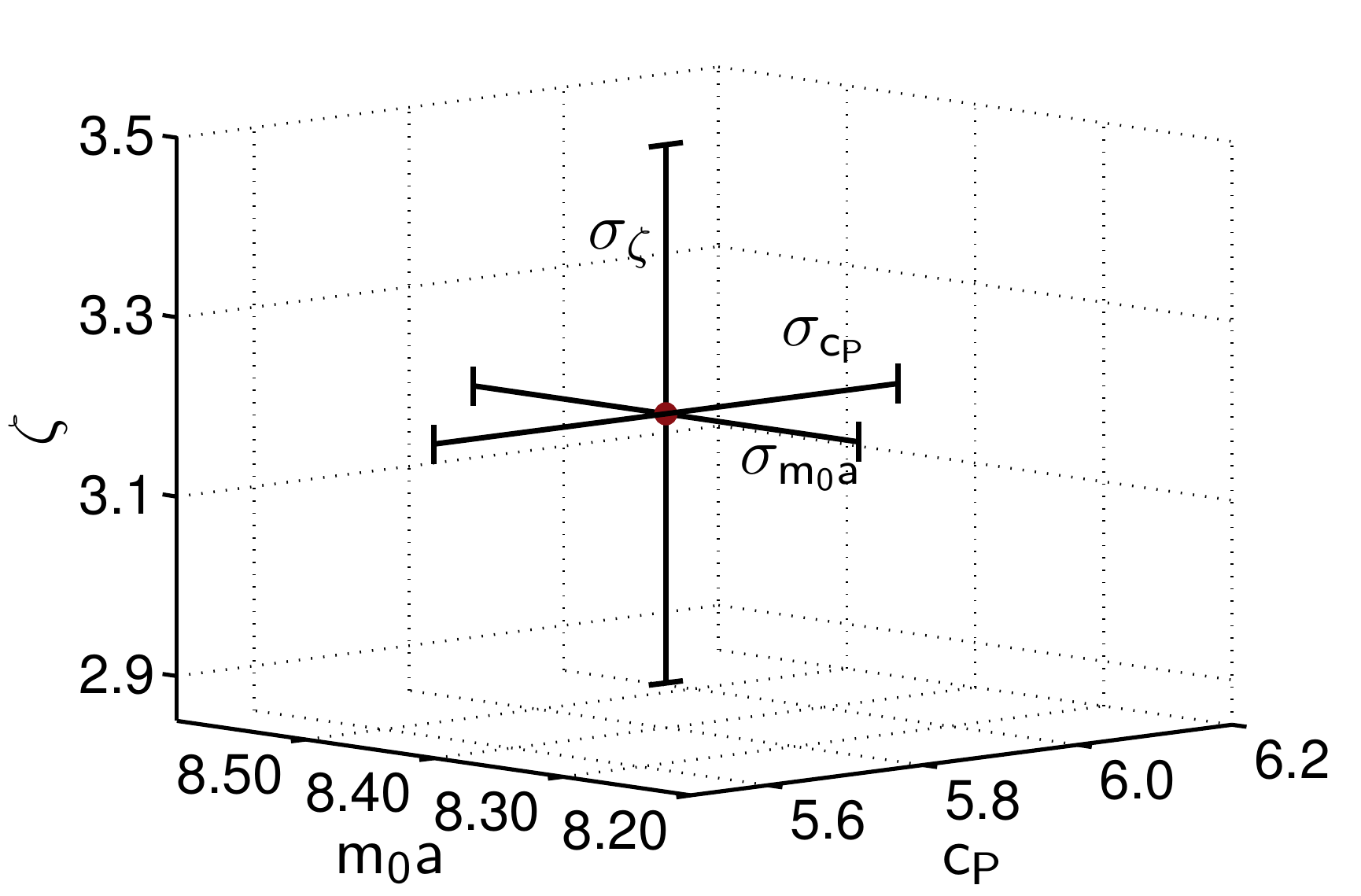}
\caption{Location of the seven sets of parameters used to obtain the tuned values of $\{m_0a, c_P, \zeta\}$.}
\label{fig:RHQBox}
\end{figure}

We determine the tuned values of  $\{m_0a, c_P, \zeta\}$ nonperturbatively using an iterative procedure.  The bottom-strange meson masses in general will have a nonlinear dependence on the RHQ parameters.  We choose to work in a region sufficiently close to the true parameters such that the following linear approximation holds:
\begin{equation}
    \left[
      \begin{array}{c}
	\bar{M}_{B_s}\\
	\Delta M_{B_s}\\
	\frac{M_1^{B_s}}{M_2^{B_s}}
      \end{array} \right] = J\cdot \left[
      \begin{array}{c}
        m_0a \\
        c_P \\
        \zeta
      \end{array}
      \right]
      + A \,,
      \label{eq:linear_approx}
\end{equation}
where $J$ is a $3\times 3$ matrix containing the linear coefficients (analogous to the slope in the $1\times 1$ case) and $A$ is a 3-element column vector containing the constants (analogous to the intercept).  For a single step of the iteration procedure we compute the quantities $\{ \bar{M}_{B_s}, \Delta M_{B_s}, M_1^{B_s}/M_2^{B_s} \}$ at seven sets of parameters (see Fig.~\ref{fig:RHQBox}) in which we vary one of the three parameters $\{m_0a, c_P, \zeta\}$ by a chosen uncertainty $\pm \sigma_{\{ m_0a, c_P, \zeta \}}$ (not to be confused with the statistical errors in the tuned parameters $\{m_0a, c_P, \zeta\}$) while holding the other two fixed:
\begin{align}
&\left[\!\!\begin{array}{c} m_0 a\\ c_P\\ \zeta\\ \end{array}\right],
\left[\!\!\begin{array}{c}m_0 a -\sigma_{m_0 a}\\ c_P\\\zeta \\ \end{array}\!\!\right],\;
\left[\!\!\begin{array}{c}m_0 a +\sigma_{m_0 a}\\ c_P\\\zeta \\ \end{array}\!\!\right],\;
\left[\!\!\begin{array}{c}m_0 a\\c_P-\sigma_{c_P}\\ \zeta\\ \end{array}\!\!\right],\;\nonumber \\
&\left[\!\!\begin{array}{c}m_0 a\\c_P+\sigma_{c_P}\\ \zeta\\ \end{array}\!\!\right],\;
\left[\!\!\begin{array}{c}m_0 a\\ c_P\\ \zeta-\sigma_{\zeta}\\ \end{array}\!\!\right],\;
\left[\!\!\begin{array}{c}m_0 a\\ c_P\\ \zeta+\sigma_{\zeta}\\ \end{array}\!\!\right] \,.
\label{Eq:SevenSets}
\end{align}
This allows us to test whether or not the ``box" of parameter space defined by the seven parameter sets in Fig.~\ref{fig:RHQBox} is in the linear region such that Eq.~(\ref{eq:linear_approx}) applies.  If indeed we are in the linear region, we then compute the matrix $J$ and vector $A$ via a simple finite difference approximation of the derivatives:
\begin{align}
J &= \left[\frac{Y_3- Y_2}{2\sigma_{m_0a}},\,\frac{Y_5-Y_4}{2\sigma_{c_P}},\,\frac{Y_7 - Y_6}{2\sigma_\zeta}\right] \,, \label{Eq:Jmatrix} \\
\ST A &= Y_1 - J\times \left[m_0a,\, c_P,\, \zeta\right]^T \,,
\label{Eq:Amatrix}
\end{align}
where $Y_i$ is the 3-element column vector containing the values of meson masses and splittings measured on the $i^\textrm{th}$ parameter set listed in Eq.~(\ref{Eq:SevenSets}):
\begin{align}
Y_i = \left[ \bar{M}_{B_s}, \Delta M_{B_s}, M_1^{B_s}/M_2^{B_s} \right]^T_i \,. \label{eq:Yi}
\end{align}
Finally, the tuned RHQ parameters are given by:
\begin{align}
\left[\begin{array}{c} m_0a\\ c_P \\ \zeta \end{array}\right]^\text{RHQ} = J^{-1} \times\left(\left[\begin{array}{c}\bar{M}_{B_s}\\ \Delta M_{B_s}\\ \frac{M_1^{B_s}}{M_2^{B_s}}\end{array}\right]^\text{PDG} - A\right) \,.
\label{Eq:RHQDetermination}
\end{align}

We consider the RHQ parameters $\{m_0a, c_P, \zeta\}$ to be ``tuned" when all three of the values obtained via Eq.~(\ref{Eq:RHQDetermination}) are within the ``box" defined by the seven parameter sets in Fig.~\ref{fig:RHQBox}.  This condition ensures that we are interpolating, rather than extrapolating, to the tuned point.  If the result for any of the parameters $\{m_0a, c_P, \zeta\}$ lies outside the box, we re-center the box around the result of Eq.~(\ref{Eq:RHQDetermination}) and perform another iteration step.  We repeat this procedure until all three tuned RHQ parameters lie inside the box. 

Once the RHQ parameters have been tuned, we can use them to predict the masses of other heavy-light and heavy-heavy meson states, and ultimately to compute heavy-light meson weak matrix elements.  We compute the desired quantities on the same seven sets of parameters used for the final iteration of the tuning procedure.  We then propagate the statistical errors in the tuned RHQ parameters to these quantities using the jackknife method;  this accounts for correlations between the parameters $m_0$, $c_P$, and $\zeta$.

\section{Lattice calculation of RHQ parameters for bottom}
\label{sec:bottom_RHQ}

\subsection{Lattice simulation parameters}
\label{sec:LatParams}

The parameters of the RHQ action suitable for describing $b$-quarks depend upon the choice of actions for the gauge fields and sea quarks.  In this work we perform our numerical lattice computations on the ``2+1" flavor domain-wall fermion ensembles generated by the LHP, RBC, and UKQCD Collaborations~\cite{Allton:2008pn,Aoki:2010dy}.  These lattices include the effects of three light dynamical quarks;  the lighter two sea quarks are degenerate and we denote their mass by $m_l$, while the heavier sea quark, whose mass we denote by $m_h$, is a little heavier than the
 physical strange quark.  The RBC/UKQCD lattices combine the Iwasaki action for the gluons~\cite{Iwasaki:1983ck} with the five-dimensional domain-wall action for the fermions~\cite{Shamir:1993zy,Furman:1994ky}.  Use of the Iwasaki gauge action in combination with domain-wall sea quarks allows for adequate tunneling between topological sectors~\cite{Antonio:2008zz}, and in combination with domain-wall valence quarks reduces chiral symmetry breaking and the size of the residual quark mass as compared to the Wilson gauge action~\cite{Wilson:1974sk}.

We compute the RHQ $b$-quark parameters on several ensembles with different light sea-quark masses; this allows us to study the sea-quark mass dependence, which we find to be statistically insignificant.  We also determine the parameters at two lattice spacings;  we refer to the coarser ensembles with $a \approx 0.11$~fm as the ``$24^3$" ensembles and the finer ensembles with $a \approx 0.086$~fm as the ``$32^3$" ensembles.  Use of two lattice spacings allows us to take a na\"ive continuum limit of physical quantities such as meson masses and splittings, although we still include a conservative power-counting estimate of the residual $\CO(|\vec{p}a|^2)$ discretization errors from the RHQ action that may not be removed with this approach.  Table~\ref{tab:lattices} shows the parameters of the ensembles used for the RHQ parameter tuning and bottomonium spectroscopy presented in this work.  On the finer lattice spacings we double the statistics by performing two fermion inversions per gauge configuration with the origins of the quark sources separated by half of the temporal lattice extent.

\begin{table*}
\caption{Lattice simulation parameters used in our determination of the RHQ parameters for $b$-quarks and in our predictions for the bottomonium masses and mass-splittings.  The columns list the lattice volume, approximate lattice spacing, light ($m_l$) and strange ($m_h$) sea-quark masses, unitary pion mass, and number of configurations and time sources analyzed.}
\vspace{3mm}
\label{tab:lattices}
\begin{tabular}{cccccrc} \hline\hline
  &         &         &         &            & &\# time\\
$\left(L/a\right)^3 \times \left(T/a\right)$ \qquad & $\approx a$(fm) & ~~$am_l$ & ~~$am_h$ & \quad $M_\pi$(MeV) \quad & \# configs.&sources\\[0.5mm] \hline
$24^3 \times 64$ &  0.11 &  0.005 & 0.040 & 329 & 1636~~~~&1\\
$24^3 \times 64$ &  0.11 &  0.010 & 0.040 & 422 & 1419~~~~&1\\ \hline
$32^3 \times 64$ &  0.086 &  0.004 & 0.030 & 289 & 628~~~~&2\\ 
$32^3 \times 64$ &  0.086 &  0.006 & 0.030 & 345 & 889~~~~&2\\
$32^3 \times 64$ &  0.086 &  0.008 & 0.030 & 394 & 544~~~~&2\\ \hline\hline
\end{tabular}\end{table*}

The ensembles listed in Table~\ref{tab:lattices} have already been utilized to study the light pseudoscalar meson sector; we can therefore take advantage of many results from this earlier work.  The amount of chiral symmetry breaking in the light-quark sector can be parameterized in terms of an additive shift to the bare domain-wall quark mass called the residual quark mass.  At the values of $M_5 = 1.8$ and $L_s = 16$ used by RBC/UKQCD, the size of the residual quark mass is quite small; $am_\textrm{res} = 0.003152(43)$ on the $24^3$ ensembles and $am_\textrm{res} = 0.0006664(76)$ on the $32^3$ ensembles~\cite{Aoki:2010dy}.  In order to compute the masses of $B_s$ and $B_s^*$ mesons for the tuning procedure we also need the value of the physical strange-quark mass on these ensembles.  This was already determined in Ref.~\cite{Aoki:2010dy}; $am_s = 0.0348(11)$ on the $24^3$ ensembles and $am_s =  0.0273(7)$ on the $32^3$ ensembles.  (In practice we use slightly different values of the strange-quark mass --- $am_s = 0.0343$ on the $24^3$ ensembles and $am_s = 0.0272$ on the $32^3$ ensembles --- because we began this work before the light pseudoscalar meson analysis in Ref.~\cite{Aoki:2010dy} was finalized.  These values, however, are within the stated statistical errors.)  Finally, we must convert lattice meson masses into physical units for the tuning procedure and for comparison between predictions and experiment.  The lattice scale was determined from the $\Omega$ mass to be $a^{-1} = 1.729(25)$~GeV on the $24^3$ ensembles and $a^{-1} = 2.281(28)$~GeV on the $32^3$ ensembles~\cite{Aoki:2010dy}.  These values are consistent with an independent determination of the $24^3$ and $32^3$ lattice spacings using the $\Upsilon(2S)-\Upsilon(1S)$ mass-splitting by Meinel~\cite{Meinel:2010pv}.

\subsection{Heavy-light meson correlator fits}
\label{sec:HLFits}

We extract the $B_s$ and $B_s^*$ meson energies from the exponential behavior of the following 2-point correlation functions:
\begin{align}
C_{B_s}(t,t_0; \vec{p}) &= \sum_{\vec{y}}  e^{ip \cdot \vec{y}} \langle \CO^\dagger_{P}(\vec{y},t)  \tilde{\CO}_{P} (\vec{0},t_0) \rangle \label{eq:C_Bs} \,, \\
C_{B_s^*}(t,t_0; \vec{p}) &= \frac{1}{3} \sum_i \sum_{\vec{y}} e^{ip \cdot \vec{y}} \langle  \CO^\dagger_{V_i}(\vec{y},t) \tilde{\CO}_{V_i} (\vec{0},t_0) \rangle \,, \label{eq:C_BsStar}
\end{align}
where $\CO_P$ and $\CO_{V_i}$ are the pseudoscalar and vector heavy-strange meson interpolating operators, respectively:
\begin{align}
\CO_{P} = \bar{b} \gamma_5 s \,, \qquad \CO_{V_i} = \bar{b} \gamma_i s \,,
\end{align}
and the index ``$i$" denotes the three spatial directions.  We will explain the meaning of the tilde on some of the operators in Eqs.~(\ref{eq:C_Bs}) and (\ref{eq:C_BsStar}) later in this section.  At sufficiently large times, excited-state contributions to these correlators will die away and the correlators will fall off as an exponential function of the meson ground-state energy exp[$-E(\vec{p})(t-t_0)$].  We can therefore obtain the ground-state energy from the following ratio of correlators:
\begin{align}
E_\textrm{eff}(\vec{p}) = \lim_{t \gg t_0} \textrm{cosh}^{-1} \left[ \frac{C(t,t_0; \vec{p}) + C(t+2,t_0; \vec{p})}{2C(t+1,t_0; \vec{p})} \right] \,,
\label{eq:E_Eff}
\end{align}
which we refer to as the ``effective energy".  In the above equation and throughout the remainder of this work, meson masses and energies are given in lattice units (where the factor of ``$a$" is implied) unless other units ({\it e.g.} GeV) are specified.

We use the Chroma lattice QCD software system to compute the heavy and strange-quark propagators, as well as the 2-point correlation functions~\cite{Edwards:2004sx}.  In order to minimize autocorrelations between data on nearby configurations, we translate the gauge field by a randomly chosen 4-dimensional vector before computing the strange-quark and $b$-quark propagators.  We generate the domain-wall light-quark propagators with a local (point) source; this allows them to be re-used for a future computation of $B$-meson decay constants and mixing matrix elements.  In order to suppress excited-state contamination we generate the $b$-quark propagators with a gauge-invariant Gaussian source for the spatial wavefunction~\cite{Alford:1995dm,Lichtl:2006dt}:
\begin{eqnarray}
	 \tilde{b}(\vec{x},t) & = &  \sum_{\vec{y}} S(\vec{x},\vec{y}; \sigma, N) b(\vec{y},t) \,,
\end{eqnarray}
where the smearing function $S(\vec{x},\vec{y})$ depends upon the width $\sigma$ and the number of smearing iterations $N$:
\begin{align}
	&S(\vec{x},\vec{y}; \sigma, N) = \left( 1 + \frac{\sigma^2}{4N} \nabla^2_{\vec{x},\vec{y}} \right)^N \,, \label{eq:ChromaSmear} \\
	&\nabla^2_{\vec{x},\vec{y}} = \sum_{k=1}^{3} \left( U_k(x) \delta_{\vec{x}+\hat{k},\vec{y}} + U^\dagger_k(\vec{x}-\hat{k})\delta_{\vec{x}-\hat{k},\vec{y}} - 2 \delta_{\vec{x},\vec{y}} \right) \,.
\end{align}
As long as the parameters $(\sigma, N)$ satisfy the criteria $N > 3 \sigma^2 / 2$, the source is spatially smooth and a good approximation to a Gaussian.  For the free-field case ($U=1$) with large $N$ and small $\sigma$, the root-mean-squared (rms) radius $r_\textrm{rms} \approx \sqrt{3} \sigma / 2$ independent of $N$.  Heavy-light meson interpolating operators with a Gaussian-smeared $b$-quark are labeled with a tilde in Eqs.~(\ref{eq:C_Bs}) and (\ref{eq:C_BsStar}).  We use a point sink, however, for both the strange and $b$-quark in the sink meson interpolating field because we find that this source-sink combination minimizes statistical errors in the correlators.  

\begin{figure*}[t]
\centering
\includegraphics[scale=0.58,clip]{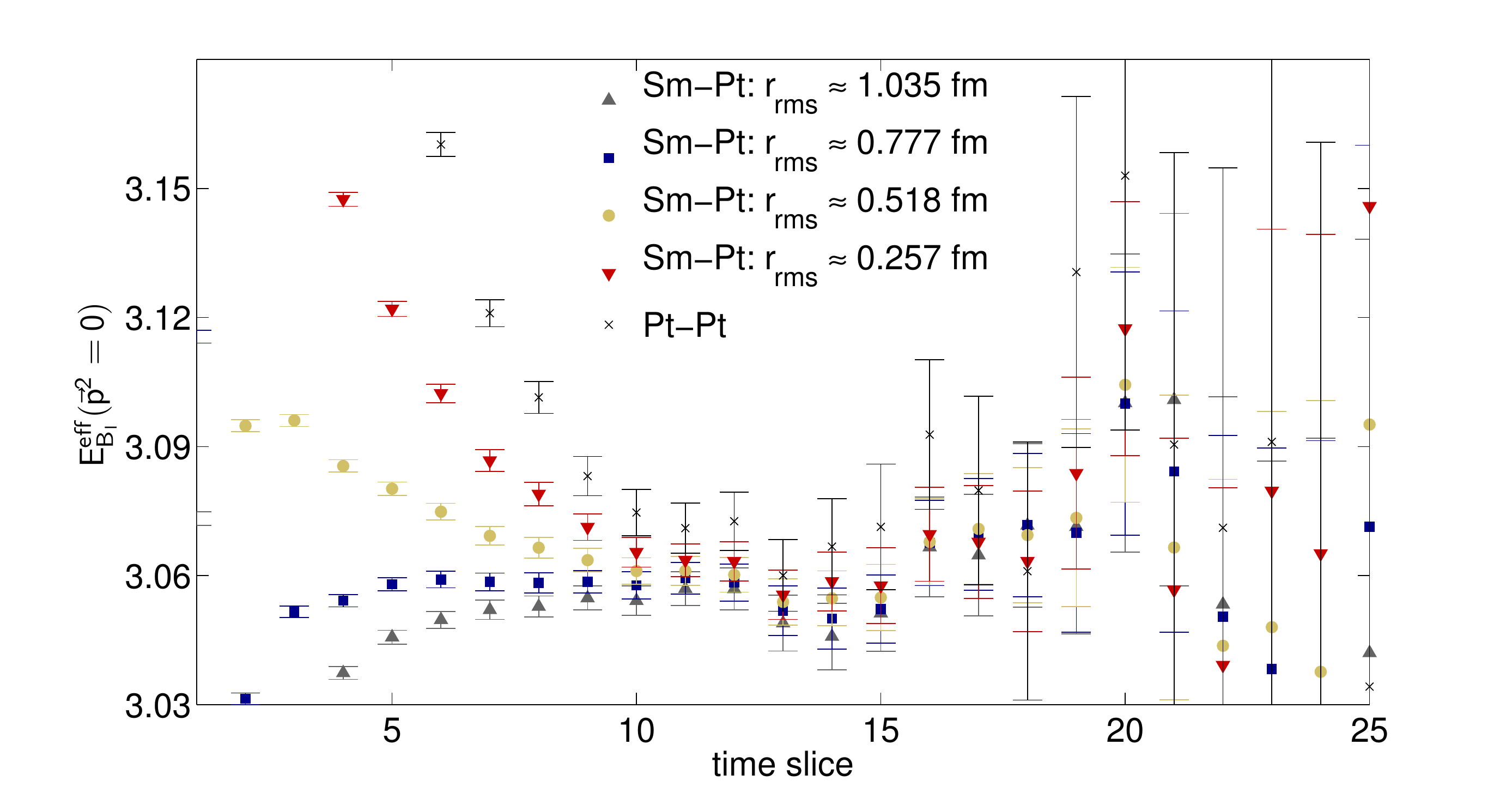}
\caption{Pseudoscalar meson effective mass for several choices for the Gaussian radius of the $b$-quark in the heavy-light meson interpolating operator.  Results are shown for the unitary point on the $am_l = 0.005$ $24^3$ ensemble with RHQ parameters $\{ m_0 a, c_P, \zeta \} = \{ 7.38, 3.89, 4.19\}$.}
\label{fig:24cSmear}
\end{figure*}

\begin{figure*}[t]
\centering
\includegraphics[scale=0.58,clip]{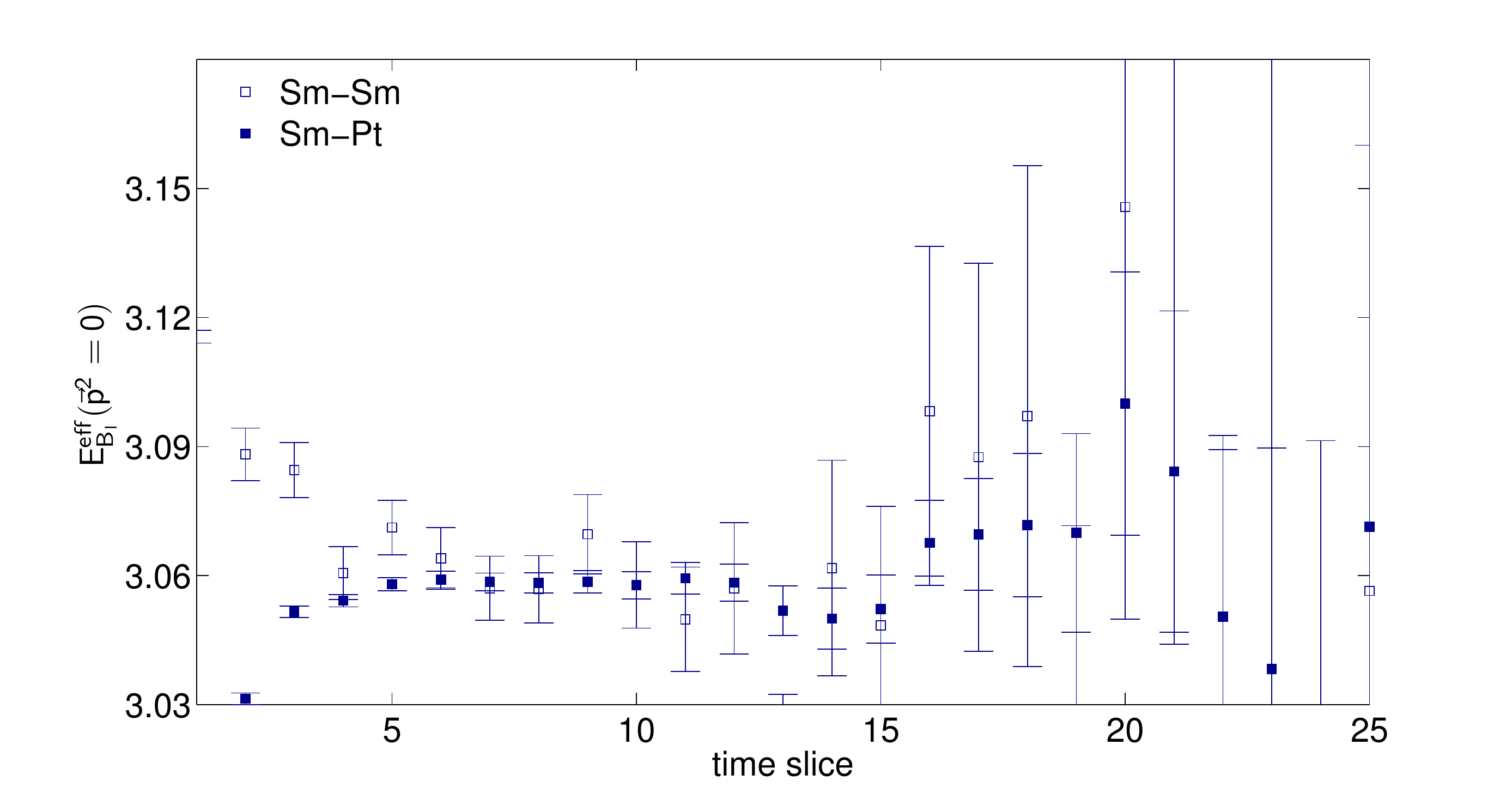}
\caption{Pseudoscalar meson effective mass for the $b$-quark Gaussian radius $r_\textrm{rms} = \textcolor{black}{0.777}$~fm.  The full symbols correspond to correlators in which the $b$-quark is generated with a Gaussian spatial wavefunction but has a point sink;  the open points correspond to correlators in which the $b$-quark has a Gaussian spatial wavefunction at both the source and sink.  The effective masses agree, but the smeared-point data has smaller statistical errors.}
\label{fig:SPvsSS}
\end{figure*}

Before beginning the iterative procedure to tune the RHQ parameters described in Sec.~\ref{sec:Method} we compute the zero-momentum heavy-light meson pseudoscalar correlator [Eq.~(\ref{eq:C_Bs}) with $B_s \to B_l$] for several values of the Gaussian radius; these are given in Table~\ref{tab:GaussParams}.  Because we expect both the light-quark and $b$-quark mass-dependence of the optimal smearing choice to be mild, for each lattice spacing we analyze data on a single sea-quark ensemble and with a single light-quark mass and set of RHQ parameters $\{ m_0 a, c_P, \zeta \}$.  For the smearing study on the $24^3$ ensembles we use the preliminary results for the RHQ parameters in the chiral limit from Ref.~\cite{MinPhDThesis}, $\{ m_0 a, c_P, \zeta \} = \{ 7.38, 3.89, 4.19\}$, which are similar to the earlier values presented at Lattice 2008~\cite{Li:2008kb}.  We analyze the unitary point on the $am_l = 0.005$ ensemble.  Figure~\ref{fig:24cSmear} shows the heavy-light pseudoscalar meson effective mass [$E_\textrm{eff}(\vec{p} = 0)$] for several choices of the Gaussian radius (including the limit of a point source).  The correlator generated with a $b$-quark spatial wavefunction with a root-mean-squared (rms) radius of $r_\textrm{rms} = 0.777$~fm clearly has the longest plateau with the earliest onset;  we therefore choose to use this spatial wavefunction for the RHQ parameter tuning on the $24^3$ ensembles.  One might worry that the extremely long plateau in Fig.~\ref{fig:24cSmear} is due to cancellations between excited states with positive and negative amplitudes, and does not correspond to the true ground-state mass.  Figure~\ref{fig:SPvsSS} therefore shows a comparison of the pseudoscalar meson effective mass in which the $b$-quark  has a smeared source and point sink and one in which the $b$-quark has both a smeared source and sink.  The two effective masses agree within statistical errors, suggesting that we have obtained the true plateau.  

\begin{table}
\caption{Root-mean-squared radii and corresponding Chroma Gaussian smearing parameters [defined in Eq.~(\ref{eq:ChromaSmear})] considered here.  The parameters shown in bold are used to obtain the RHQ parameters in the following subsection.}
\vspace{3mm}
  \begin{tabular}{lccccc}\hline\hline
    &  \multicolumn{2}{c}{$a\approx$~0.086 fm} &~~~& \multicolumn{2}{c}{$a\approx$~0.11 fm}\\
     $r_\textrm{rms}$~(fm)~~ & $\sigma$ & $N$ && $\sigma$ & $N$   \\[0.5mm] \hline
     0.137 & 1.39 & 5 && 1.83 & 5 \\
     0.275 & 2.78 & 15 && 3.6 & 25 \\
     0.518 & 5.24 & 5 && 6.92 & 80 \\
     \bf{0.777} & \bf{7.86 }& \bf{100} && \bf{10.36} & \bf{170} \\
     1.035 & 10.48 & 175 \\
     1.047 & & && 13.98 & 310 \\\hline\hline
  \end{tabular}
  \label{tab:GaussParams}
\end{table}


For the smearing study on the $32^3$ ensemble we analyze the unitary point on the $am_l = 0.004$ sea-quark ensemble.  We use the RHQ parameters $\{ m_0 a, c_P, \zeta \} = \{ 3.70, 3.60, 2.20 \}$, which are close to the preliminary results on the $am_l = 0.004$ ensemble in Ref.~\cite{Peng:Lattice10}.  As in the case of the $24^3$ ensembles, the Gaussian radius of $r_\textrm{rms} = 0.777$~fm leads to the best plateau, so we use it for the RHQ parameter tuning procedure.  This is consistent with expectations that the size of the $B$-meson in physical units should be independent of the lattice spacing.


We estimate the errors in the correlation functions and in the fitted meson energies using a single-elimination jackknife procedure.  This allows us to propagate the statistical uncertainties including correlations between the parameters $\{m_0a, c_p, \zeta\}$ into subsequent steps of the RHQ parameter tuning procedure.  We find no evidence of residual autocorrelations between subsequent trajectories, as measured by comparing the errors between binned and un-binned data.  We perform the $\chi$-squared minimization including the full covariance matrix, and choose fit ranges that yield acceptable correlated confidence levels ($p$-value\footnote{We adopt the PDG convention that the $p$-value is the probability of finding a $\chi^2$ value greater than that obtained in the fit;  hence a larger $p$-value denotes a stronger compatibility between the data and the fit hypothesis~\cite{Nakamura:2010zzi}.}~$\gtapprox$~10\%). 

\begin{figure*}[t]
\centering
\includegraphics[scale=0.47,clip]{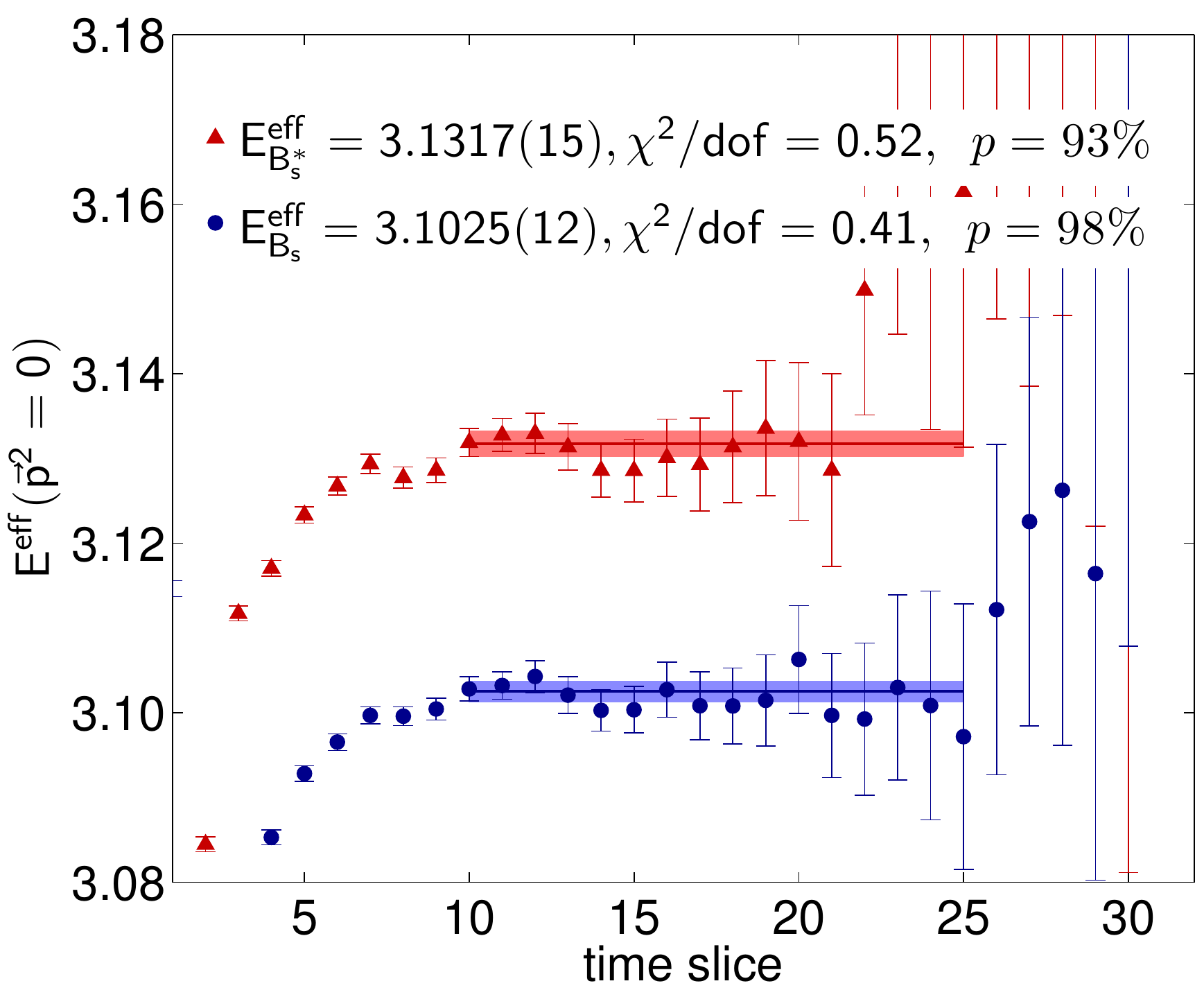}~\includegraphics[scale=0.47,clip]{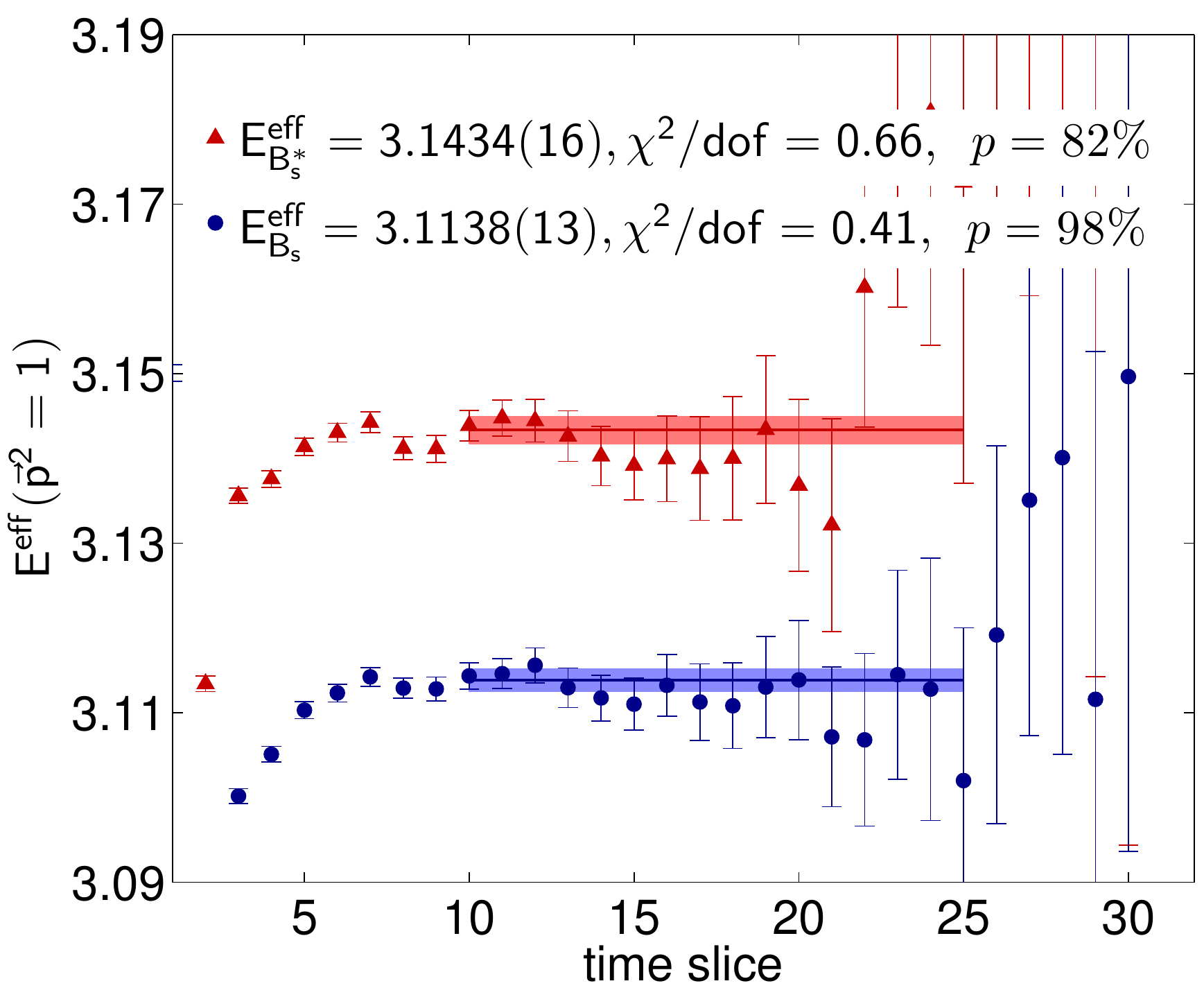}
\includegraphics[scale=0.47,clip]{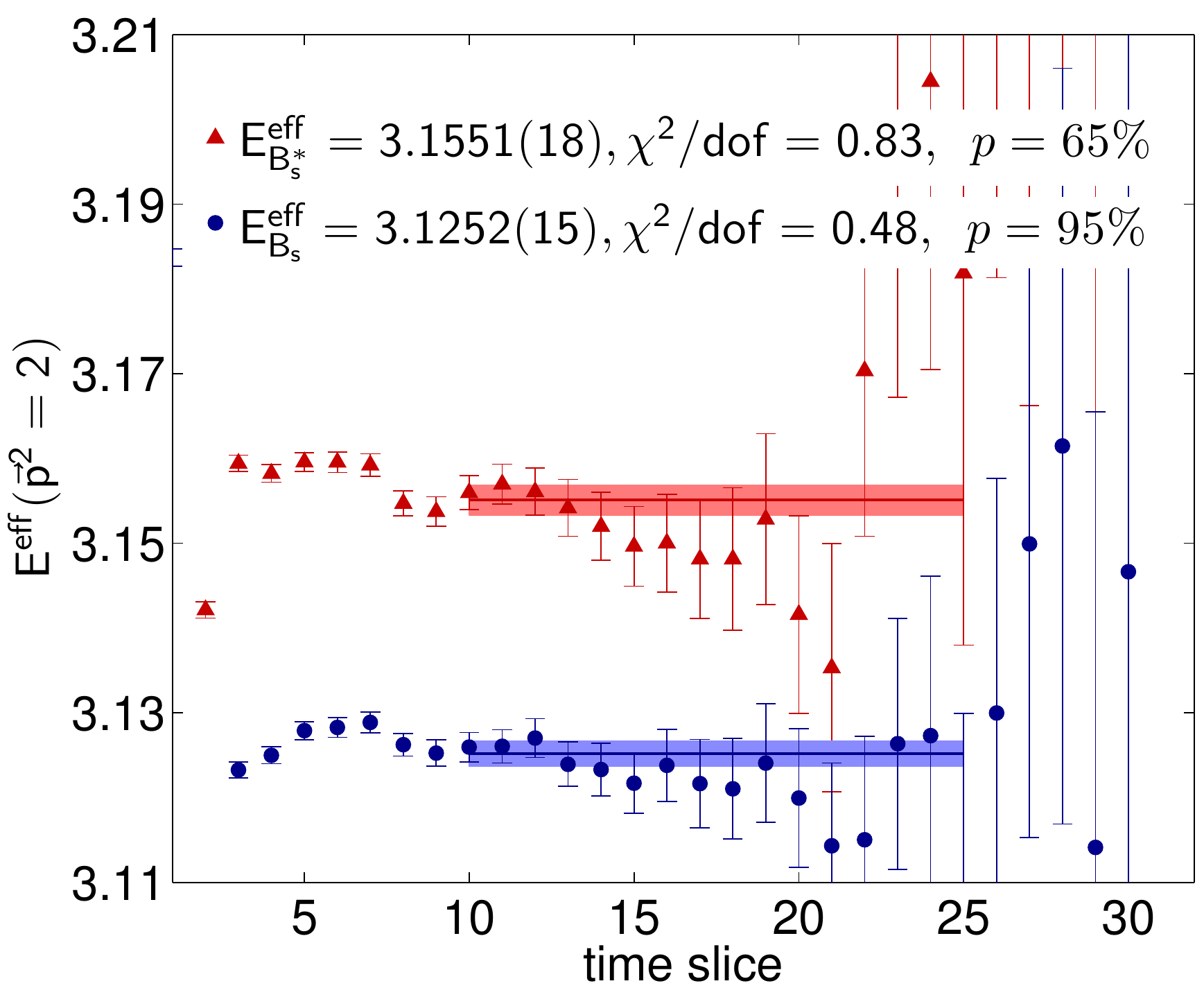}~\includegraphics[scale=0.47,clip]{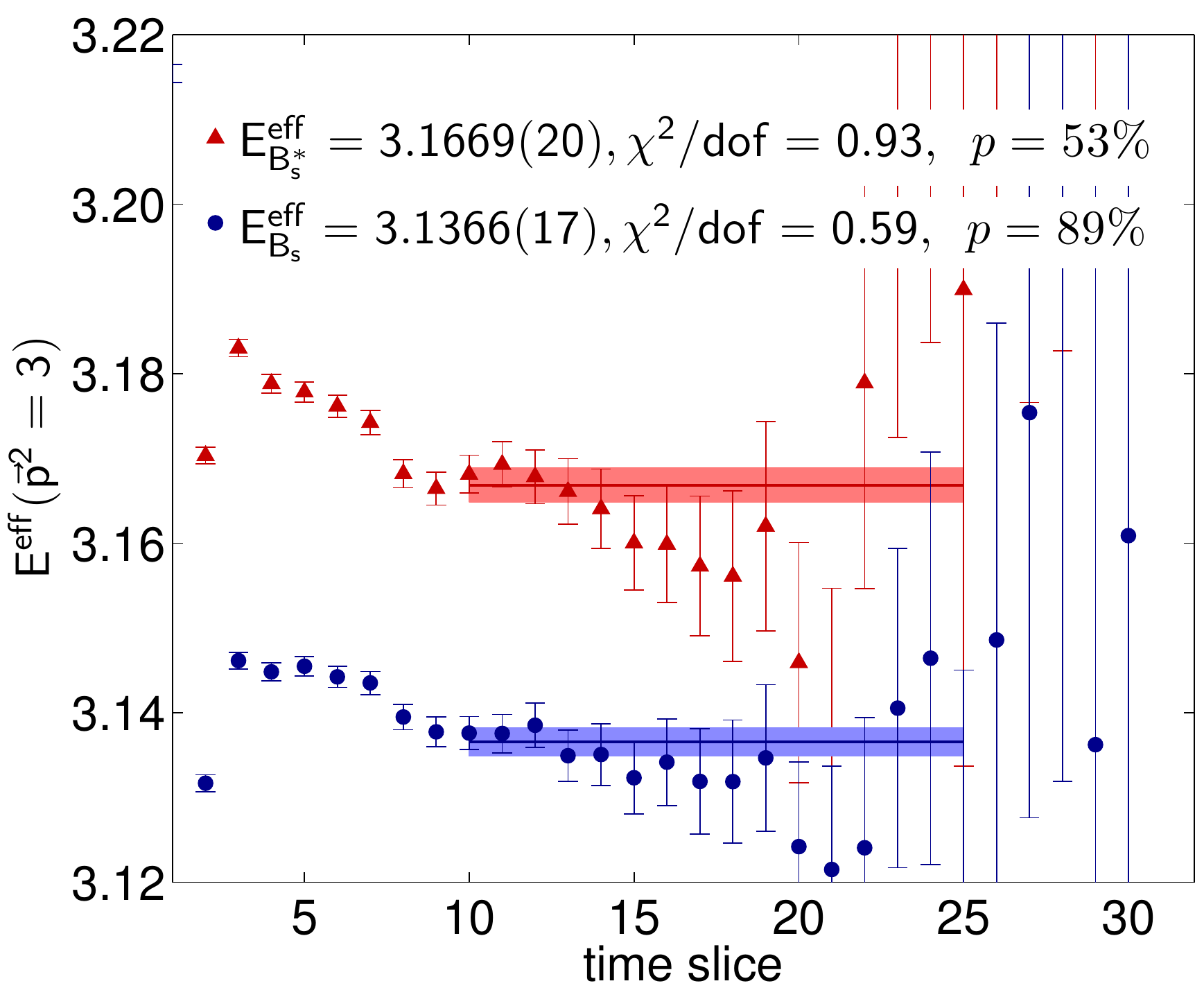}
\caption{Heavy-strange pseudoscalar meson (blue circles) and vector meson (red triangles) effective energies on the $am_l = 0.005$ $24^3$ ensemble with RHQ parameters $\{m_0a, c_P, \zeta\}=\{8.40,5.80,3.20\}$.  From upper-left to lower-right the six plots show spatial momenta $(a \vec{p})^2 = 0$ through $(a\vec{p})^2=3$. For each plot the shaded horizontal band shows the fit range used and the fit result with jackknife statistical errors.}
\label{fig:243_Bs_BsStar}
\end{figure*}

Because the $B_s$ and $B_s^*$ meson energies are largely insensitive to the sea-quark mass, we expect the excited-state contamination to die off and the onset of the ground-state plateau to occur at around the same location on all sea-quark ensembles for a given lattice spacing.  We therefore choose the same fit range for all sea-quark ensembles on a given lattice spacing.  The requirement that we obtain a constant fit to the effective energy with a good correlated confidence level using the same fitting range for all ensembles helps to ensure that we obtain the true ground-state energy, and are not misled by ``wiggles" in the plateau that are due to fluctuations in the gauge field, but are different on each ensemble.  We do not, however, expect excited-state contributions to be the same for all momenta, and, in fact, we observe an earlier onset for the plateau in the zero momentum effective energy than for the other momenta.  Table~\ref{tab:BsFitRange} shows the fitting ranges used on the $24^3$ and $32^3$ ensembles.  Figure~\ref{fig:243_Bs_BsStar} shows the $B_s$ and $B_s^*$ meson effective energies for lattice momenta up to $(a \vec{p})^2 = 3$ on the $am_l = 0.005$ $24^3$ ensemble. Effective energy plots for the other $24^3$ and $32^3$ ensembles look similar. 

\begin{table}
\caption{Time ranges used in plateau fits of the $B_s$ and $B_s^*$ effective energies.  We use different ranges for zero and nonzero momenta, but use the same range for all sea-quark masses at a given lattice spacing.}
\vspace{3mm}
  \begin{tabular}{lccc} \hline\hline
     & \multicolumn{3}{c}{fit range} \\
     & $\vec{p} = 0$ & \qquad & $\vec{p} \neq 0$  \\[0.5mm] \hline
   $a \approx 0.11$ fm~ & [10,25] & & [10,25]  \\
   $a \approx 0.086$ fm~ & [11,21] & & [14,21] \\ \hline\hline
  \end{tabular}
  \label{tab:BsFitRange}
\end{table}

\subsection{Determination of bottom-quark parameters}
\label{sec:RHQParams}

We begin our iterative tuning procedure using the preliminary values for $\{m_0 a, c_P, \zeta \}$ determined in the pilot studies of Refs.~\cite{MinPhDThesis} and~\cite{Peng:Lattice10}.  We compute the $B_s$ and $B_s^*$ meson energies for seven sets of parameters surrounding these values.  We then determine the ratio of the rest mass to the kinetic mass for these seven parameter sets by fitting the nonzero momentum data for the $B_s$ meson to the energy-momentum dispersion relation, Eq.~(\ref{eq:DispRel}).  Finally, we determine the predicted values of the RHQ parameters from Eq.~(\ref{Eq:RHQDetermination}) using the experimentally-measured meson masses  $M_{B_s} = 5.366$~GeV and $M_{B_s^*} = 5.415$~GeV~\cite{Nakamura:2010zzi}.  We find that the resulting values of $\{m_0 a, c_P, \zeta \}$ lie outside the ``box" determined by the seven parameter sets.  We therefore re-center the box around the newly-determined values and repeat the procedure.  We find that we need to iterate once or twice before the values of $\{m_0 a, c_P, \zeta \}$ settle down and remain inside the box.  Here we only show results for the final iteration, since plots for intermediate iterations look similar. The final sets of parameters used to obtain the tuned values of $\{m_0 a, c_P, \zeta \}$ on the $24^3$ and $32^3$ ensembles are given in Table~\ref{tab:BoxParams}.

\begin{table}[t]
\caption{Final ``box" of parameters used to obtain the tuned values of $\{m_0 a, c_P, \zeta \}$ (see Fig.~\ref{fig:RHQBox}).  In each column the first number is the central value of the parameter and the second number is the variation.}
\label{tab:BoxParams}
\begin{tabular}{cccc}
\hline\hline
& $m_0a$ & $c_P$ & $\zeta$ \\\hline
$a \approx 0.11$ fm~~ & 8.40 $\pm$ 0.15& 5.80 $\pm$ 0.45 & 3.20 $\pm$ 0.30 \\
$a \approx 0.086$ fm~~ & 3.98 $\pm$ 0.10& 3.60 $\pm$ 0.30 & 1.97 $\pm$ 0.15 \\
\hline\hline
\end{tabular}
\end{table}

\begin{figure}[t]
\centering
\includegraphics[scale=0.5,clip]{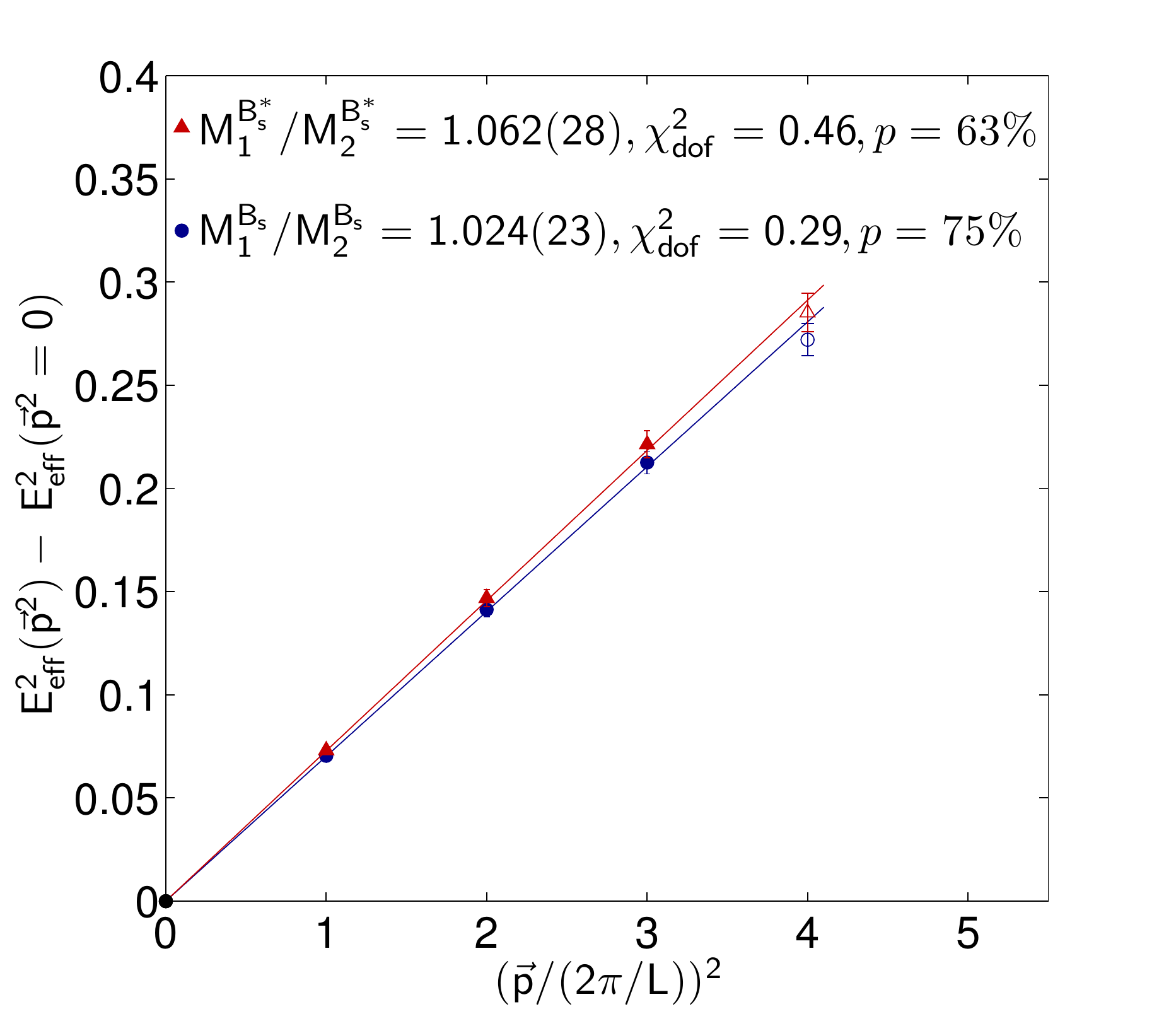}
\caption{$B_s$ (blue circles) and $B_s^*$ (red triangles) meson squared-energy difference versus spatial momentum-squared on the $am_l = 0.005$ $24^3$ ensemble for the RHQ parameter values $\{m_0 a, c_P, \zeta \} = {\{8.40, 5.80, 3.20 \}}$.  The slope of the data gives the ratio of the meson rest mass over the kinetic mass $(M_1/M_2)$. Data points shown with open symbol are not included in the fit.}
\label{fig:243_Disp}
\end{figure}

Figure~\ref{fig:243_Disp} shows the energy-momentum dispersion relation fit for both the $B_s$ and $B_s^*$ mesons on the $am_l = 0.005$ $24^3$ ensemble.  Dispersion relation plots for the other sea-quark masses, RHQ parameter sets, and lattice spacing look similar.  The slopes ($M_1/M_2$) of the $B_s$ and $B_s^*$ energy-momentum dispersion relations agree with unity (and hence with each other) within errors in the region of the parameter space near the tuned values of $\{m_0 a, c_P, \zeta \}$.  We choose to use the pseudoscalar meson data, however, for the parameter tuning because it has smaller statistical errors.  We perform a one parameter linear fit in which we fix the intercept to go through the measured value of the rest mass $E(\vec{p}=0)$ and allow the slope to vary.  We include data with lattice momenta through $(ap)^2 = 3$, and see no evidence for higher-order, e.g. $\CO([ap]^4)$, lattice discretization effects at these values of the momenta.  We account for correlations between data points by propagating the jackknife values of the energies from the 2-point fits described in the previous subsection.    As a cross-check we compare the fit result with those of a two-parameter fit in which we allow both the slope and intercept to vary; we find that the results are consistent, and choose to use the one-parameter fit because it leads to smaller statistical errors in $M_1^{B_s}/M_2^{B_s}$. 

In order to reliably determine the RHQ parameters via Eq.~(\ref{Eq:RHQDetermination}) we must be interpolating in a regime in which the bottom-strange meson observables $\{ \bar{M}_{B_s}, \Delta M_{B_s}, M_1^{B_s}/M_2^{B_s}\}$ depend linearly upon the parameters in the action $\{m_0 a, c_P, \zeta \}$.  We test this assumption and look for signs of curvature by computing the observables for three different boxes of seven parameters with sizes $\pm \sigma_{\{m_0 a, c_P, \zeta \}}$, $\pm 2\sigma_{\{m_0 a, c_P, \zeta \}}$, and $\pm 3\sigma_{\{m_0 a, c_P, \zeta \}}$ (except for the parameter $m_0a$ on the $24^3$ ensemble for which the largest box is $\pm 4\sigma_{m_0 a}$).  We then determine the predicted values of the RHQ parameters for each of the three boxes;  we find that the difference is negligible within statistical errors. 

\begin{figure}[p]
\centering
\includegraphics[scale=0.47,clip]{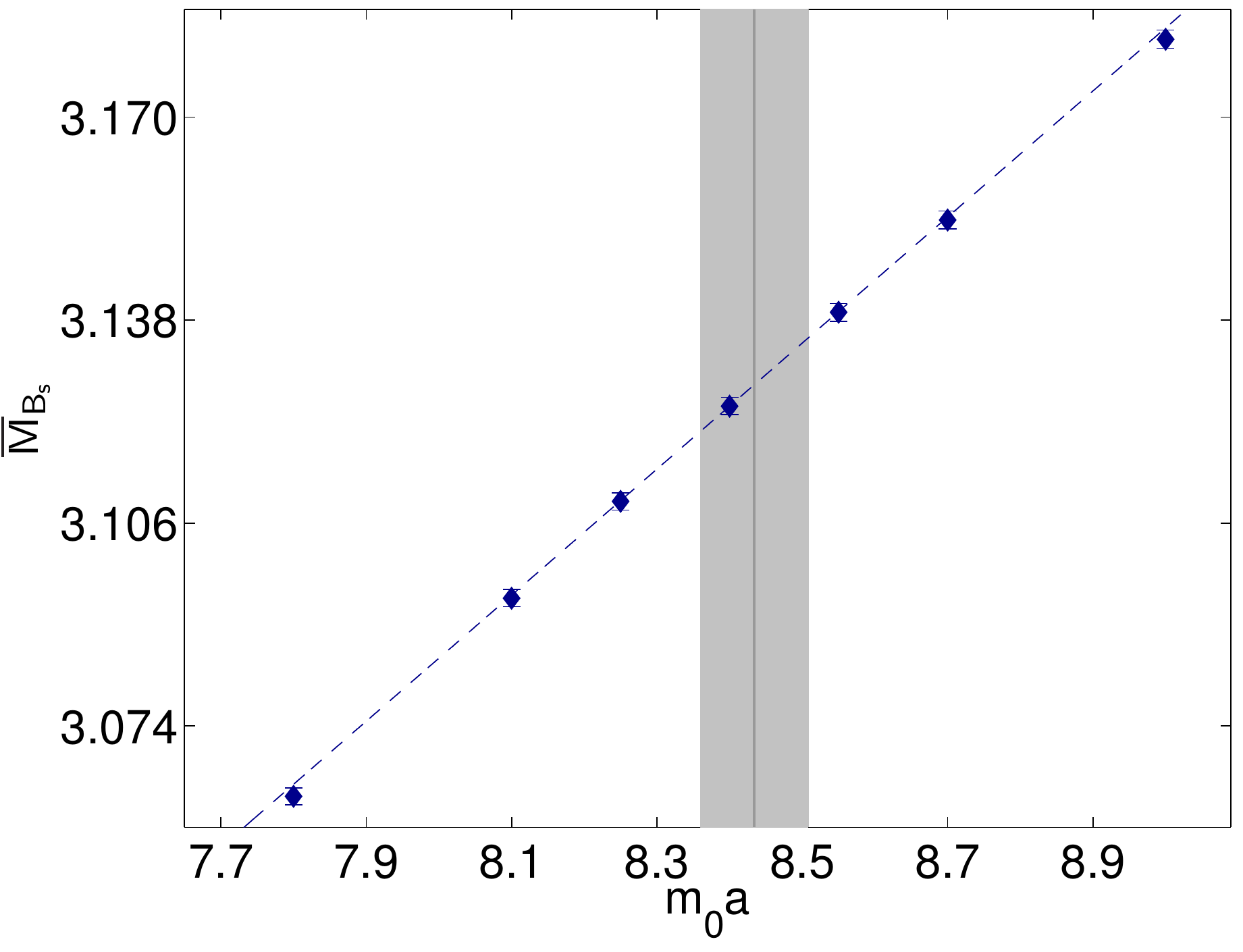}
\includegraphics[scale=0.47,clip]{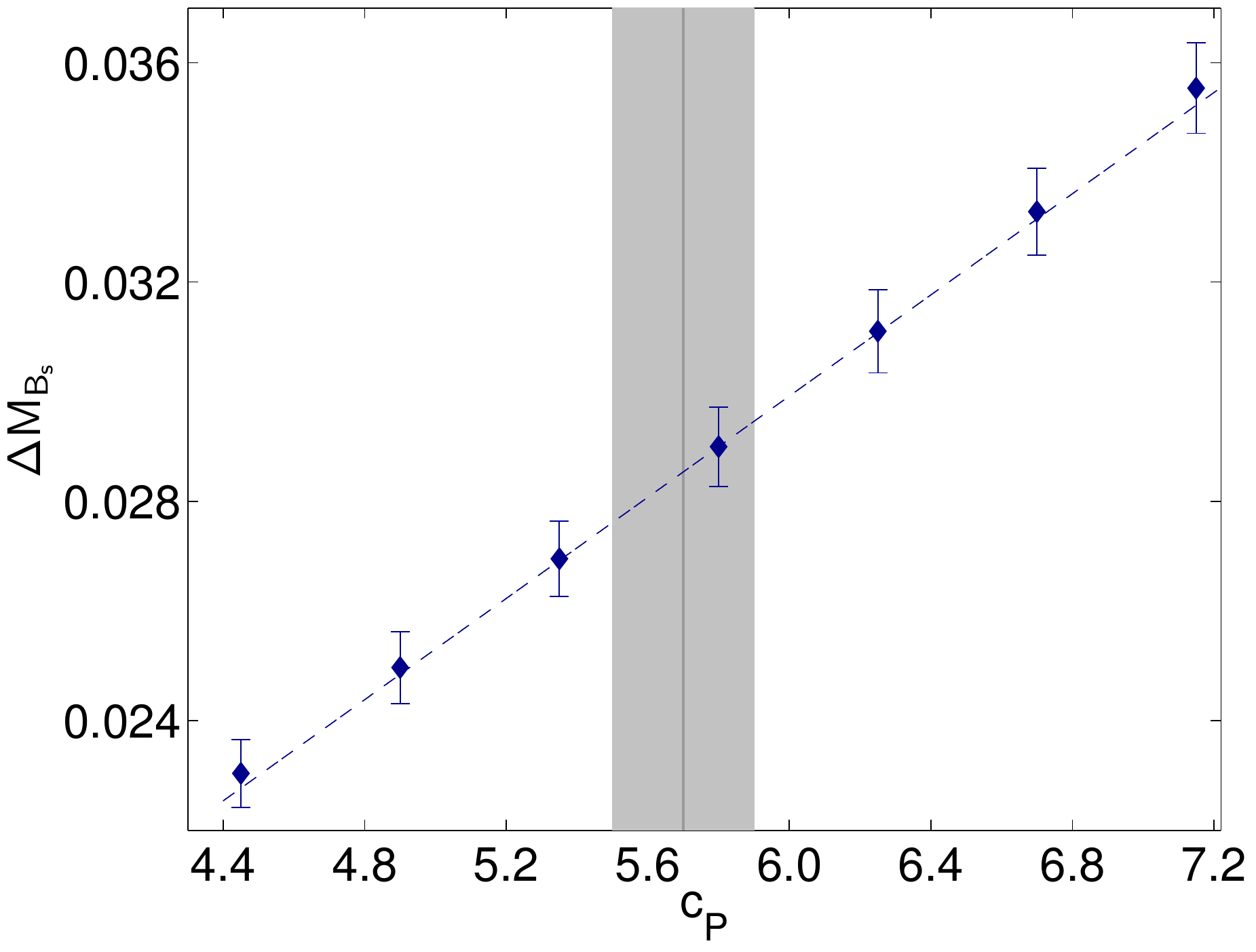}
\includegraphics[scale=0.47,clip]{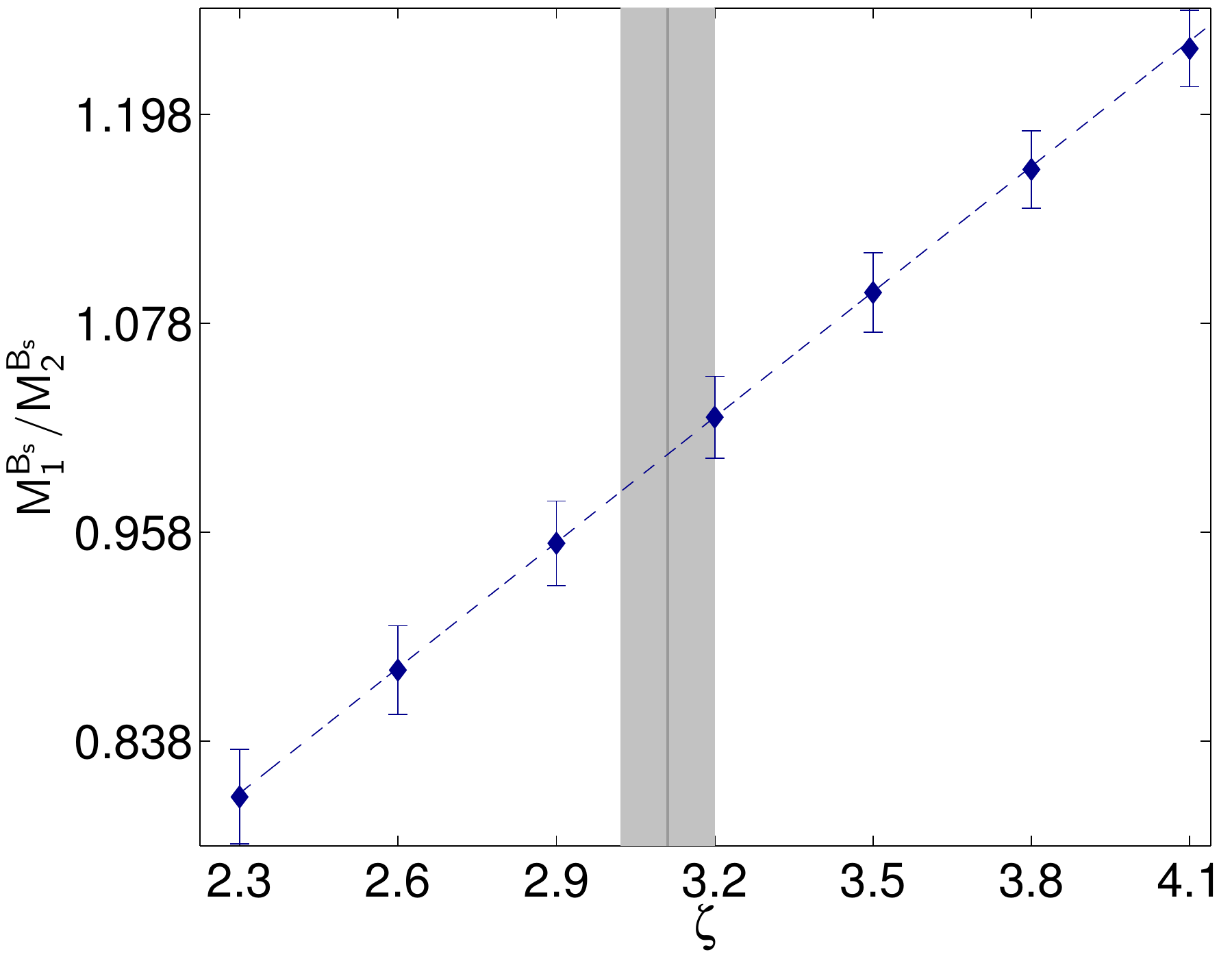}
\caption{Spin-averaged mass versus $m_0a$ (upper plot), hyperfine splitting versus $c_P$ (center plot), and rest mass over kinetic mass versus $\zeta$ (lower plot) on the $am_l = 0.005$ $24^3$ ensemble.  The solid vertical lines with shaded gray error bands denote the tuned values of the RHQ parameters with jackknife statistical errors.  For each quantity, the dashed line shows the dependence on $m_0a$, $c_P$, or $\zeta$ calculated from Eqs.~(\ref{eq:linear_approx})--(\ref{eq:Yi}).}
\label{fig:243_LinTest}
\end{figure}

Figure~\ref{fig:243_LinTest} shows the dependence of the spin-averaged mass, hyperfine splitting, and rest mass over kinetic mass on the parameters $m_0 a$, $c_P$, and $\zeta$, respectively, on the $am_l = 0.005$ $24^3$ ensemble.  We plot these dependencies because these are the parameters to which each observable is most sensitive.  The bottom-strange observables $\{ \bar{M}_{B_s}, \Delta M_{B_s}, M_1^{B_s}/M_2^{B_s}\}$ depend linearly on the parameters  $\{m_0 a, c_P, \zeta \}$ throughout the range.  The analogous plots for the other sea-quark ensembles look similar.
 
Table~\ref{tab:243_RHQParams} shows the nonperturbatively-tuned RHQ parameters $\{m_0 a, c_P, \zeta \}$ obtained on the two $24^3$ ensembles.  We do not observe any statistically-significant sea-quark mass dependence.  Hence, instead of extrapolating the RHQ parameters to the physical light-quark masses, we simply take an error-weighted average of the two values to obtain our final preferred results.  Similarly, Table~\ref{tab:323_RHQParams} shows the nonperturbatively-tuned RHQ parameters on the three $32^3$ ensembles and the corresponding weighted average.  

\begin{table}
\caption{Tuned RHQ parameter values on the $24^3$ ensembles determined using the parameter sets in Table~\ref{tab:BoxParams}.  Because we do not observe any statistically-significant sea-quark mass dependence, we obtain the final preferred values from an error-weighted average of the two sets of results.}
\vspace{3mm}
  \begin{tabular}{lccc} \hline\hline
  & $m_0 a$ & $c_P$ & $\zeta$  \\[0.5mm] \hline 
    $am_l = 0.005$~ & 8.43(7) & 5.7(2) & 3.11(9)  \\
    $am_l = 0.01$  & 8.47(9) & 5.8(2) & 3.1(1) \\\hline
    average & 8.45(6) & 5.8(1) & 3.10(7) \\ \hline\hline
  \end{tabular}
  \label{tab:243_RHQParams}
\end{table}

\begin{table}
\caption{Tuned RHQ parameter values on the $32^3$ ensembles determined using the parameter sets in Table~\ref{tab:BoxParams}.  Because we do not observe any statistically-significant sea-quark mass dependence, we obtain the final preferred values from an error-weighted average of the three sets of results.}
\vspace{3mm}
  \begin{tabular}{lccc} \hline\hline
  & $m_0 a$ & $c_P$ & $\zeta$  \\[0.5mm] \hline 
    $am_l = 0.004$~ & 4.07(6) & 3.7(1) & 1.86(8)   \\
    $am_l = 0.006$  & 3.97(5) & 3.5(1) & 1.94(6) \\
    $am_l = 0.008$  & 3.95(6) & 3.6(1) & 1.99(8) \\\hline
    average & 3.99(3) & 3.57(7) & 1.93(4) \\ \hline\hline
  \end{tabular}
  \label{tab:323_RHQParams}
\end{table}

\subsection{Comparison with perturbation theory}
\label{sec:LPT}

It is useful to compare the nonperturbatively-determined values of the RHQ parameters with those computed in lattice perturbation theory.  First, this provides a consistency check of the nonperturbative tuning procedure.  Second, this allows us to see how well the perturbative estimates are working in a case where we know the true nonperturbative value.  Reasonable agreement between the two approaches bolsters confidence in our ability to rely on lattice perturbation theory in future situations where we do not have nonperturbative matching factors available. 

We calculate the RHQ parameters $c_P$ and $\zeta$ at 1-loop in mean-field improved lattice perturbation theory~\cite{Lepage:1992xa}.  The details of the perturbative calculation will be given in a separate publication~\cite{LehnerLPT}.  The clover coefficient $c_P$ is obtained by matching the lattice quark-gluon vertex to the continuum counterpart in the on-shell limit.  At intermediate steps of the calculation infrared divergences are regulated with a nonzero gluon mass $\lambda$; the final results are obtained in the limit $\lambda\to0$.  Similarly, the anisotropy parameter $\zeta$ is obtained by requiring that the lattice heavy-quark dispersion relation, extracted from the momentum dependence of the pole in the heavy-quark propagator at one loop, agrees with the continuum.  We implement the mean-field improvement in two ways.  First we use the nonperturbative value of the fourth root of the plaquette, $u_0 = P^{1/4}$, to resum tadpole contributions as in Ref.~\cite{Lepage:1992xa}.   We also use the value of the spatial link field in Landau gauge to estimate $u_0$.  A comparison of these two approaches is useful for ascertaining the systematic uncertainty due to the ambiguity in how to implement the tadpole resummation.  The lattice perturbation theory calculations of $c_P$ and $\zeta$ also use the nonperturbatively-determined values of the bare-quark mass $m_0a$ and the $2\times 1$ rectangle $R$ as inputs.   The latter allows for a refined resummation of tadpole contributions in improved gauge actions~\cite{Ali_Khan:2001tx}.    

\begin{figure*}[t]
\centering
\includegraphics[scale=0.68,clip]{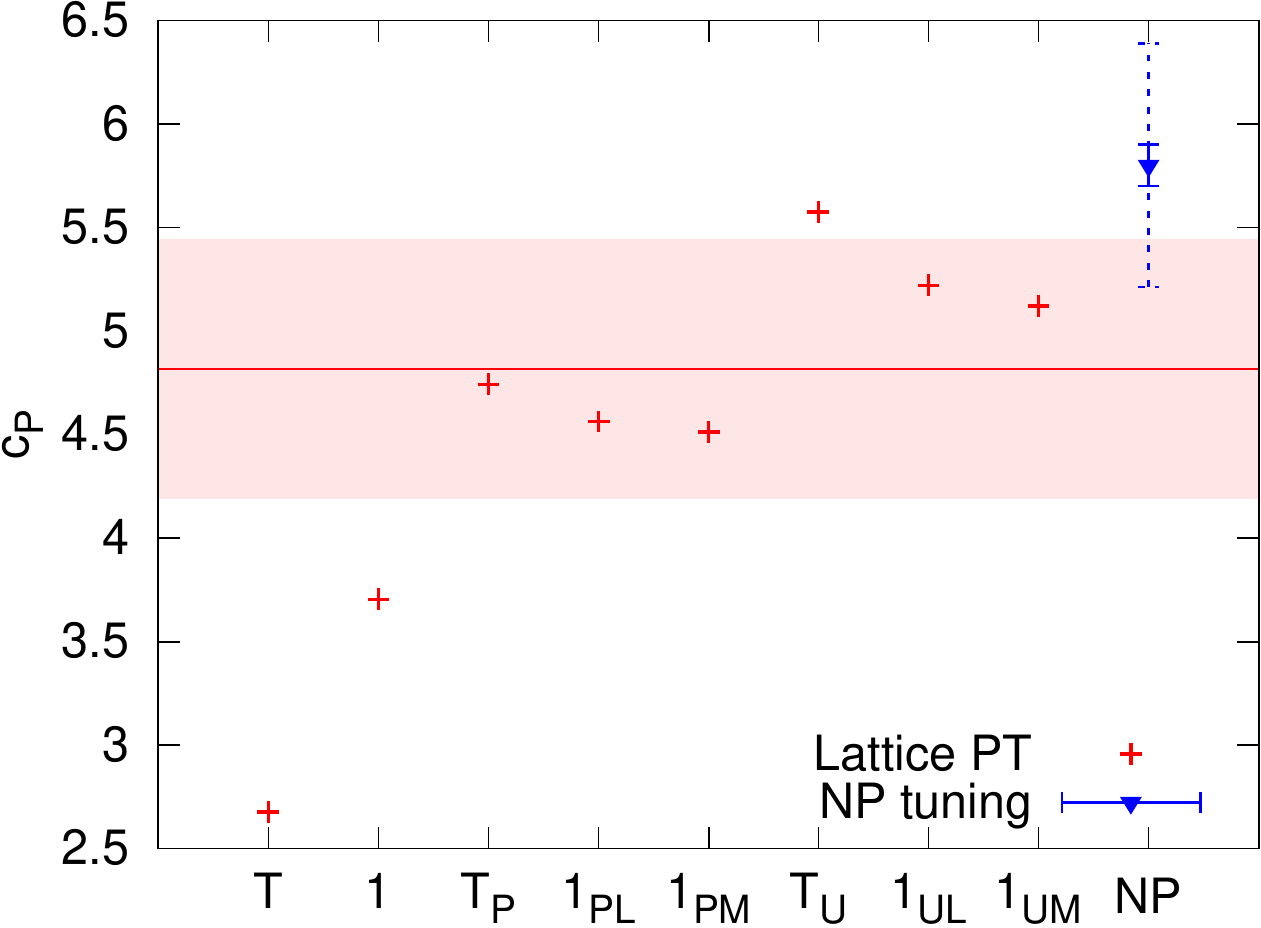}
\includegraphics[scale=0.68,clip]{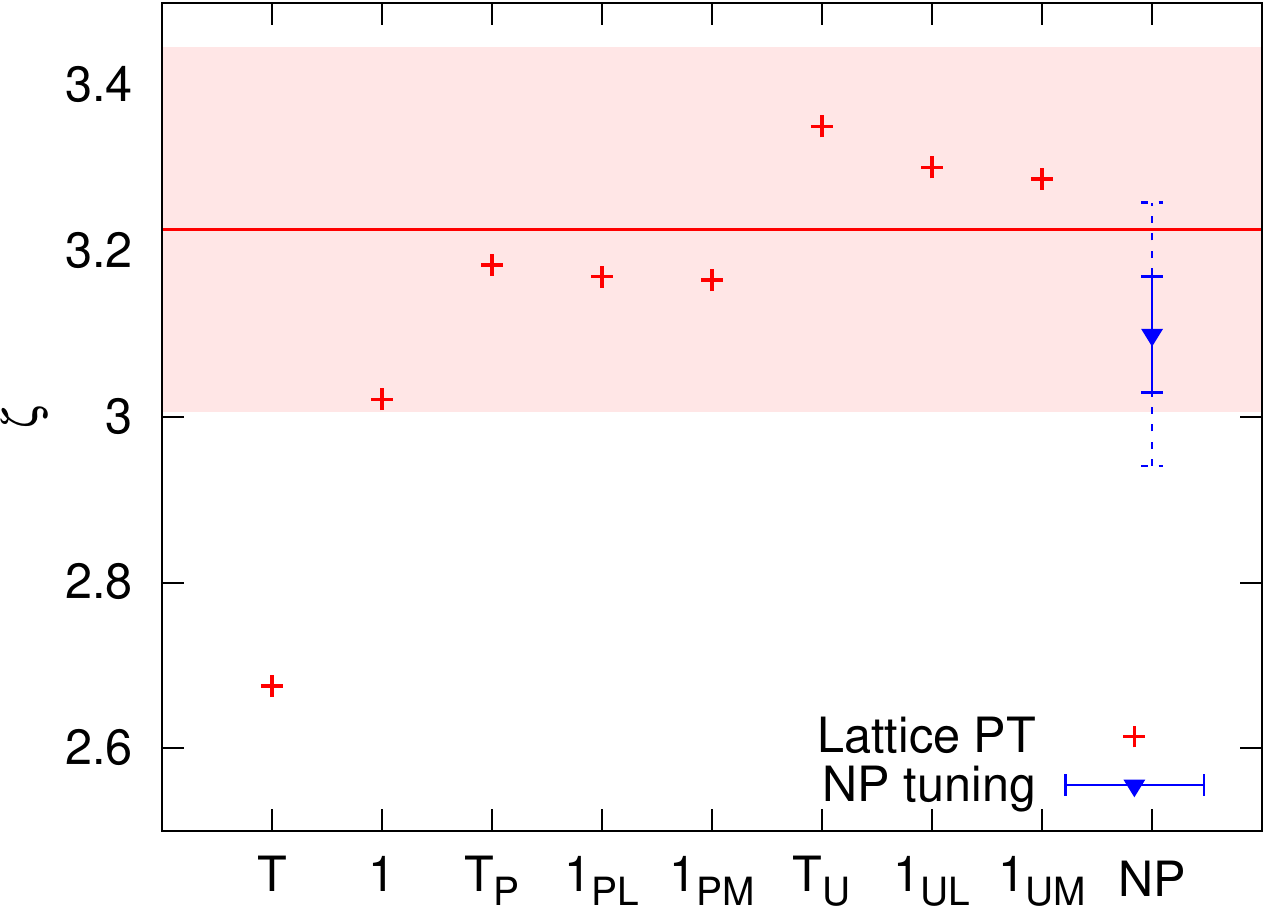}
\caption{Lattice perturbation theory calculations of $c_P$ (left plot) and $\zeta$ (right plot) on the $24^3$ ensembles~\cite{LehnerLPT}.  From left to right, the perturbative calculations shown are (T) unimproved tree-level, (1) unimproved 1-loop, $({\rm T}_{\rm P})$ mean-field improved tree-level using the plaquette to estimate the tadpole factor $u_0$, ($1_{\rm PL}$), 1-loop mean-field improved value using the lattice coupling and the plaquette, ($1_{\rm PM}$) 1-loop mean-field improved value using the $\overline{{\rm MS}}$ coupling and the plaquette, (${\rm T_U}$), mean-field improved tree-level using the spatial link in Landau gauge to estimate $u_0$, ($1_{\rm UL}$), 1-loop mean-field improved value using the lattice coupling and the spatial link, and ($1_{\rm UM}$) 1-loop mean-field improved value using the $\overline{{\rm MS}}$ coupling and the spatial Landau link.  In each plot, the horizontal line indicates our choice of central value for $c_P$ or $\zeta$ while the solid horizontal band denotes our estimate of the uncertainty with errors due to the truncation of the perturbative series and errors due to the uncertainty in $m_0a$ added in quadrature.  For comparison, the nonperturbatively-determined values are shown at the far right with statistical errors (solid inner error bar) and statistical and systematic errors added in quadrature (dashed outer error bar).}
\label{fig:24c_LPT}
\end{figure*}

Figure~\ref{fig:24c_LPT} compares results on the $24^3$ ensembles in both unimproved and mean-field improved lattice perturbation theory with the nonperturbatively-determined values.  The results on the $32^3$ ensembles look qualitatively similar.  The use of mean-field improved lattice perturbation theory brings the perturbative results into better agreement with the nonperturbative values.  It also reduces the size of the one-loop corrections, thereby appearing to improve the convergence of the perturbative series, although one cannot be entirely sure that this trend persists to higher orders.   In the case of $c_P$, the unimproved one-loop corrections are very large (approximately a factor of 1.5) but are reduced to a more sensible level by resumming tadpole contributions, whereas in the case of $\zeta$ the unimproved one-loop corrections are already close to the na\"ive power-counting estimate of $\alpha_S^{\bar{\rm MS}}(1/a_{24^3}) \sim 23\%$ and the mean-field improved one-loop correctons are even smaller.

We can use the results shown in Fig.~\ref{fig:24c_LPT} to estimate the uncertainties in the values of $c_P$ and $\zeta$ calculated in lattice perturbation theory.  We consider two approaches for obtaining the error.  A na\"ive power-counting estimate of the size of the neglected 2-loop corrections would lead to a predicted error of $\alpha_S^2 \sim 5\%$.  As mentioned earlier, however, there is an ambiguity in how to estimate the tadpole factor $u_0$ used in the resummation procedure.  This is not strictly a measure of the size of higher-order corrections, but taking the difference between the values of $c_P$ and $\zeta$ computed at one-loop using $u_0$ from the plaquette and from the spatial Landau link gives a larger estimate of the error in $c_P$ ($\sim$10--12.5\%) than the na\"ive power-counting approach.  We therefore take this difference to be the error in the perturbatively-calculated value of $c_P$, but take $\alpha_S^2 \sim 5\%$ to be the error in the perturbatively-calculated value of $\zeta$.  For the central values we quote the average of the one-loop mean-field improved values expanded in the $\overline{{\rm MS}}$ coupling at scale $a^{-1}$ and computed with $u_0$ obtained from the plaquette and from the spatial Landau link.    

Our final perturbative estimates for $c_P$ and $\zeta$ on the $24^3$ and $32^3$ ensembles are given in Table~\ref{tab:LPT_RHQParams}.  They agree with the nonperturbatively-determined values given in Table~\ref{tab:RHQParamErr} in all cases.  In order to provide a fair comparison, we include an estimate of systematic errors for both the perturbatively-calculated and nonperturbatively-computed values.  The largest source of uncertainty in the lattice perturbation theory determinations is the error due to neglected terms in the coupling-constant expansion of $\CO(\alpha_S^2)$ and higher.  In contrast, the largest source of uncertainty in the nonperturbative determinations of $c_P$ and $\zeta$ is heavy-quark discretization errors from neglected operators in the action of $\CO(a^2p^2)$ and higher (for $m_0a$ the uncertainty in the lattice scale dominates).   The good agreement between lattice perturbation theory and the nonperturbative tuning procedure suggests that one-loop mean-field improved lattice perturbation theory is sufficiently reliable that it can be used in situations where the nonperturbative matching factors are not available, such as in our future computations of decay constants and mixing matrix elements.

\begin{table}
\caption{One-loop mean-field improved lattice perturbation theory predictions for the RHQ parameters $c_P$ and $\zeta$ (right panel)~\cite{LehnerLPT}.  The nonperturbative inputs used in the calculation -- the bare heavy-quark mass $m_0a$, the plaquette $P$, the $2\times 1$ rectangle $R$, and the spatial link in Landau gauge $L$ -- are given in the center panel.  The errors in $c_P$ and $\zeta$ are due to the truncation of lattice perturbation theory and the uncertainty in $m_0a$, respectively.  The jackknife statistical errors in $P$, $R$, and $L$ are negligible.  }
\vspace{3mm}
  \begin{tabular}{l|cccc|cc} \hline\hline
  & \multicolumn{4}{c|}{nonperturbative } & \multicolumn{2}{c}{perturbative} \\
    & \multicolumn{4}{c|}{inputs} & \multicolumn{2}{c}{estimates} \\
    & $m_0a$& $P$ & $R$ & $L$ & $c_P$ & $\zeta$  \\[0.5mm] \hline 
    $a \approx 0.11$~fm  & 8.45 & 0.588 & 0.344 & 0.844 & 4.8(6)(2) & 3.2(2)(1) \\ 
    $a \approx 0.086$~fm & 3.99 & 0.616 & 0.380 & 0.861 & 3.04(28)(7) & 2.10(11)(5) \\ \hline\hline
  \end{tabular}
  \label{tab:LPT_RHQParams}
\end{table}

\begin{table*}
\caption{Tuned values of the RHQ parameters on the $24^3$ and $32^3$ ensembles.   The central values and statistical errors are from Tables~\ref{tab:243_RHQParams} and~\ref{tab:323_RHQParams}.   The systematic error estimates are obtained using the same approach as for the bottomonium masses and mass-splittings described in Sec.~\ref{sec:Error}.  The errors listed in $m_0a$, $c_P$, and $\zeta$ are from left to right:  statistics, heavy-quark discretization errors, the lattice scale uncertainty, and the uncertainty in the experimental measurement of the $B_s$ meson hyperfine splitting, respectively.  Errors that were considered but were found to be negligible are not shown.  For the scale uncertainty we quote smaller errors on the $32^3$ ensembles because the lattice-spacing is determined more precisely than on the $24^3$ ensembles.}
\vspace{3mm}
  \begin{tabular}{lr@{.}lr@{.}lr@{.}l} \hline\hline
    & \multicolumn{2}{c}{$m_0a$} &  \multicolumn{2}{c}{$c_P$} & \multicolumn{2}{c}{$\zeta$}  \\[0.5mm] \hline 
    $a \approx 0.11$~fm  &   8&45(6)(13)(50)(7) & 5&8(1)(4)(4)(2) & 3&10(7)(11)(9)(0) \\ 
    $a \approx 0.086$~fm &  3&99(3)(6)(18)(3) & 3&57(7)(22)(19)(14) & 1&93(4)(7)(3)(0) \\ \hline\hline
  \end{tabular}
  \label{tab:RHQParamErr}
\end{table*}

\section{Bottomonium mass predictions}
\label{sec:MassPred}

Given the determinations of the RHQ parameters described in the previous section, we can now make predictions for other states involving $b$-quarks, such as bottomonium masses and splittings.  Comparison of the results with experiment then provides a check of the relativistic heavy quark framework and tuning methodology.

\subsection{Heavy-heavy meson correlator fits}
\label{sec:HHFits}

We extract the bottomonium meson masses from the following zero-momentum meson 2-point correlation functions:
\begin{align}
C_{\bar{b}b}(t,t_0) &= \sum_{\vec{y}}  \langle {\CO^{\Gamma}_{\bar{b}b}}^\dagger(\vec{y},t)  \tilde{\CO}_{\bar{b}b}^{\Gamma} (\vec{0},t_0) \rangle \,,\label{eq:C_bbar}
\end{align}
where $\CO^{\Gamma}_{\bar{b}b}$ is the $b\bar{b}$ meson interpolating operator for the state with spin structure $\Gamma$:
\begin{align}
	\CO^{\Gamma}_{\bar{b}b} = \bar{b} \Gamma b \,.
\end{align}
Table~\ref{tab:IntOp} shows the interpolating operators used in the computation of the bottomonium 2-point functions.  Again, the tilde over the interpolating operator in Eq.~(\ref{eq:C_bbar}) denotes that the $b$-quark in the operator was generated with a Gaussian-smeared source.  

\begin{table}
\caption{Interpolating operators used to compute the $\bar{b}b$ 2-point correlation functions.   We average correlators over equivalent directions for the vector, axial-vector, and tensor states.}
\vspace{3mm}
  \begin{tabular}{lr} \hline\hline
    meson & \quad operator  \\[0.5mm] \hline
    $\eta_b$ & $\gamma_5$  \\
    $\Upsilon$ & $\gamma_i$  \\
    $\chi_{b0} $ & $1$  \\
    $\chi_{b1} $ & $\gamma_i \gamma_5$  \\
    $h_b$ & $\gamma_i \gamma_j$ \\ \hline\hline
  \end{tabular}
  \label{tab:IntOp}
\end{table}   

\begin{figure*}[t]
\centering
\includegraphics[scale=0.58,clip]{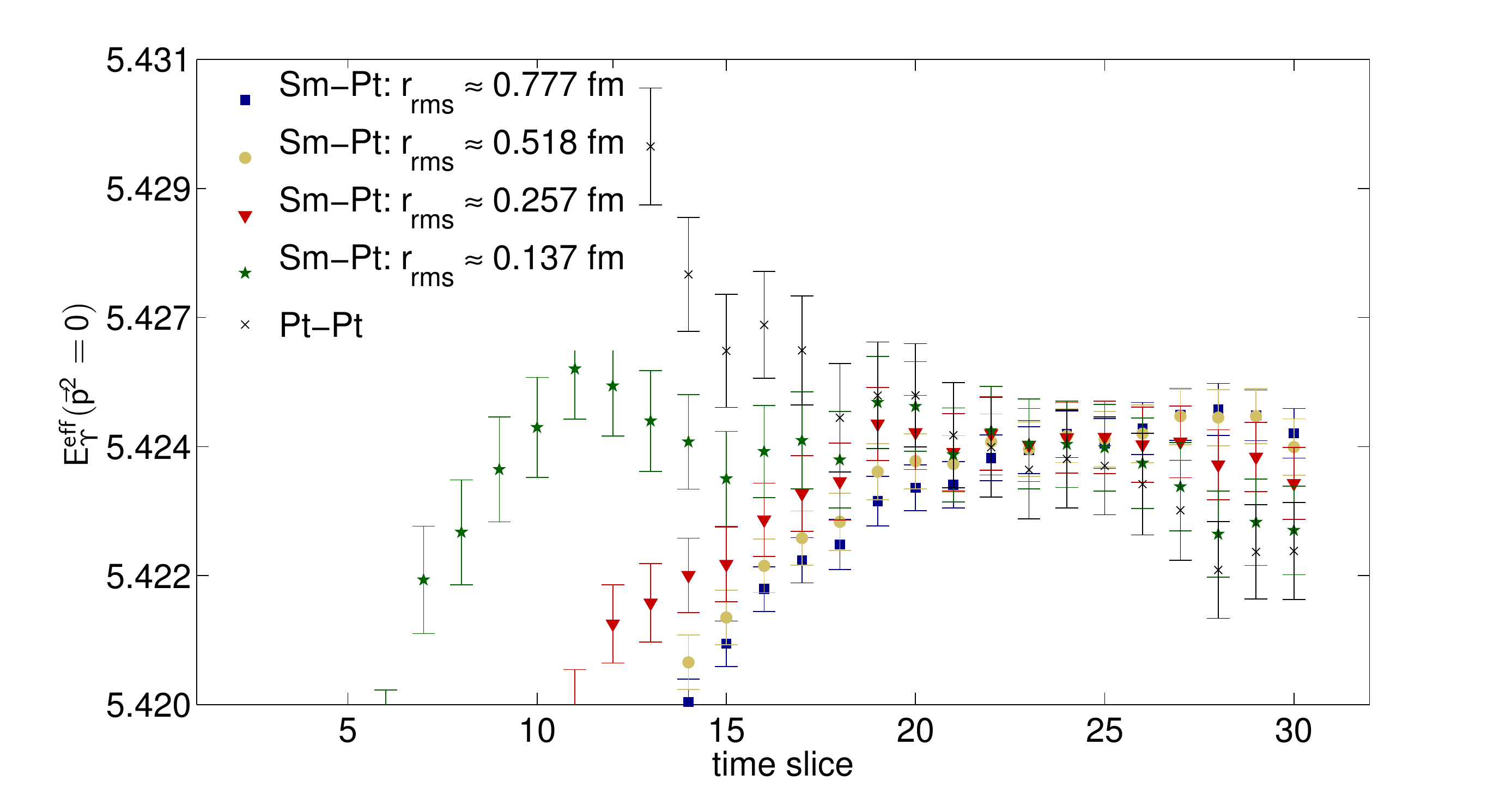}
\includegraphics[scale=0.58,clip]{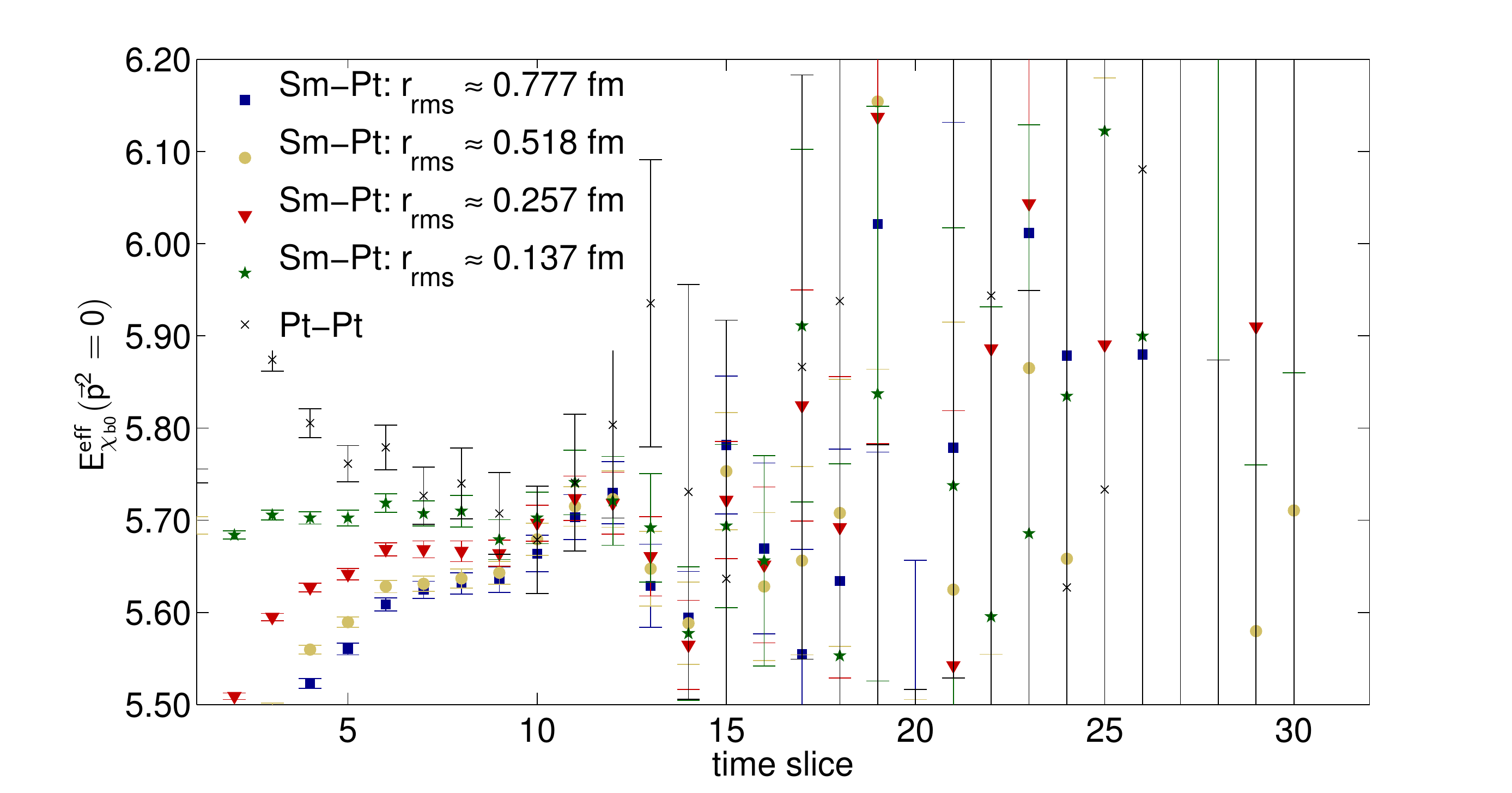}
\caption{$\Upsilon$ (upper plot) and $\chi_{b0}$ (lower plot) effective mass for several choices for the Gaussian radius of the $b$-quark in the heavy-light meson interpolating operator.  Results are shown for the $am_l = 0.005$ $24^3$ ensemble with RHQ parameters $\{ m_0 a, c_P, \zeta \} = \{8.40, 5.80, 3.20\}$.}
\label{fig:bbbar_24cSmear}
\end{figure*}

Plots of the effective energy, Eq.~(\ref{eq:E_Eff}), for the bottomonium correlators show that excited-state contamination is significant for the choice of smearing that we used to obtain the RHQ parameters.  In fact, on the $32^3$ ensembles excited-state contamination appears to persist over the entire time range up to the temporal mid-point of the lattice, making a clean determination of the ground-state mass difficult.  We therefore choose to use a different smearing for the $b$ quarks in the bottomonium correlators than for those in the bottom-strange correlators.  We perform a similar smearing study to that described for bottom-strange states in Sec.~\ref{sec:HLFits}.  Figure~\ref{fig:bbbar_24cSmear} shows the $\Upsilon$ (vector) and $\chi_{b0}$ (scalar) meson effective masses on the $am_l = 0.005$ $24^3$ ensemble for several choices of the Gaussian radius and values of the RHQ parameters $\{ m_0a, c_P, \zeta\} = \{8.40, 5.80, 3.20\}$.  The correlator generated with a $b$-quark spatial wavefunction with $r_\textrm{rms} = 0.137$~fm has the longest plateau with the earliest onset; we therefore choose to use this spatial wavefunction to compute the bottomonium masses and mass-splittings on the $24^3$ ensembles.  We perform an analogous smearing test on the $am_l = 0.004$ $32^3$ ensemble with RHQ parameters $\{ m_0a, c_P, \zeta\} = \{3.70, 3.60, 2.20\}$.  Again, we find that the Gaussian spatial wavefunction with $r_\textrm{rms} = 0.137$~fm is best.  Physically one expects a $\bar{b}b$ meson to have a narrower spatial wavefunction than a $\bar{b}s$ meson, and this is consistent with our observations.  We find an optimal wavefunction that is approximately half as wide as the bottomonium  rms radius $r_\textrm{rms}^\textrm{Richardson} = 0.224(23)$~fm computed from the Richardson potential model~\cite{Menscher:2005kj}.  

\begin{figure*}[p]
\includegraphics[scale=0.47]{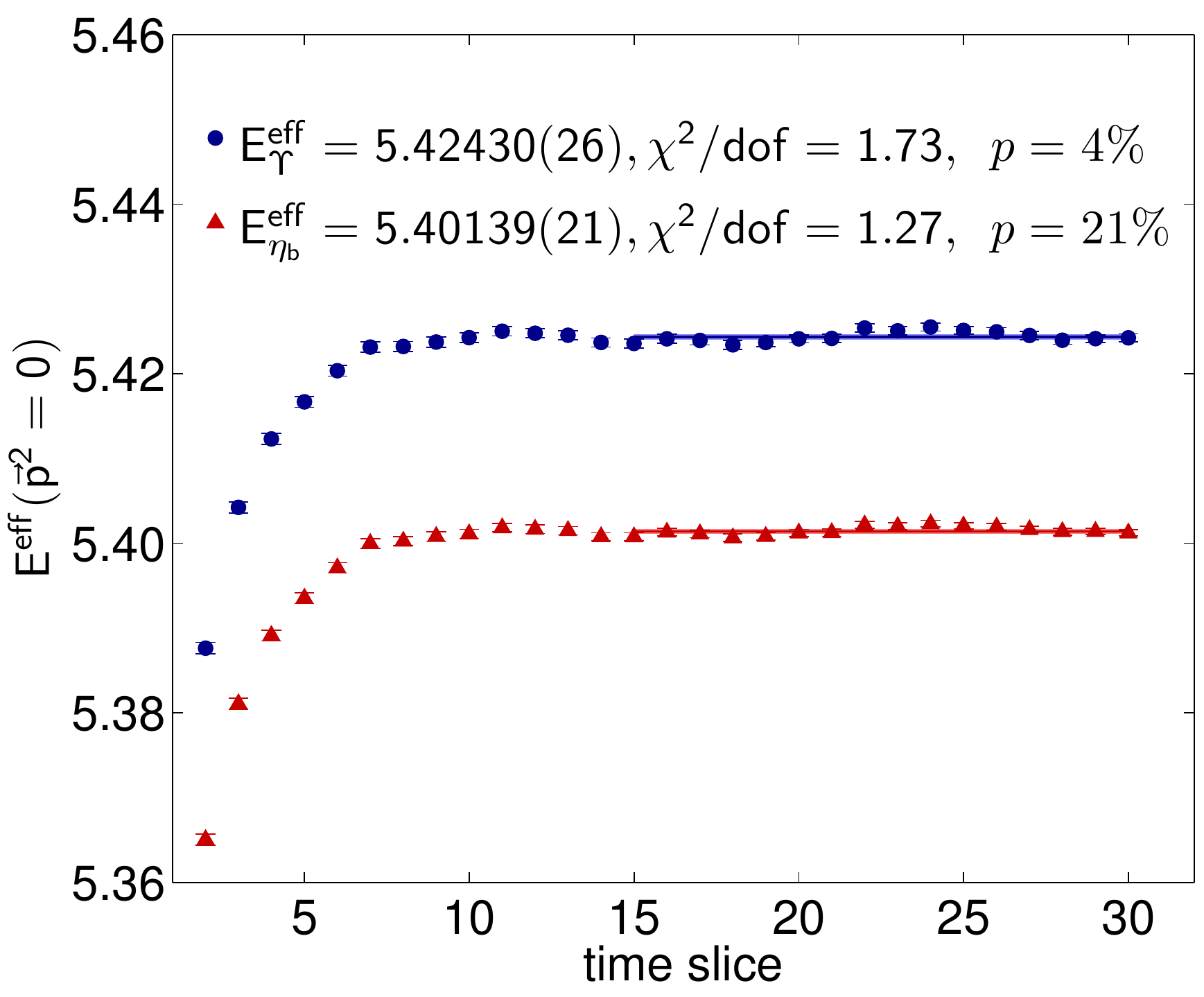}
\includegraphics[scale=0.47]{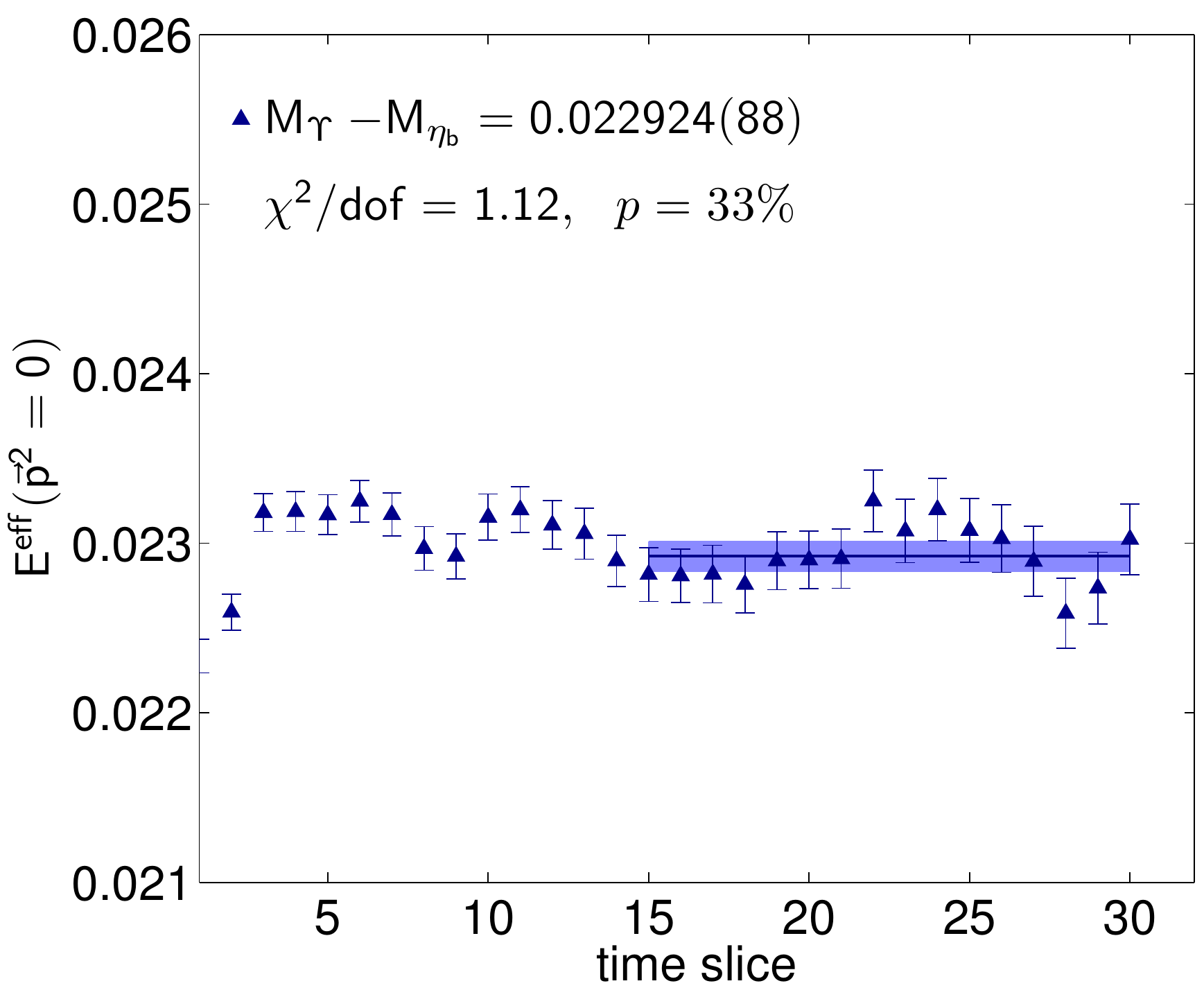}
\includegraphics[scale=0.47]{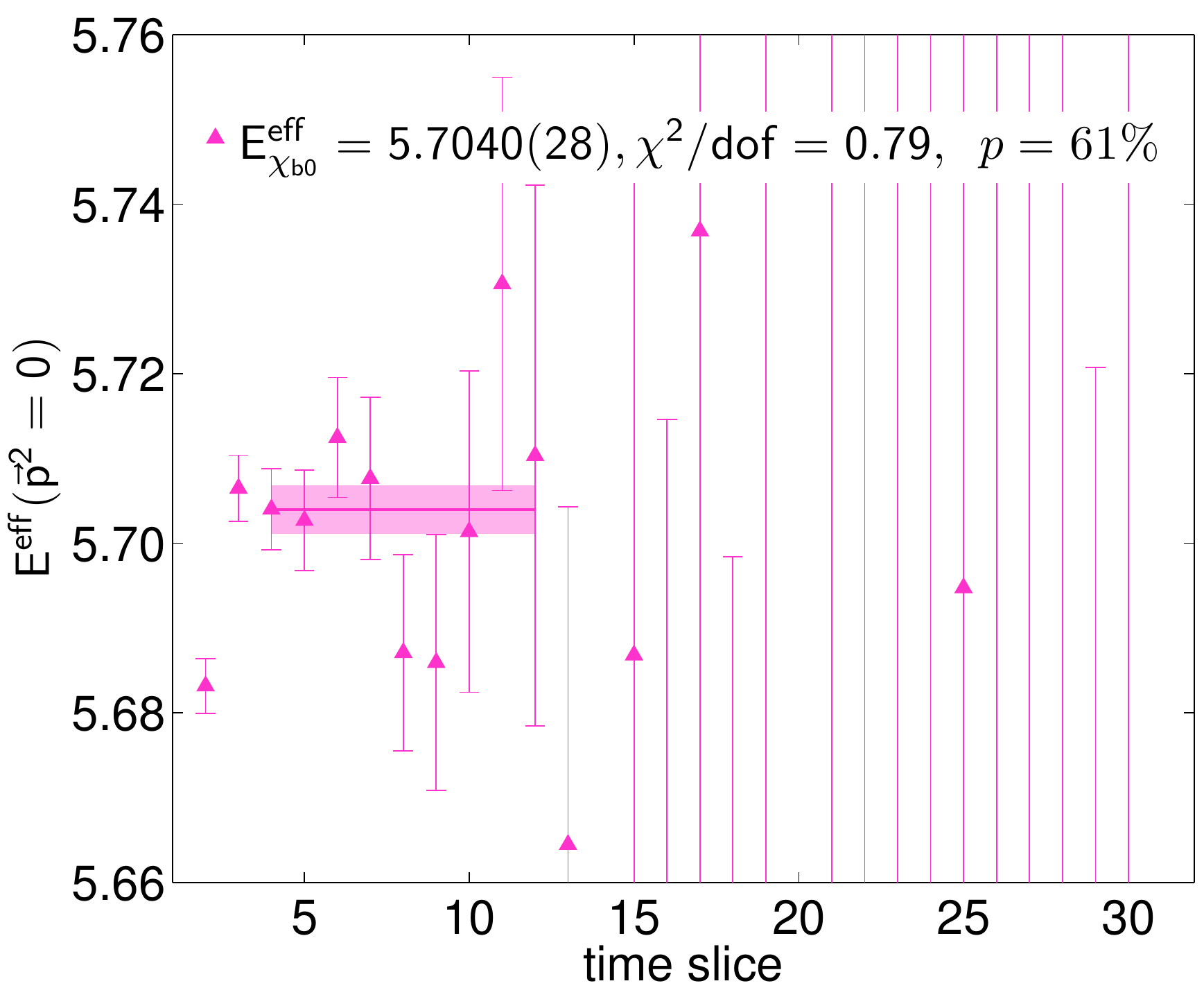}
\includegraphics[scale=0.47]{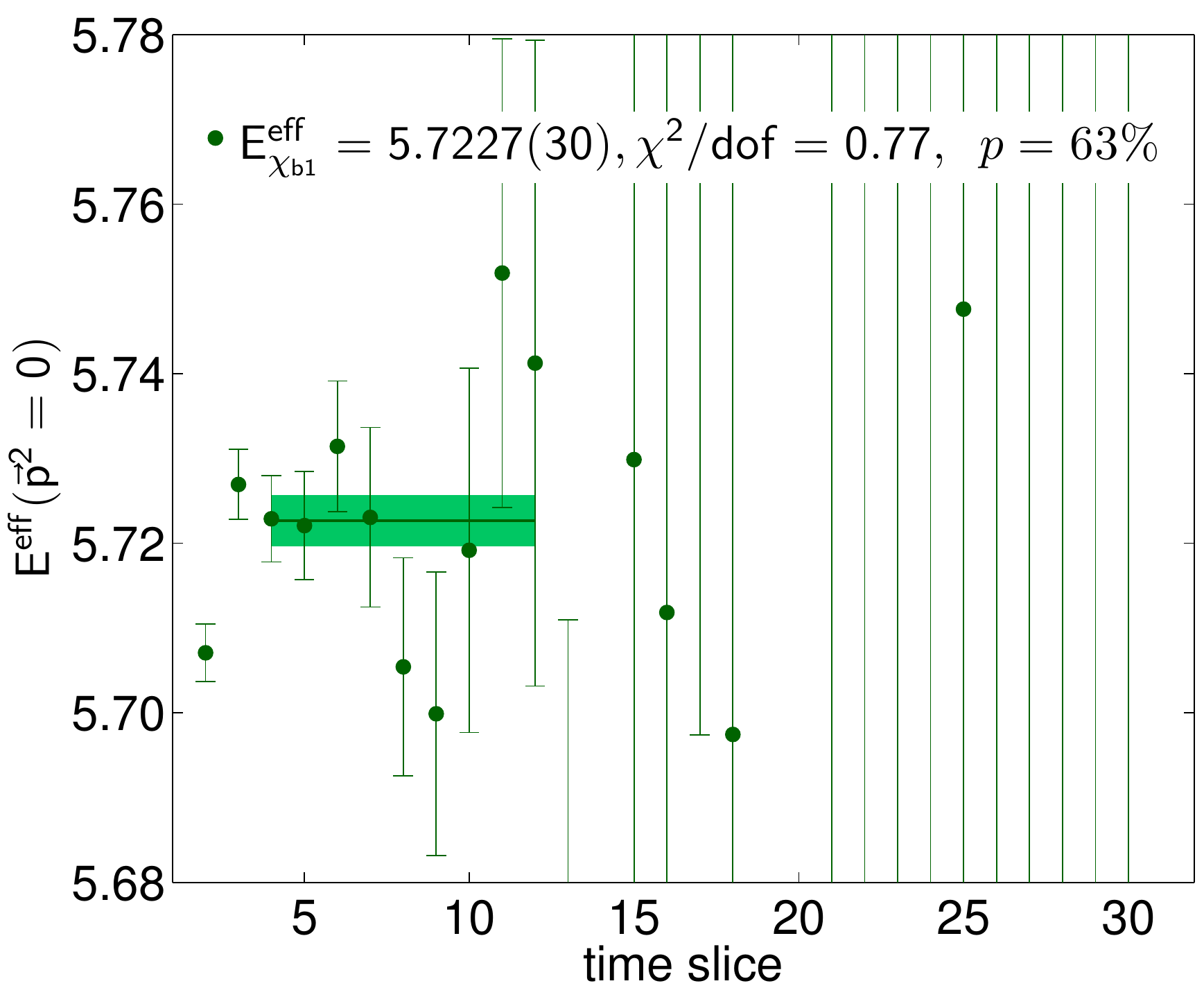}
\includegraphics[scale=0.47]{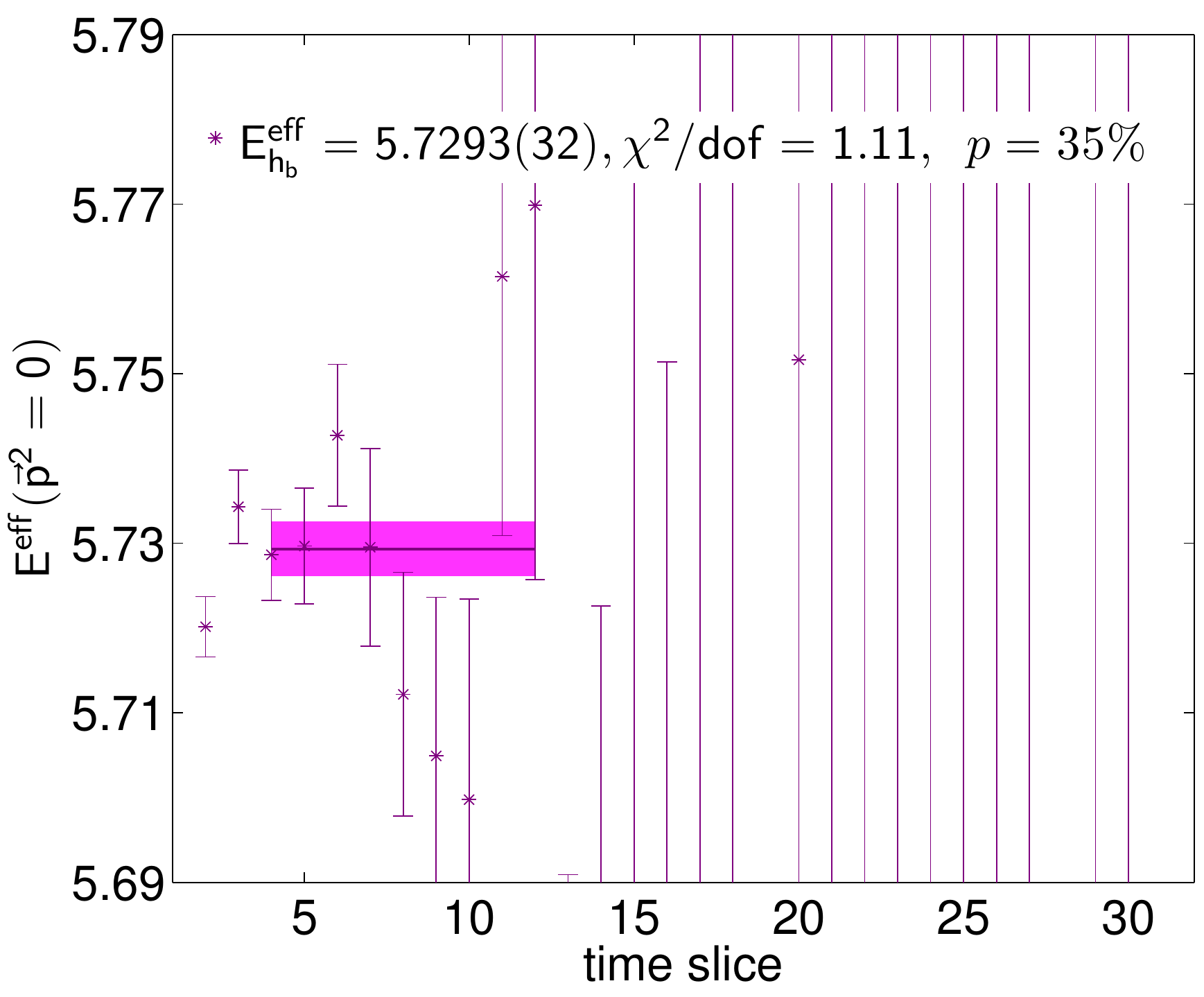}
\includegraphics[scale=0.47]{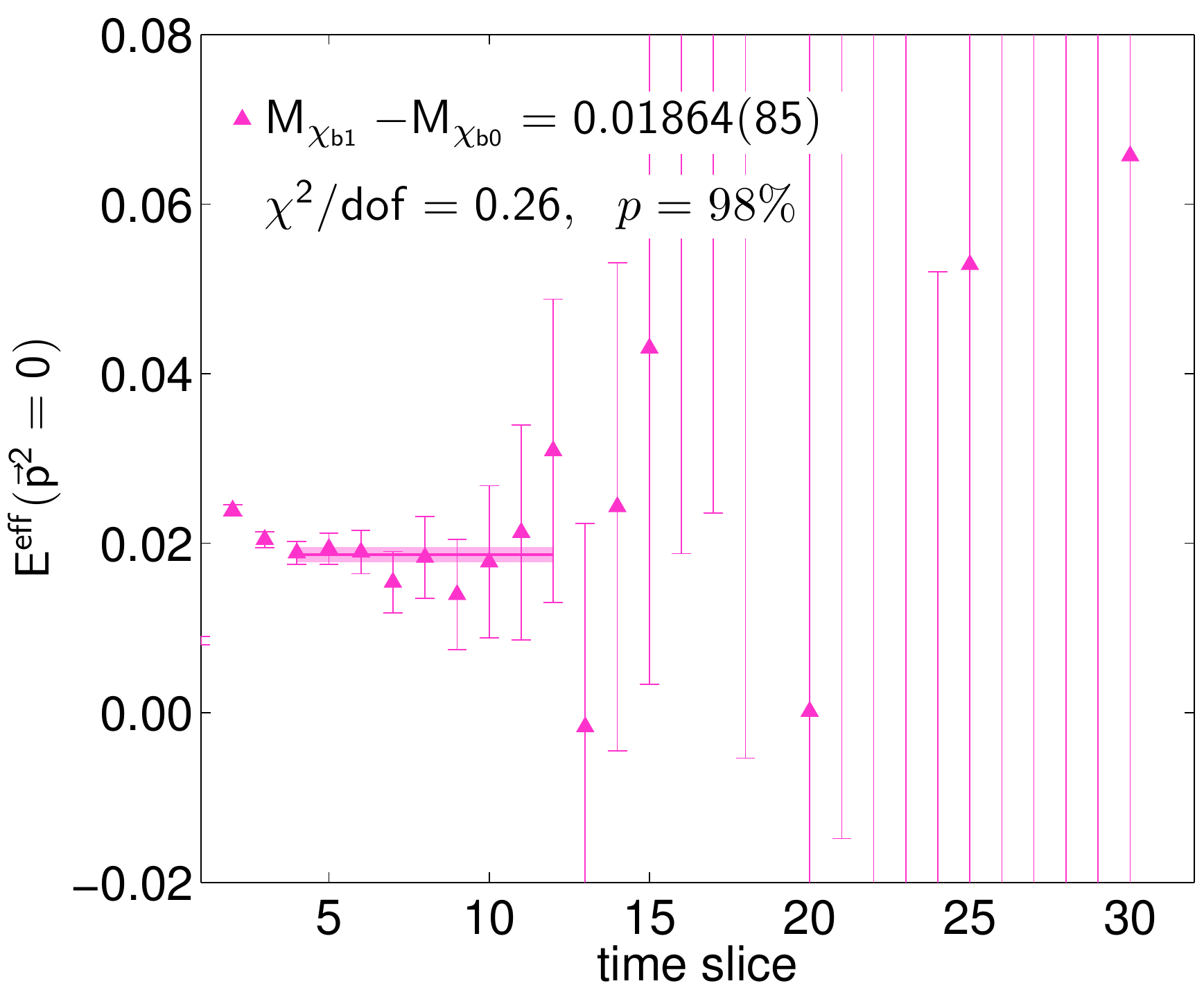}
\caption{Bottomonium masses and mass-splittings on the $am_l = 0.005$ $24^3$ ensemble with RHQ parameters $\{m_0a, c_P, \zeta\}=\{8.40,5.80,3.20\}$.  The meson states shown in each plot are specified in the legend.  For each plot the shaded horizontal band shows the fit range used and the fit result with jackknife statistical errors.}
\label{fig:24c_MEff_Bott}
\end{figure*}

Using the optimized $b$-quark smearing, we then compute the bottomonium correlators, Eq.~(\ref{eq:C_bbar}), on each ensemble for the final set of seven RHQ parameters used in the iterative tuning procedure.  This enables us to propagate the statistical uncertainties in the RHQ parameters from the tuning procedure into our determinations of the bottomonium masses and mass-splittings.  We determine the ground-state meson masses from constant fits to the effective mass.  We observe similar excited-state contamination in the $\eta_b$ and $\Upsilon$ states, so we choose a fit range that yields a good correlated confidence level for fits to both effective masses. Similarly, we use the same fit range for the $\chi_{b0}$, $\chi_{b1}$, and $h_b$ states.  Finally, because we do not expect any significant sea-quark mass dependence, we use the same fit range for all sea-quark ensembles with the same lattice spacing.  These constraints help to ensure that we are not fooled by false plateaus due to fluctuations in the gauge field, which will differ among uncorrelated ensembles.  Table~\ref{tab:bbar_FitRange} gives the fit ranges to determine the various meson masses on the two lattice spacings.  Figure~\ref{fig:24c_MEff_Bott} shows sample bottomonium effective masses and mass-splittings on the $am_l = 0.005$ $24^3$ ensemble.  Plots for other sea-quark ensembles (including at the finer lattice spacing) and other values of the RHQ parameters look similar.

\begin{table}
\caption{Time ranges used in plateau fits of the bottomonium effective masses.  We use different ranges for the $\eta_b$ and $\Upsilon$ states than for the $\chi$ and $h$ states, but use the same range for all sea-quark masses at a given lattice spacing. }
\vspace{3mm}
  \begin{tabular}{lccc} \hline\hline
     & \multicolumn{3}{c}{fit range} \\
     &  $\eta_b$ \& $\Upsilon$ & ~~ & $\chi_{b0}$, $\chi_{b1}$, \& $h_b$  \\[0.5mm] \hline
    $a \approx 0.11$ fm~ & [15,30] & & [4,12]  \\
    $a \approx 0.086$ fm & [13,30] & & [7,20] \\ \hline\hline
  \end{tabular}
  \label{tab:bbar_FitRange}
\end{table}   

\subsection{Determination of bottomonium masses and fine-structure splittings}
\label{sec:BBbarMass}

We determine the predicted values of the bottomonium masses at the tuned RHQ parameters using equations similar to Eqs.~(\ref{Eq:Jmatrix})--(\ref{Eq:RHQDetermination}):
\begin{align}
M_{\bar{b}b}^\text{RHQ} = J_M \times \left[\begin{array}{c} m_0a\\ c_P \\ \zeta \end{array}\right]^\text{RHQ}+ A_M \,,
\label{Eq:RHQprediction}
\end{align}
where the $1 \times 3$ matrix $J_M$ and constant $A_M$ are determined from a finite difference approximation of the derivatives:
\begin{align}
J_M &= \left[\frac{M_3-M_2}{2\sigma_{m_0a}},\, \frac{M_5-M_4}{2\sigma_{c_P}},\, \frac{M_7-M_6}{2\sigma_{\zeta}}\right] \,,\label{eq:JM} \\
A_M &= M_1 - J_M \times \left[m_0a,\, c_P,\,\zeta \right]^T \label{eq:AM}
\end{align}
and $M_i$ is the $\bar{b}b$ meson mass measured on the $i^\textrm{th}$ parameter set listed in Eq.~(\ref{Eq:SevenSets}).  (Note that the values of $M_i$, $J_M$, and $A_M$ are different for each bottomonium state.)  For each jackknife set we use the values of the tuned RHQ parameters $\{ m_0 a, c_P, \zeta \}^\textrm{RHQ}$ determined on that jackknife set, thereby preserving correlations between the three parameters $m_0a$, $c_P$, and $\zeta$.  Hence the jackknife statistical errors in the $\bar{b}b$ meson masses determined via Eq.~(\ref{Eq:RHQprediction}) already include the uncertainty due to the statistical errors in the tuned RHQ parameters.

\begin{figure*}[p]
\includegraphics[scale=0.47]{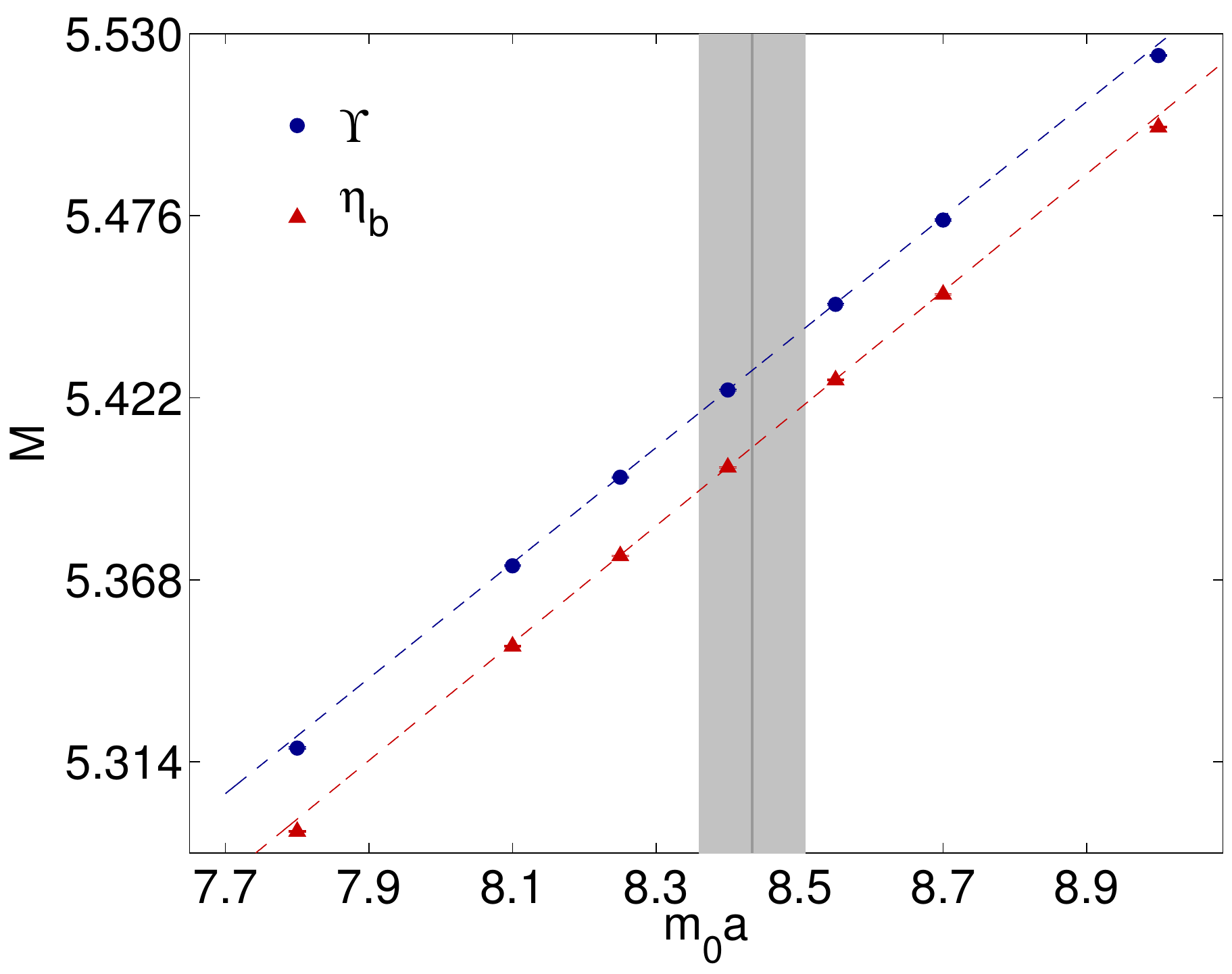}
\includegraphics[scale=0.47]{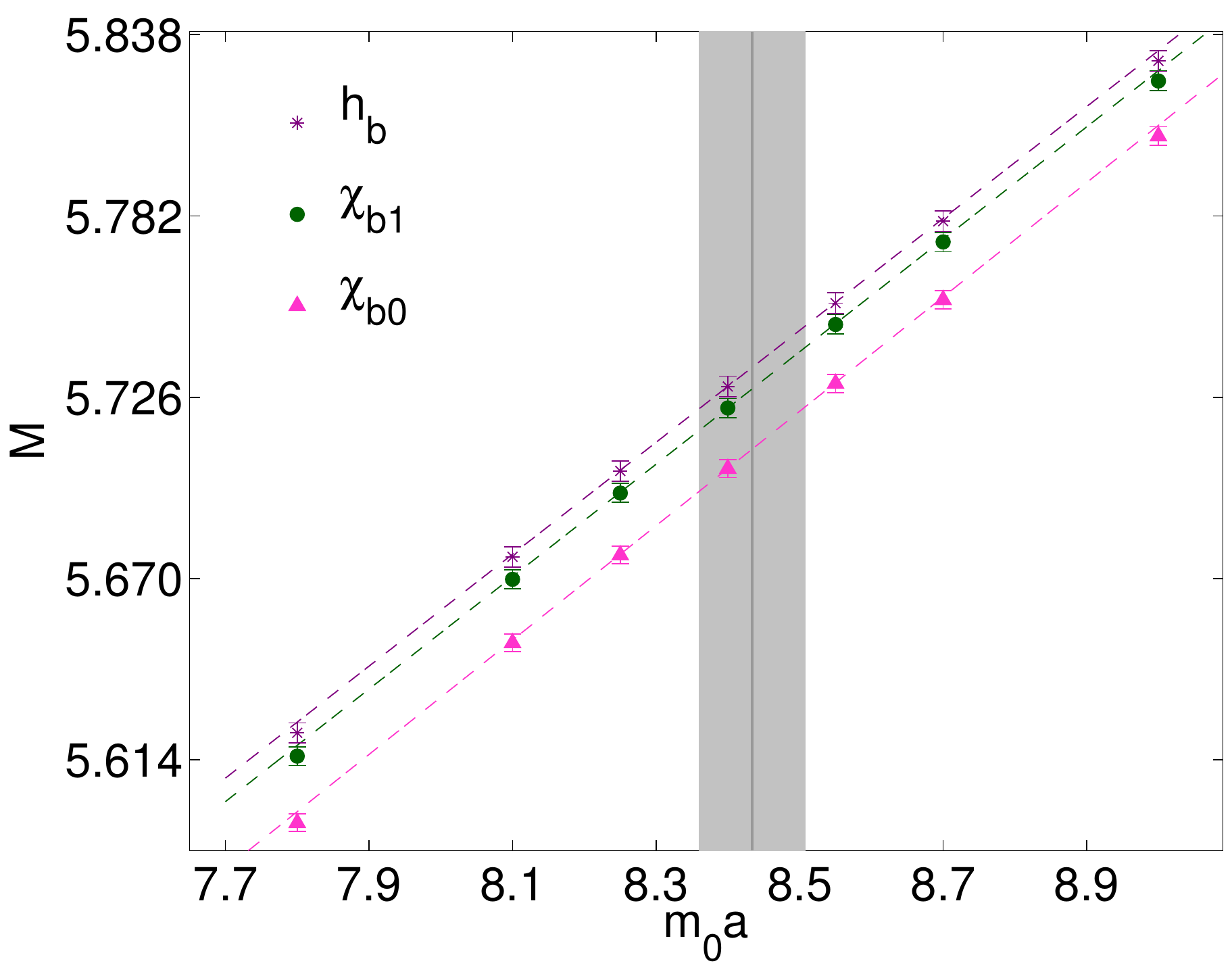}
\includegraphics[scale=0.47]{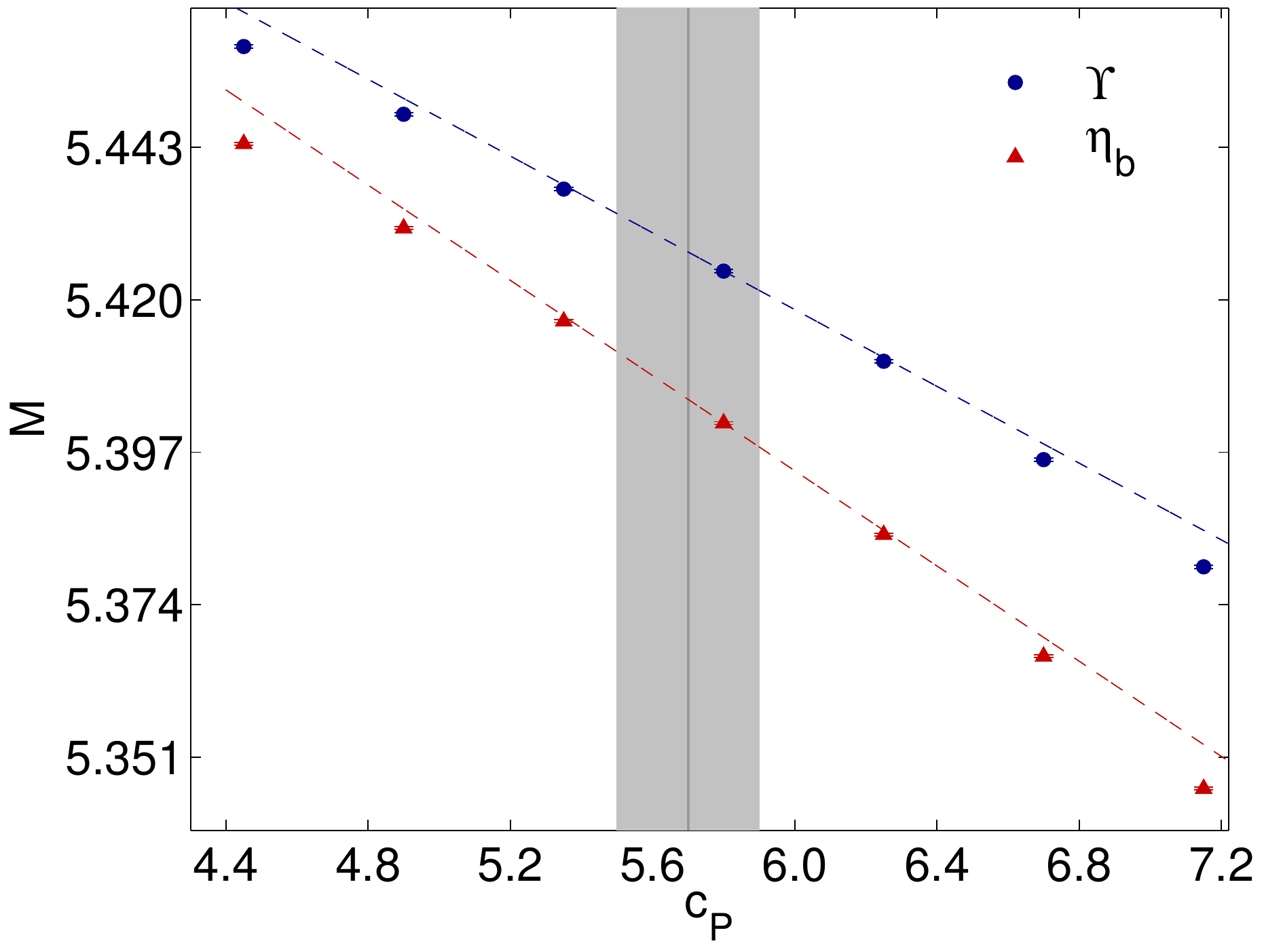}
\includegraphics[scale=0.47]{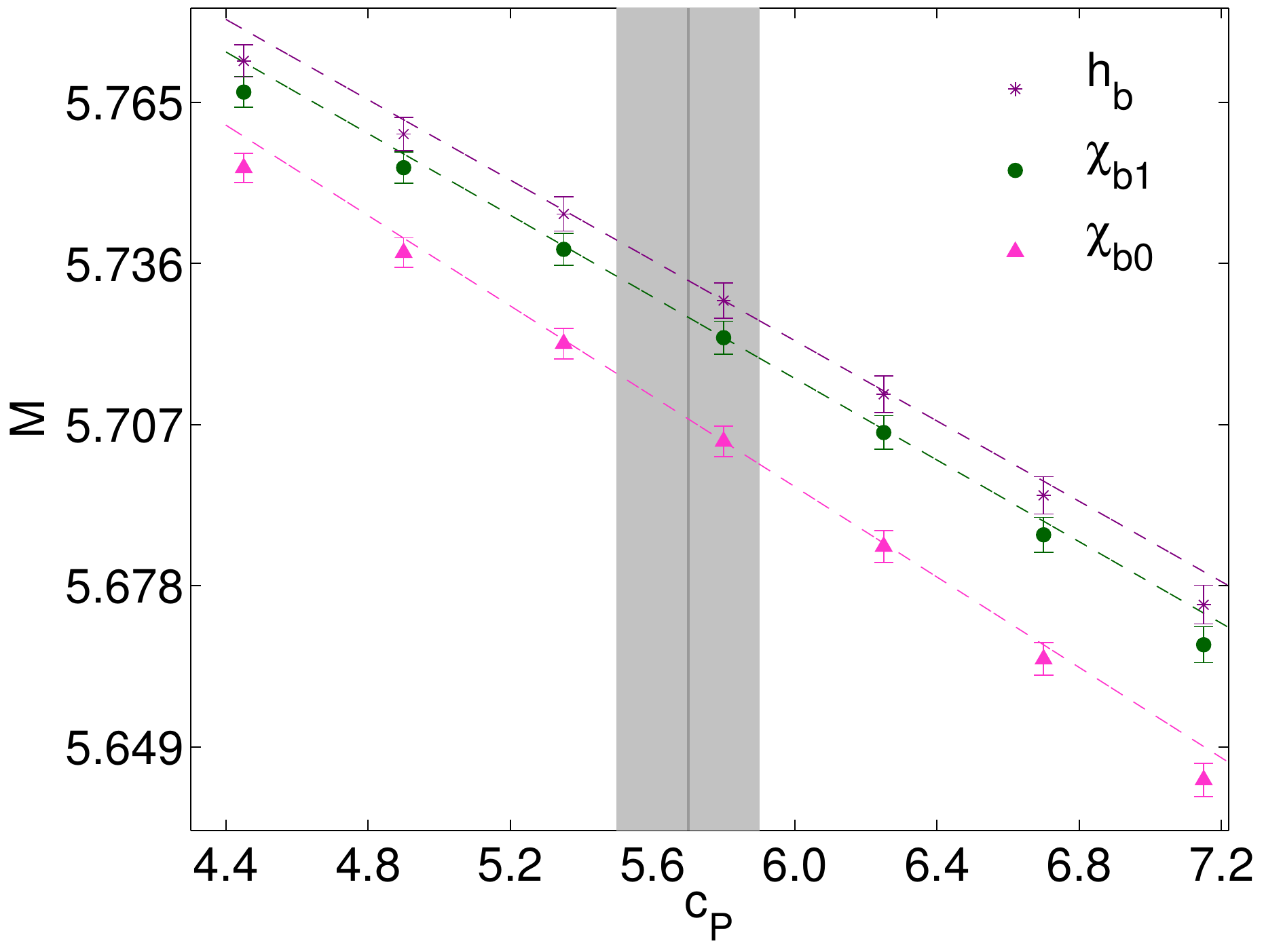}
\includegraphics[scale=0.47]{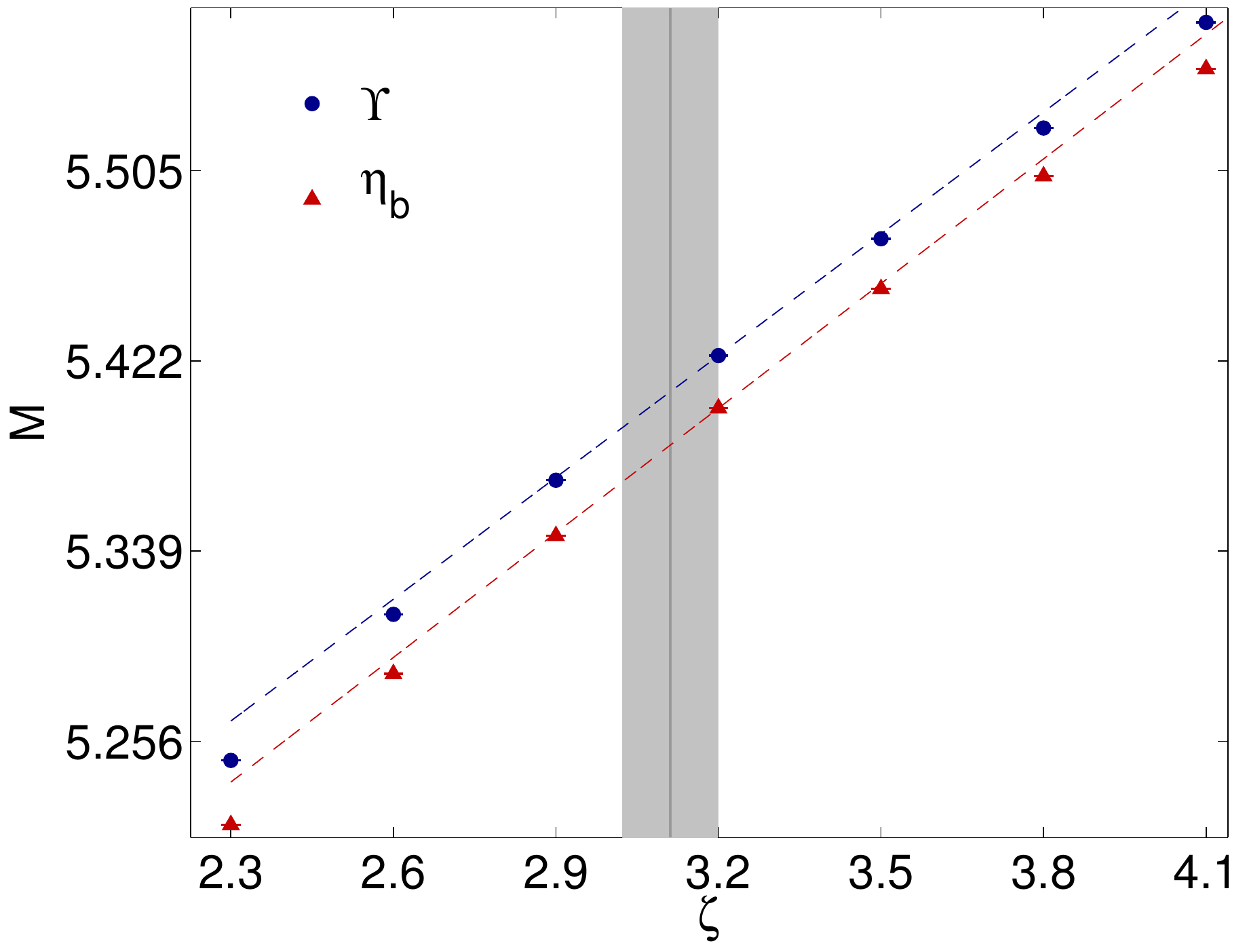}
\includegraphics[scale=0.47]{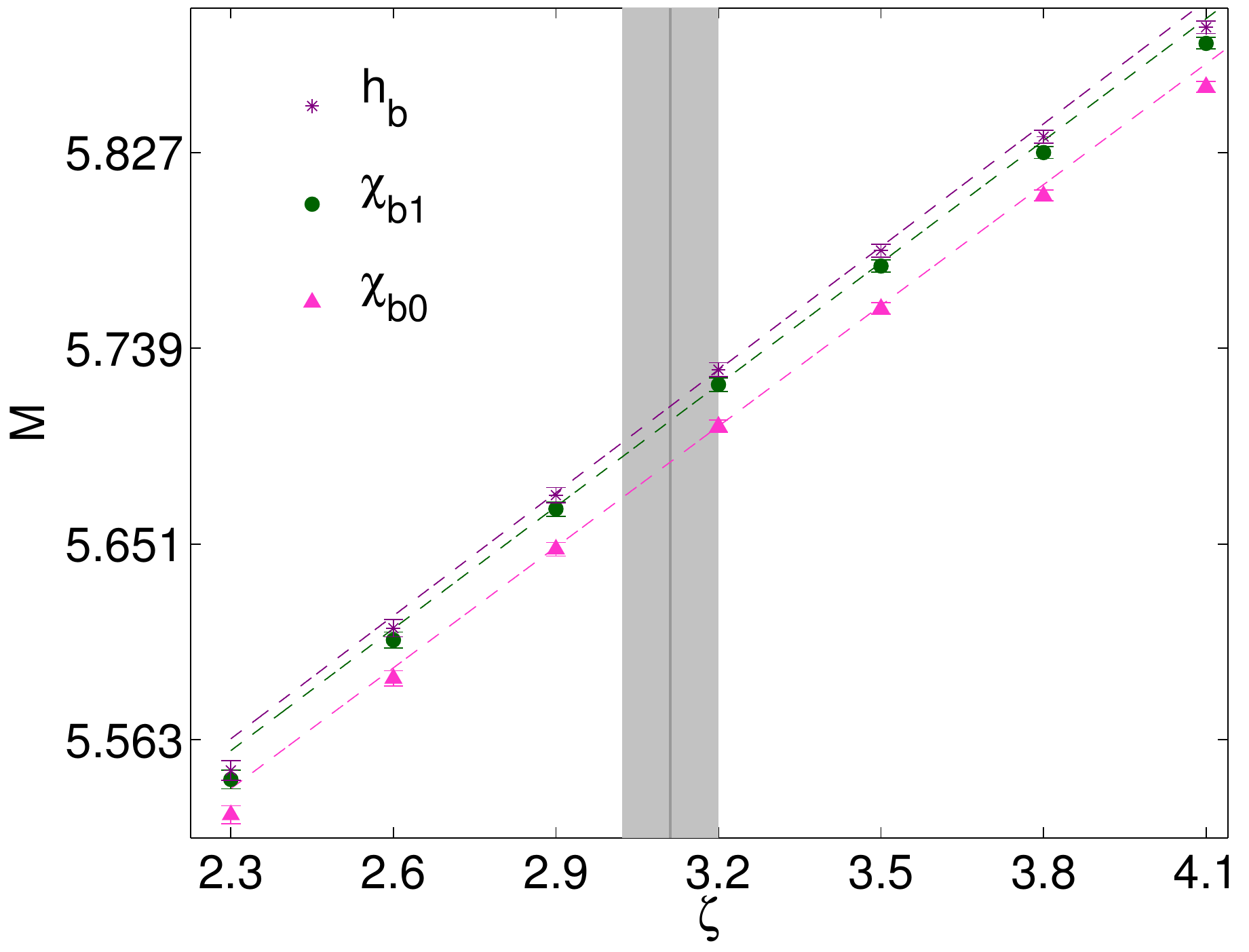}
\caption{Bottomonium masses versus $m_0a$ (upper plots), $c_P$ (center plots), and $\zeta$ (lower plots) on the $am_l = 0.005$ $24^3$ ensemble.  The meson states shown in each plot are specified in the legend.  The solid vertical lines with shaded gray error bands denote the tuned values of the RHQ parameters with jackknife statistical errors.  For each quantity, the dashed line in the same color as the plotting symbol shows the dependence on the RHQ parameters calculated from Eqs.~(\ref{Eq:RHQprediction})--(\ref{eq:AM}).}
\label{fig:24c_Bott_LinTest1}
\end{figure*}

\begin{figure*}[p]
\includegraphics[scale=0.47]{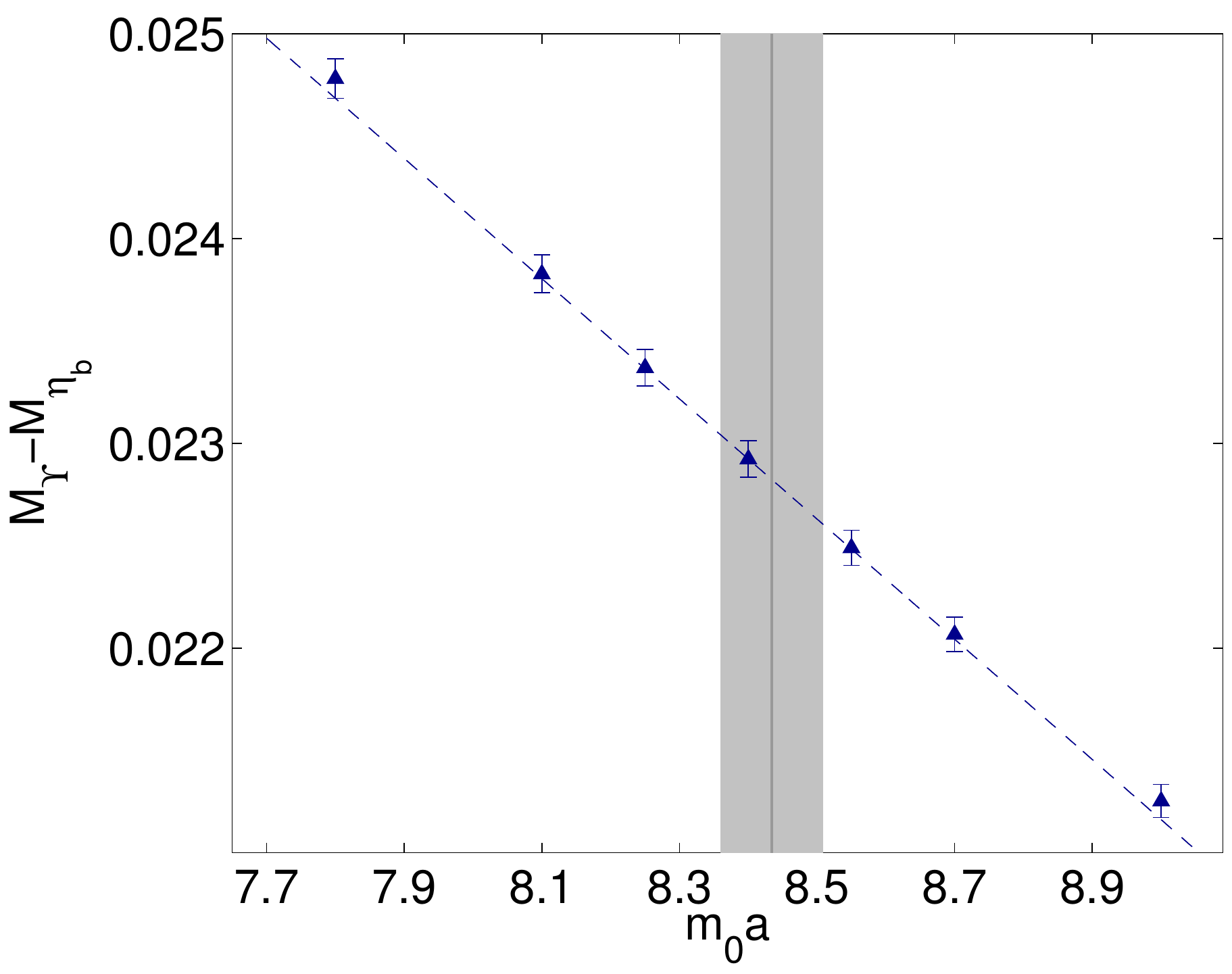}
\includegraphics[scale=0.47]{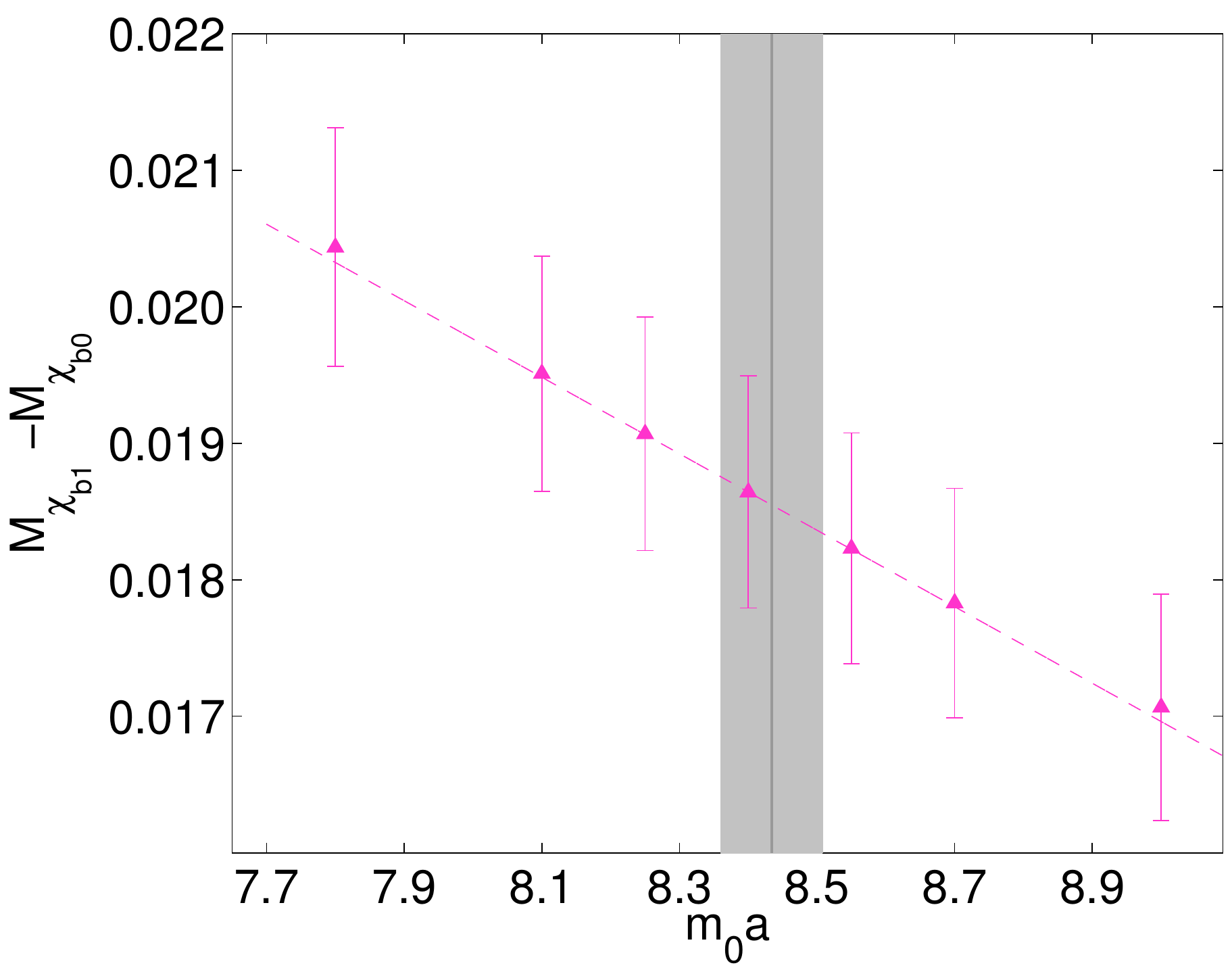}
\includegraphics[scale=0.47]{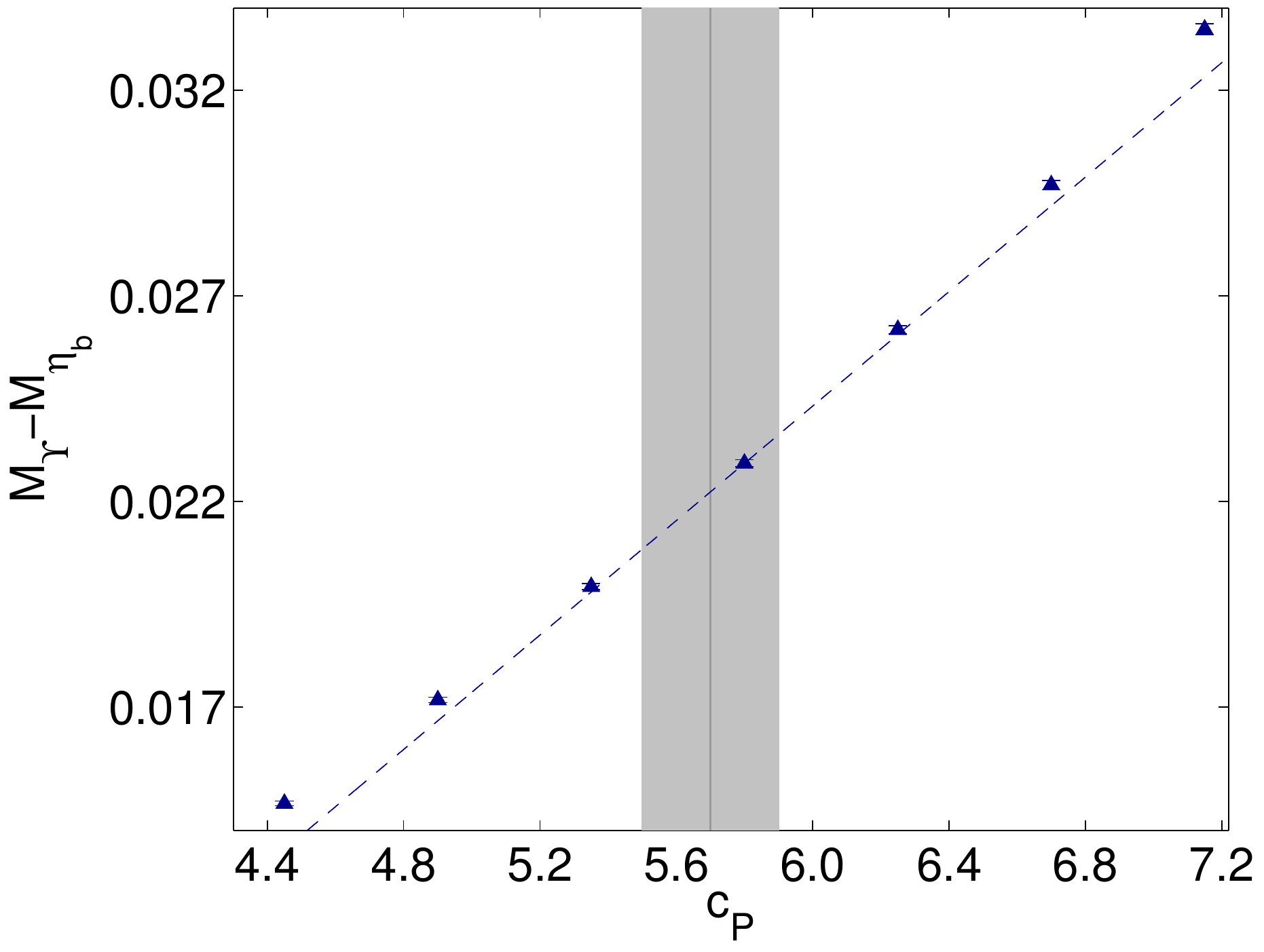}
\includegraphics[scale=0.47]{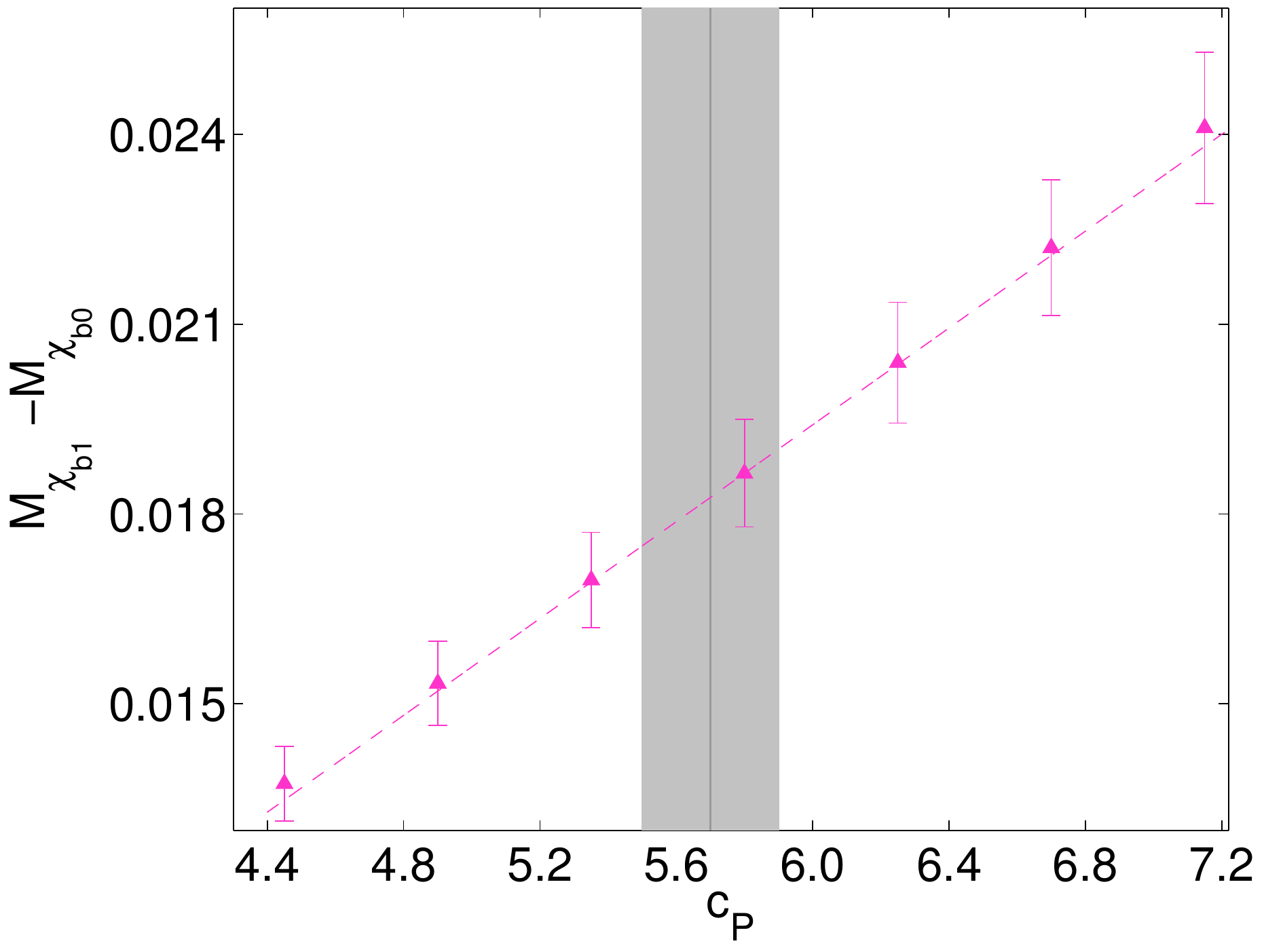}
\includegraphics[scale=0.47]{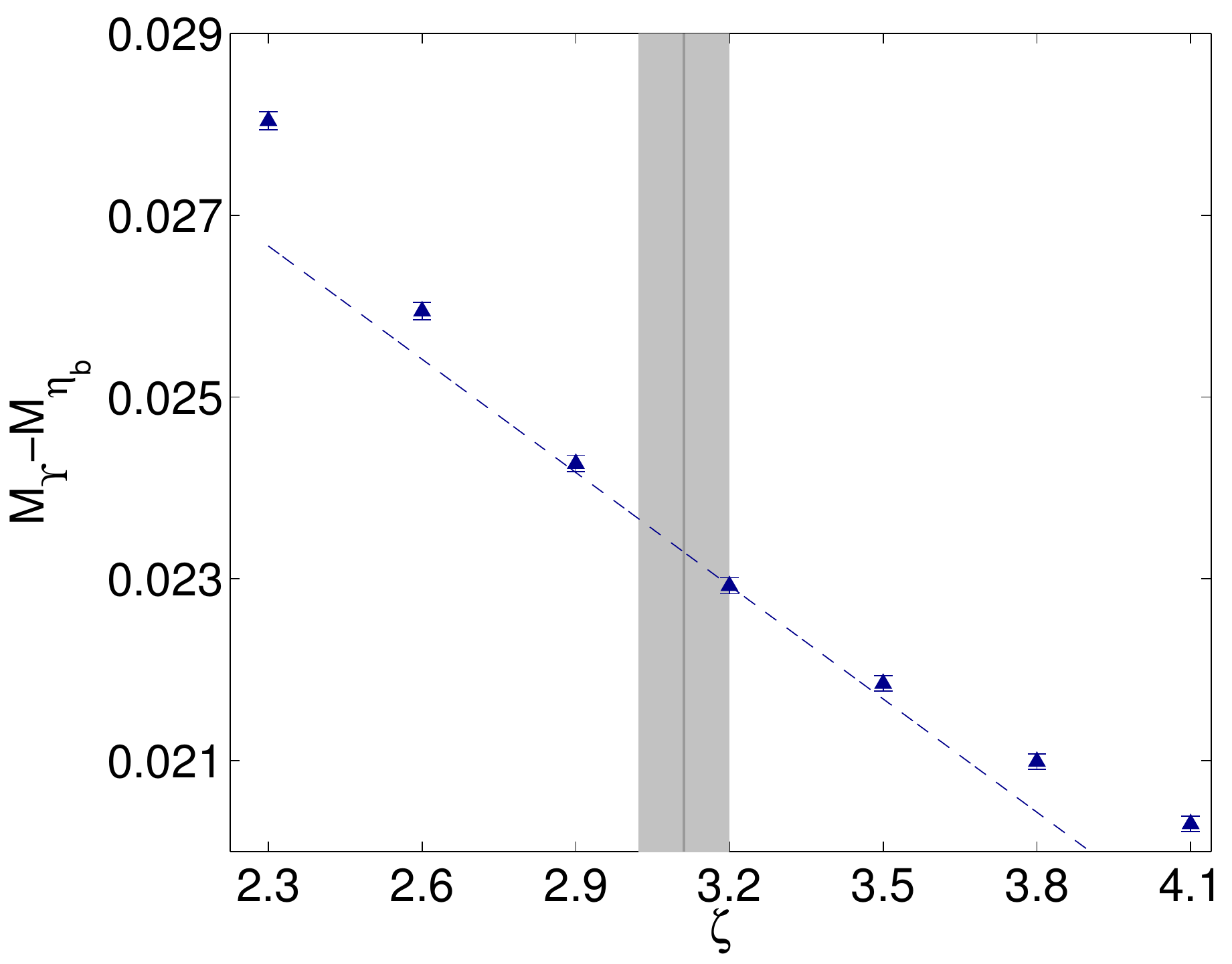}
\includegraphics[scale=0.47]{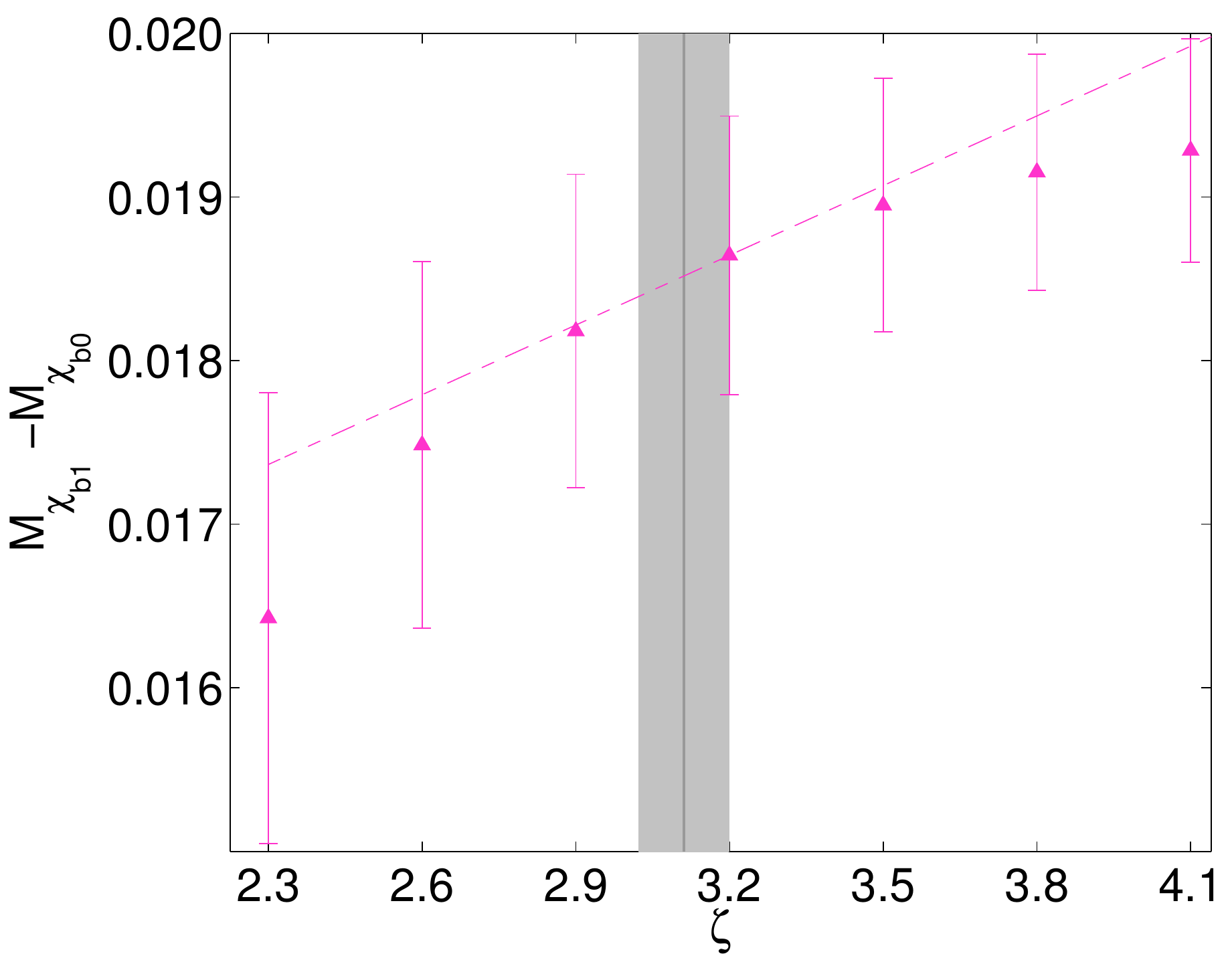}
\caption{Bottomonium mass-splittings versus $m_0a$ (upper plots), $c_P$ (center plots), and $\zeta$ (lower plots) on the $am_l = 0.005$ $24^3$ ensemble.  The hyperfine splitting $M_{\Upsilon} - M_{\eta_b}$ is shown on the left and the splitting $M_{\chi_{b1}}-M_{\chi_{b0}}$ is shown on the right.  The solid vertical lines with shaded gray error bands denote the tuned values of the RHQ parameters with jackknife statistical errors.  For each quantity, the dashed line shows the dependence on the RHQ parameters calculated from Eqs.~(\ref{Eq:RHQprediction})--(\ref{eq:AM}).}
\label{fig:24c_Bott_LinTest2}
\end{figure*}

The use of Eqs.~(\ref{eq:JM}) and (\ref{eq:AM}) requires that we are in a regime in which the bottomonium masses depend linearly on the RHQ parameters.  We test this assumption and look for signs of curvature by computing the bottomonium masses for three different boxes of seven parameters with sizes $\pm \sigma_{\{m_0a,c_P,\zeta\}}$,  $\pm 2\sigma_{\{m_0a,c_P,\zeta\}}$, and $\pm 3\sigma_{\{m_0a,c_P,\zeta\}}$.  Figures~\ref{fig:24c_Bott_LinTest1} and \ref{fig:24c_Bott_LinTest2} show the seven bottomonium masses and splittings --- $M_{\eta_b}$, $M_\Upsilon$, $M_\Upsilon-M_{\eta_b}$, $M_{\chi_{b0}}$, $M_{\chi_{b1}}$, $M_{\chi_{b1}}-M_{\chi_{b0}}$, and $M_{h_b}$ --- versus $m_0a$, $c_P$, and $\zeta$ on the $am_l = 0.005$ $24^3$ ensemble; plots for the $am_l = 0.004$ $32^3$ ensemble look similar.  The statistical errors in the $\bar{b}b$ meson masses are approximately ten times smaller than those of the bottom-strange meson masses, and we can resolve a nonlinear dependence of the $\bar{b}b$ meson masses on the RHQ parameters within statistical errors.  This curvature is most pronounced in the hyperfine splitting $M_\Upsilon-M_{\eta_b}$, and the dependence is strongest upon the parameter $\zeta$.  The nonlinear dependence is mild, however, within the region of parameter space defined by the inner-most box of parameters.  Hence we expect that the use of Eq.~(\ref{Eq:RHQprediction}) in this region will lead to only a small error in the bottomonium mass predictions.  Nevertheless, we will include a systematic uncertainty in our predictions for the bottomonium meson masses due to quadratic and higher-order corrections to Eq.~(\ref{Eq:RHQprediction}).

Once we have the results for the bottomonium masses and mass-splittings at fixed sea-quark mass and lattice spacing, we must extrapolate to the physical light-quark masses and the continuum limit.  Because the $\bar{b}b$ states contain no valence light quarks, we expect only a weak light-quark mass dependence and a correspondingly mild chiral extrapolation.  In practice, as shown in Table~\ref{tab:BottomPred}, we do not observe any statistically significant dependence of the observables on the light sea-quark masses at either lattice spacing.  We therefore compute the error-weighted average of each mass and mass-splitting over the different sea-quark ensembles at the two lattice spacings.

\begin{table*}
\caption{Bottomonium masses and mass-splittings on the five sea-quark ensembles and averaged for each lattice spacing.  For the masses, we extrapolate the results on the two lattice spacings to the continuum limit linearly in $a^2$ as described in the text.  Errors shown are statistical only, but include the uncertainty due to the statistical errors on the tuned RHQ parameters.}
\vspace{3mm}
  \begin{tabular}{l|lll|llll|r} \hline\hline
     & \multicolumn{3}{c|}{$a \approx 0.11$ fm} & \multicolumn{4}{c|}{$a \approx 0.086$ fm} & continuum \\
     mass [MeV] & $am_l = 0.005$  & $am_l = 0.01$ & average  & $am_l = 0.004$ & $am_l = 0.006$ & $am_l = 0.008$ &  average \\[0.5mm] \hline
$M_{\eta_b}$    & 9328(14) & 9327(18) & 9328(11) & 9326(18) & 9341(15) & 9347(18) & 9338(10) & 9350(33)\\
$M_{\Upsilon}$  & 9367(14) & 9367(17) & 9367(11) & 9379(16) & 9388(13) & 9395(16) & 9388(9) & 9410(30)\\
$M_{\Upsilon}-M_{\eta_b}$ & 38.8(2.3) & 40.6(2.5) & 39.6(1.7) & 53.1(3.0) & 47.3(2.4) & 48.2(3.4) & 49.2(1.6) & --- \\
$M_{\chi_{b0}}$ & 9853(15) & 9848(18) & 9851(12) & 9816(19) & 9836(15) & 9837(20) & 9831(10) & 9808(35)\\
$M_{\chi_{b1}}$ & 9884(15) & 9882(19) & 9883(12) & 9853(19) & 9873(15) & 9875(20) & 9868(10) & 9851(35)\\
$M_{\chi_{b1}}-M_{\chi_{b0}}$ & 31.2(1.8) & 33.5(2.0) & 32.3(1.3) & 37.8(2.7) & 36.6(2.2) & 38.8(2.6) & 37.5(1.4) & --- \\
$M_{h_b}$       & 9895(16) & 9894(19) & 9895(12) & 9866(19) & 9884(16) & 9887(21) & 9879(10) & 9862(36)\\[.5mm]
\hline\hline
  \end{tabular}
  \label{tab:BottomPred}
\end{table*}

\begin{figure*}[t]
\centering
\includegraphics[scale=0.45,clip]{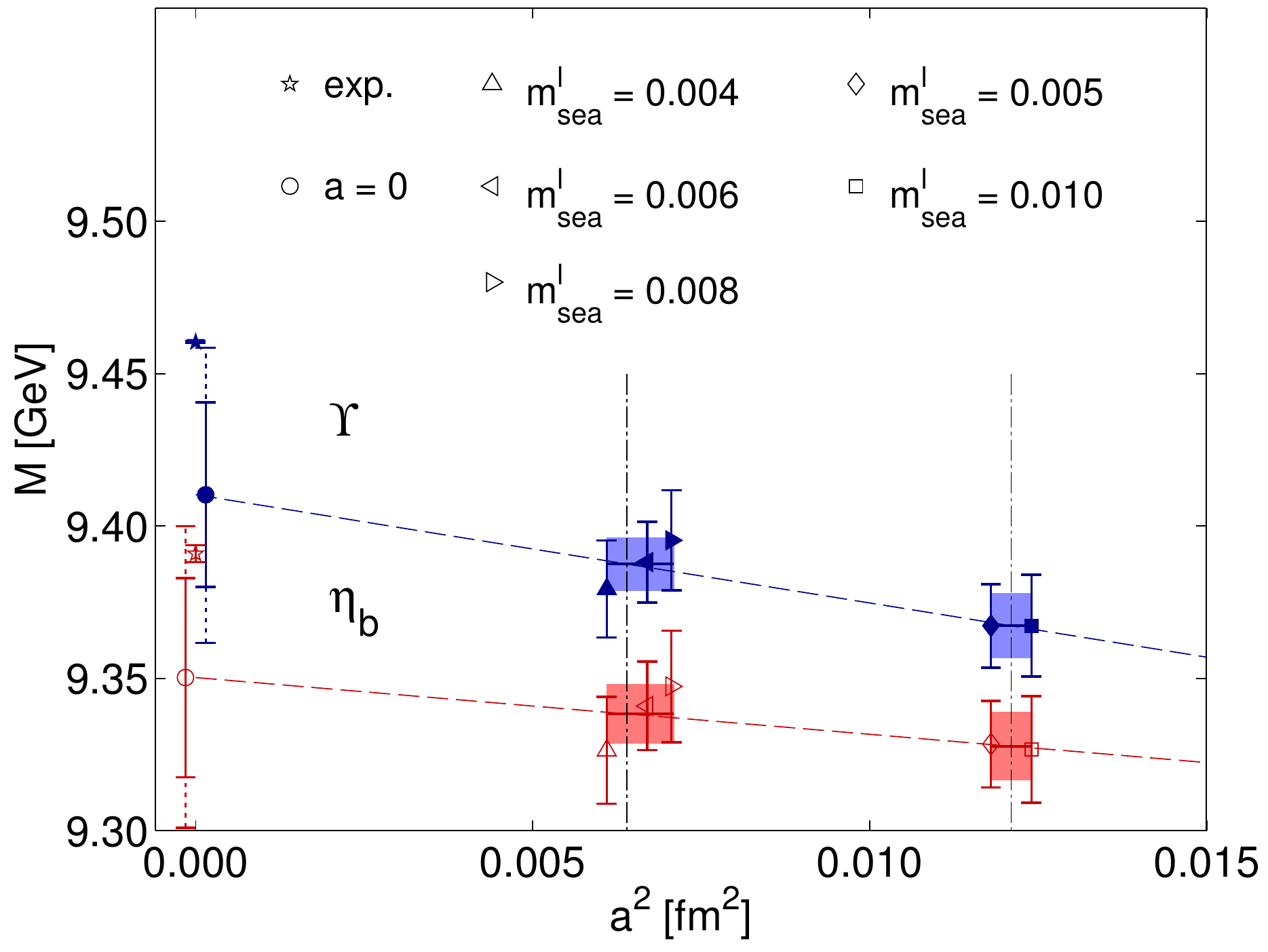}
\includegraphics[scale=0.45,clip]{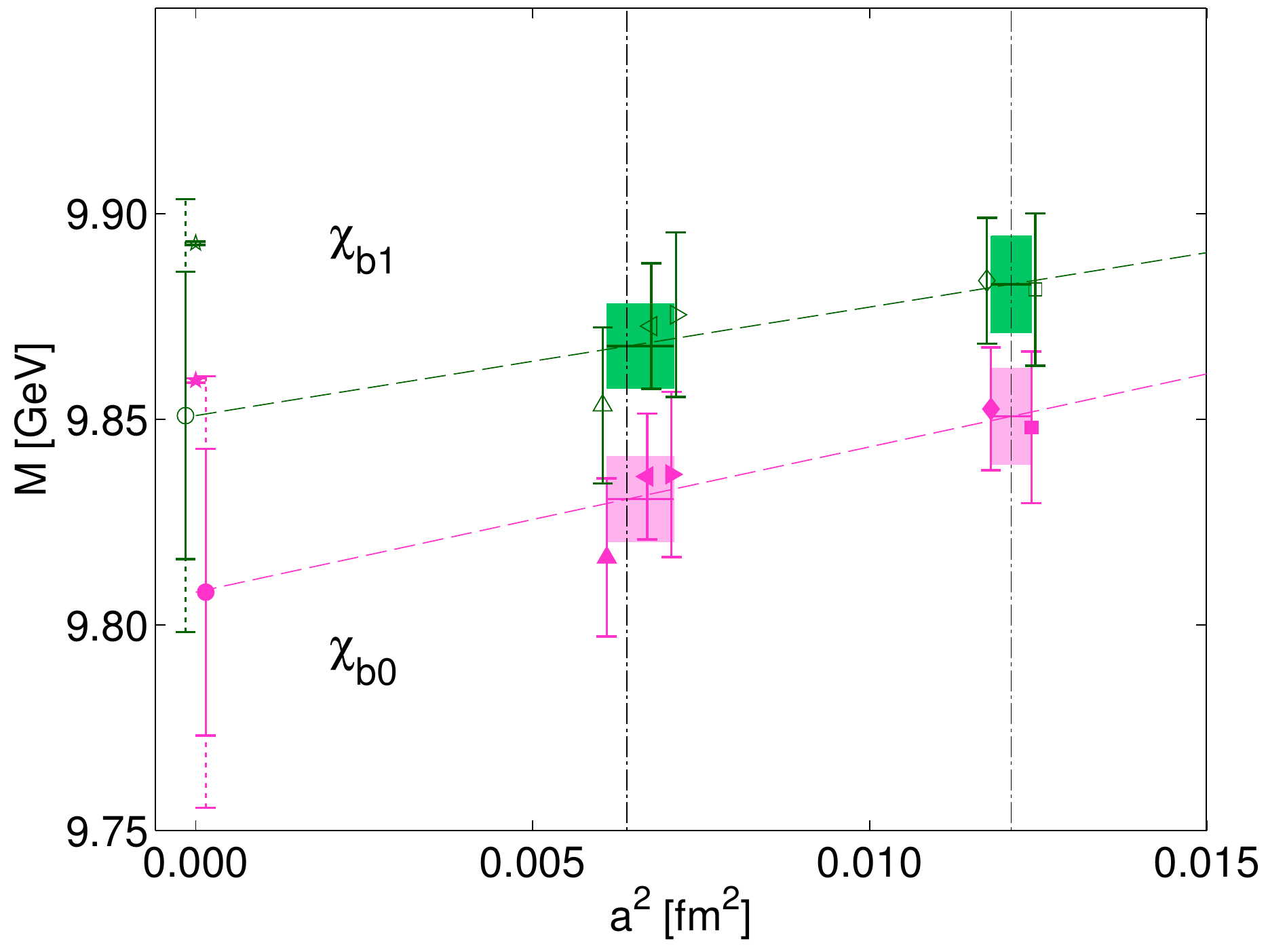}
\includegraphics[scale=0.45,clip]{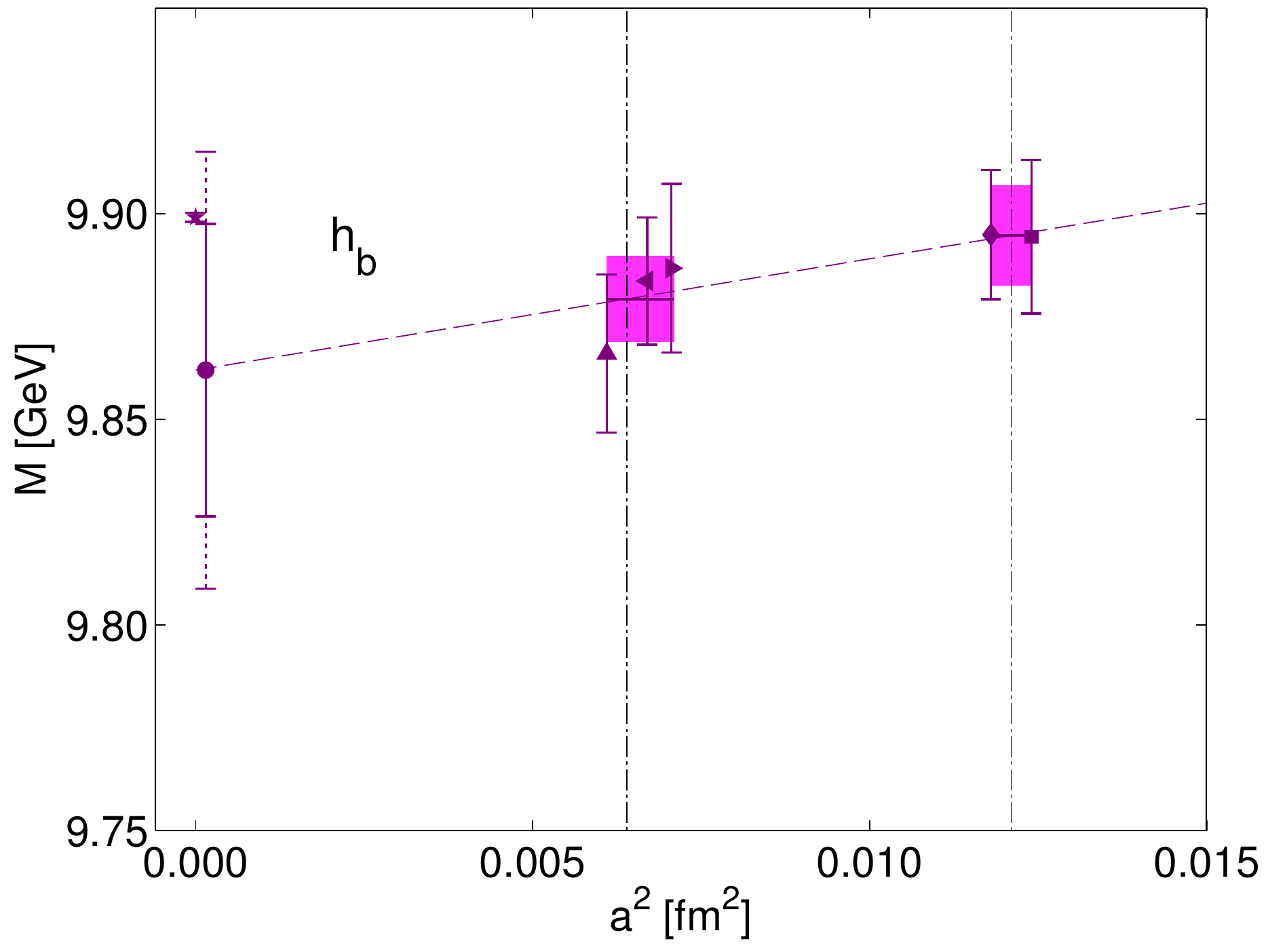}
\caption{Continuum extrapolation of bottomonium masses and mass-splittings.
Upper left plot:  $\Upsilon$ (filled blue symbols) and $\eta_b$ (open red symbols) masses versus squared lattice spacing.  Upper right plot: $\chi_{b1}$ (open green symbols) and $\chi_{b0}$ (filled pink symbols) masses versus squared lattice spacing.   Lower plot: $h_b$ mass versus squared lattice spacing.  On each plot the two lattice spacings $a \approx 0.086$~fm and $a \approx 0.11$~fm are indicated by vertical black dash-dotted lines.  Data points at different light sea quark masses but the same lattice spacing are shown with an offset for clarity.  The average values at each lattice spacing are given as shaded error bands in the same color as the symbols, and a linear extrapolation in $a^2$ of the averaged values leads to the continuum limit results denoted by circles.  On the data points we show statistical errors only.   On the continuum-extrapolated values we denote the statistical errors with solid error bars and the total statistical plus systematic errors with additional dashed error bars.  For comparison we show the experimentally-measured values as stars.}
\label{fig:ChiralExtrap}
\end{figure*}

Because the domain-wall fermion action is $\CO(a)$-improved, the leading lattice discretization effects from the light-quark and gluon sector are proportional to $a^2$.  With the relativistic heavy-quark formalism, heavy-quark discretization errors depend on the lattice spacing as unknown functions of $m_0 a$ [with coefficients of $\CO(1)$] whose behavior is only known in the asymptotic limits of very large and very small $m_0 a$; hence they do not have to scale as $a^2$.  As discussed in the following section, however, we estimate that gluon discretization errors in the bottomonium masses are larger than both light-quark and heavy-quark discretization errors, and consequently dominate the scaling behavior of the masses.  We therefore extrapolate the bottomonium masses to the continuum linearly in $a^2$ in order to remove gluon discretization errors.  We estimate the remaining systematic uncertainty from heavy-quark discretization errors using power-counting, discussed below.  Figure~\ref{fig:ChiralExtrap} shows the continuum extrapolation of the five bottomonium masses along with the experimentally-measured values for comparison.   

In contrast, light-quark and gluon discretization errors largely cancel in the fine-structure splittings, so the scaling behavior is dominated by the heavy-quark discretization errors.  With data at only two lattice spacings, however, we cannot resolve quadratic or more complicated $m_0 a$ dependence.  We therefore choose not to extrapolate the fine-structure splittings, and instead quote the results obtained on the finer $32^3$ ensembles as our central values.  Again, we estimate the residual systematic uncertainty from heavy-quark discretization errors using power-counting.


\subsection{Estimation of systematic errors}
\label{sec:Error}

We now discuss the sources of systematic uncertainty in the bottomonium masses and splittings.  Table~\ref{tab:BottomErr} presents the total statistical and systematic error budget for each quantity.

\subsubsection{Statistics}

We propagate the statistical errors through the entire multi-step analysis procedure via a single-elimination jackknife procedure.  Hence the statistical errors include the uncertainty due to the statistical errors in the tuned RHQ parameters, including correlations between $m_0 a$, $c_P$, and $\zeta$.  

\subsubsection{Heavy-quark discretization errors}
\label{sec:HQDiscErr}

The RHQ action gives rise to nontrivial lattice-spacing dependence in physical quantities in the region $m_0 a \sim 1$.  Thus, instead of including additional functions of $m_0 a$ in the combined chiral-continuum extrapolation, we estimate the size of discretization errors from the heavy-quark sector with power-counting.  We follow the method outlined by Oktay and Kronfeld in Ref.~\cite{Oktay:2008ex}, in which they outline a general framework that applies to both heavy-heavy and heavy-light systems.

We consider a  nonrelativistic description of the heavy-quark action because both the lattice and continuum theories can be described by effective Lagrangians built from the same operators.  Discretization errors arise due to mismatches between the short-distance coefficients of higher-dimension operators in the two theories.  More precisely, for each operator $\CO_i $ in the heavy-quark effective Lagrangian, the associated discretization error is given by
\bea
\textrm{error}_i^{HQ} = \left( \CC_i^\textrm{lat} - \CC_i^\textrm{cont} \right)  \langle \CO_i^{HQ} \rangle \,. \label{eq:mismatch}
\eea
The ``mismatch functions" $f_i \equiv \CC_i^\textrm{lat} - \CC_i^\textrm{cont}$ are functions of the parameters of the lattice heavy-quark action.  They have been calculated at tree-level for the anisotropic clover-improved Wilson action in Ref.~\cite{Oktay:2008ex}.  The operators $\CO_i$ in Eq.~(\ref{eq:mismatch}) specify the $\CO(a^2)$ errors present in the heavy-quark action and their expectation values $\langle \CO_i \rangle$ depend on the physical quantity of interest.  When the sizes of operators in the heavy-quark action are estimated with power-counting appropriate to heavy-light meson systems, this framework leads to HQET.  Similarly, when the sizes of operators in the nonrelativistic heavy-quark action are estimated with power-counting suitable for heavy-heavy meson systems, it leads to NRQCD.

We consider two sources of heavy-quark discretization errors in the bottomonium system.  The first is directly from operators that contribute to bottomonium masses and fine-structure splittings.  The second is indirect contributions from discretization errors in the RHQ parameters;  these are due to heavy-quark discretization errors in the $B_s$ and $B_s^*$ energies used in the tuning procedure.  We discuss each source briefly in turn and present the final error estimates here.  Details are provided in the appendices~\ref{sec:Mismatch}--\ref{sec:HSErrs}.  

\bigskip

To estimate the ``direct" heavy-quark discretization errors, we compute the values of the mismatch functions for our lattice simulation parameters and estimate the sizes of the matrix elements of the higher-dimension operators $\CO_i$ in Eq.~(\ref{eq:mismatch}) with power-counting appropriate to heavy-heavy meson systems.   We use $a^{-1} = 2.281$~GeV~\cite{Aoki:2010dy}, which is the lattice scale on our finer $32^3$ ensembles, and $m_b = 4.2$~GeV~\cite{Nakamura:2010zzi}.  The RHQ parameters on the $32^3$ lattices are given by $\{m_0a, c_P, \zeta \} = \{3.99, 3.57, 1.93 \}$.  We also need an estimate for the $b$-quark velocity $v$ in the $\bar{b}b$ mesons.  Following Ref.~\cite{Thacker:1990bm}, we expect that the mass difference between the $\Upsilon(1S)$ and $\Upsilon(2S)$ states, which is roughly 500~MeV, should be of the same size as the average kinetic energy, $E \sim m_b v^2$.  Taking the quark mass to be half the meson mass gives an estimate for the $b$-quark velocity squared of $v^2 \sim 0.1$.  

The numerical estimates of the relevant mismatch functions are given in Appendix~\ref{sec:Mismatch}.  Because the $b$ quarks in the $\bar{b}b$ mesons are nonrelativistic, we estimate the size of operators using the ``NRQCD" power-counting formulated in Ref.~\cite{Lepage:1992tx}:
\beq
\vec{D} \sim m_b v\,, \quad g\vec{E} \sim m_b^2 v^3, \quad g\vec{B} \sim m_b^2 v^4, \quad g^2 \sim v \,, \label{eq:NRQCD_PC}
\eeq
where the expansion parameter $v$ is the spatial velocity of the $b$ quarks.  Thus, in NRQCD, an operator's numerical importance is determined by the order in the heavy-quark velocity $v$, rather than the dimension.  Within the NRQCD power-counting framework, $b\bar{b}$ meson masses are approximately $M \sim 2m_b$, generic mass splittings such as $M_{\Upsilon}(2S)-M_{\Upsilon}(1S)$ are $\sim m_b v^2$ and fine-structure splittings such as the hyperfine, spin-orbit, and tensor splittings are $\sim m_b v^4$.  

In the RHQ approach we tune the coefficients of the dimension five operators in the Symanzik effective theory nonperturbatively; hence the leading heavy-quark discretization errors come from operators of dimensions 6 and 7 in the Symanzik effective theory (or alternatively the heavy-quark effective Lagrangian) that are omitted from the lattice action.  The dominant errors in the $b\bar{b}$ meson masses come from operators that are of $\CO(v^4)$ in the NRQCD power-counting.  In Appendix~\ref{sec:HHErrs}, we estimate the size of their contributions to bottomonium masses to be $\sim 0.34\%$.  Contributions from operators of $\CO(v^4)$ cancel in the fine-structure splittings, such that the dominant errors come from operators that are of $\CO(v^6)$.  In Appendix~\ref{sec:HHErrs}, we estimate the size of their contributions to hyperfine splittings to be $\sim 32\%$ and to $\chi$-state splittings to be $\sim 43\%$.  The errors in the hyperfine splittings are smaller because they only come from operators containing the term $\vec{\sigma} \cdot \vec{B}$ (and permutations thereof), where $\vec{B}$ is the chromomagnetic field.

\bigskip

To estimate the ``indirect" heavy-quark discretization errors from the bottom-strange mesons used in the RHQ tuning procedure, we use the same values of the mismatch functions but estimate the sizes of operator matrix elements with power-counting appropriate to heavy-light meson systems.   We consider separately heavy-quark discretization errors in the three input quantities:  the spin-averaged rest mass $\bar{M}_{B_s}$, the hyperfine splitting $\Delta M_{B_s}$, and the ratio of rest-to-kinetic masses $M_1^{B_s}/M_2^{B_s}$.

The $b$-quarks in $B$ hadrons typically carry a spatial momentum $|\vec{p}| \approx \Lambda_{\rm QCD}$, the scale of the strong interactions.  Therefore we estimate the size of operators using HQET power-counting, which in the continuum is an expansion in $|\vec{p}|/m_b$.  The lattice introduces an additional scale, $a$.  Following Ref~\cite{Oktay:2008ex}, we therefore expand in powers of $\lambda$, where $\lambda$ is either of the small parameters
\beq
	\lambda \sim a \Lambda_{\rm QCD}, \Lambda_{\rm QCD}/m_Q \,.  \label{eq:HQET_PC}
\eeq
Within the HQET power-counting framework, $\bar{b}l$ meson masses are approximately $M \sim m_b$ and hyperfine splittings are $\sim \Lambda_{\rm QCD}^2 / 2m_b$.

As for the estimates above, we use the lattice-spacing and RHQ parameters on the $32^3$ ensembles along with the experimentally-measured $b$-quark mass.  We also need an estimate for the $b$-quark momentum $\Lambda_{\rm QCD}$ in the heavy-strange mesons.   We choose $\Lambda_{\rm QCD} = 500$~MeV because fits to moments of inclusive $B$-decays using the heavy-quark expansion suggest that the typical QCD scale that enters heavy-light quantities tends to be larger than for light-light quantities~\cite{Buchmuller:2005zv}.  

The dominant errors in the $B_s$ and $B_s^*$ meson rest masses come from operators that are of $\CO(\lambda^2)$ in the HQET power-counting.  In Appendix~\ref{sec:HSErrs}, we estimate the size of their contributions to $M_1^{B_s^{(*)}}$ to be $\sim 0.05\%$.  This is comparable to the size of the statistical errors in the effective masses computed in our numerical simulations (see the example fits in Fig.~\ref{fig:243_Bs_BsStar}).  As can be seen from Figs.~\ref{fig:243_LinTest}, such a small variation in the spin-averaged mass leads to a statistically-negligible shift in the tuned value of $m_0a$ ({\it i.e.} well within the vertical gray error band).  Hence we neglect heavy-quark discretization effects in $\bar{M}_{B_s}$ when determining the size of heavy-quark discretization errors in the tuned RHQ parameters.  

The dominant errors in the $B_s$ hyperfine splitting come from operators that are of $\CO(\lambda^3)$ in the HQET power-counting.    In Appendix~\ref{sec:HSErrs}, we estimate the size of their contributions to $\Delta M_{B_s}$ to be $\sim 4.4\%$.  This is approximately twice as large as the statistical errors in the hyperfine splittings computed in our numerical simulations.  As can be seen from Figs.~\ref{fig:243_LinTest}, a variation of this size leads to a statistically-significant shift in the tuned value of $c_P$, so we must  propagate it to an uncertainty in the tuned RHQ parameters.  We estimate this error by varying the value of $\Delta M_{B_s}$ used in the RHQ parameter-tuning procedure by $\pm 4.4\%$ and then re-computing the bottomonium masses and mass-splittings.   For each mass or mass-splitting we take the largest variation observed on any of the sea-quark ensembles.  We find that a $\sim 4.4\%$ error $\Delta M_{B_s}$ leads to a $\sim$ 0.0--0.1\% error in the bottomonium masses, a $\sim 8.8\%$ error in the hyperfine splitting, and a $\sim 6.2\%$ error in the $\chi$-state splittings.

Discretization errors in the $B_s$ kinetic meson mass arise from both the constituent quarks' kinetic energies and the binding energy.  In Appendix~\ref{sec:HSErrs}, we estimate their size to be $\sim 2.6\%$ following the method of Ref.~\cite{Bernard:2010fr}.   This is comparable to the size of the statistical errors in the $B_s$ meson kinetic masses computed in our numerical simulations (see the example fits in Fig.~\ref{fig:243_Disp}).  As can be seen from Figs.~\ref{fig:243_LinTest}, a variation of this size leads to a statistically-significant shift in the tuned value of $\zeta$, so we must  propagate it to an uncertainty in the tuned RHQ parameters.  To estimate the resulting error we follow the same procedure as described above for the discretization errors in the hyperfine splitting.  We find that a $\sim 2.6\%$ error $M_1^{B_s}/M_2^{B_s}$ leads to a $\sim$ 0.1--0.2\% error in the bottomonium masses, a $\sim 3.6\%$ error in the hyperfine splitting, and a $\sim 1.0\%$ error in the $\chi$-state splittings.

\bigskip

To obtain the total heavy-quark discretization errors in the bottomonium masses and mass-splittings, we add the direct errors and the indirect errors in quadrature.  The resulting estimates are given in Table~\ref{tab:BottomErr}.  Numerically, the indirect errors due to discretization errors in the RHQ parameters turn out to be smaller than the direct errors for the $\bar{b}b$-meson masses, and significantly smaller than the direct errors for the fine-structure splittings.

\subsubsection{Light-quark and gluon discretization errors}

We estimate the size of light-quark and gluon discretization errors following the same approach as described for heavy-quark errors in the previous subsection.  In this case, the dimension 6 and higher-order light-quark and gluon operators in the Symanzik effective Lagrangian have no counterpart in the continuum QCD Lagrangian.  (There are no dimension 5 operators because both the light-quark and gluon actions are $\CO(a)$-improved.)  Thus the coefficients of the continuum operators in the ``mismatch functions" defined in Eq.~(\ref{eq:mismatch}) are $\CC_i^\textrm{cont} = 0$.  Further, the coefficients of the lattice operators are not expected to be suppressed by any powers of the heavy-quark mass $1/m_Q$.  Thus we take them to be $\CC_i^\textrm{lat} = 1$.  The light-quark and gluon discretization errors are then given by expectation values of light-quark and gluon operators between heavy-heavy ($\bar{Q}Q$) meson states, i.e.:
\bea
\textrm{error}_i^{LQ,g} =  \langle \CO_i^{LQ,g}  \rangle \,, \label{eq:LQmismatch}
\eea
where we estimate their size using the NRQCD power-counting, Eq.~(\ref{eq:NRQCD_PC}).

The largest discretization errors in bottomonium masses from the light-quark and gluon sector will arise from operators with only gluons.  This is because any operators containing light-quark fields must extract light quarks from the sea, and their expectation values between $\bar{Q}Q$ meson states will be suppressed by at least $\alpha_s^2$.  A typical dimension 6 gluon operator in the Symanzik effective Lagrangian is
\beq
	\CO_{\rm glue} = \rm{tr}[ F_{\mu\nu} D^2  F_{\mu\nu}] \,.
\eeq
Within the NRQCD power-counting we expect its size to be
\beq
	\langle \CO_{\rm glue} \rangle^{\rm NRQCD} \sim a^2 m^3 v^4 \,,
\eeq 
where two powers of $mv$ come from the derivative operators, and we estimate the size of $F^2$ to be the typical kinetic energy $mv^2$.  On the $24^3$ ($32^3)$ ensembles the corresponding errors in the bottomonium masses are
\beq
	\textrm{error}_{\rm glue} \sim a^2 m^3 v^4 / 2m_b = 3.0\%\, (1.7\%) \,,
\eeq
which are several times larger than the estimated sub-percent contributions of heavy-quark discretization errors.  Thus we conclude that, for bottomonium masses, the $\CO(a^2)$ light-quark and gluon discretization errors will dominate the scaling behavior, and we can remove them by extrapolating to the continuum limit in $a^2$.  The statistical errors in the continuum-limit values reflect the uncertainty on the slope in $a^2$.

Contributions from light-quark and gluon operators will largely cancel in the bottomonium fine-structure splittings, and we expect their contributions to these quantities to be negligible as compared to the heavy-quark discretization errors estimated previously.

\subsubsection{Input strange-quark mass}

We tune the parameters of the RHQ action from the bottom-strange system using the determination of the bare strange-quark mass on the two lattice-spacings from RBC/UKQCD's analysis of the light-pseudoscalar meson sector in Ref.~\cite{Aoki:2010dy}.  Hence the uncertainty in the bare strange-quark mass leads to a systematic error in the RHQ parameters, and consequently in the bottomonium masses and mass-splittings.  We estimate this error by varying the valence strange-quark mass in the $B_s$ and $B_s^*$ meson correlators used for the tuning procedure, Eqs.~(\ref{eq:C_Bs}) and~(\ref{eq:C_BsStar}), and then re-computing the bottomonium masses and mass-splittings.

\begin{figure}[p]
\centering
\includegraphics[scale=0.43,clip]{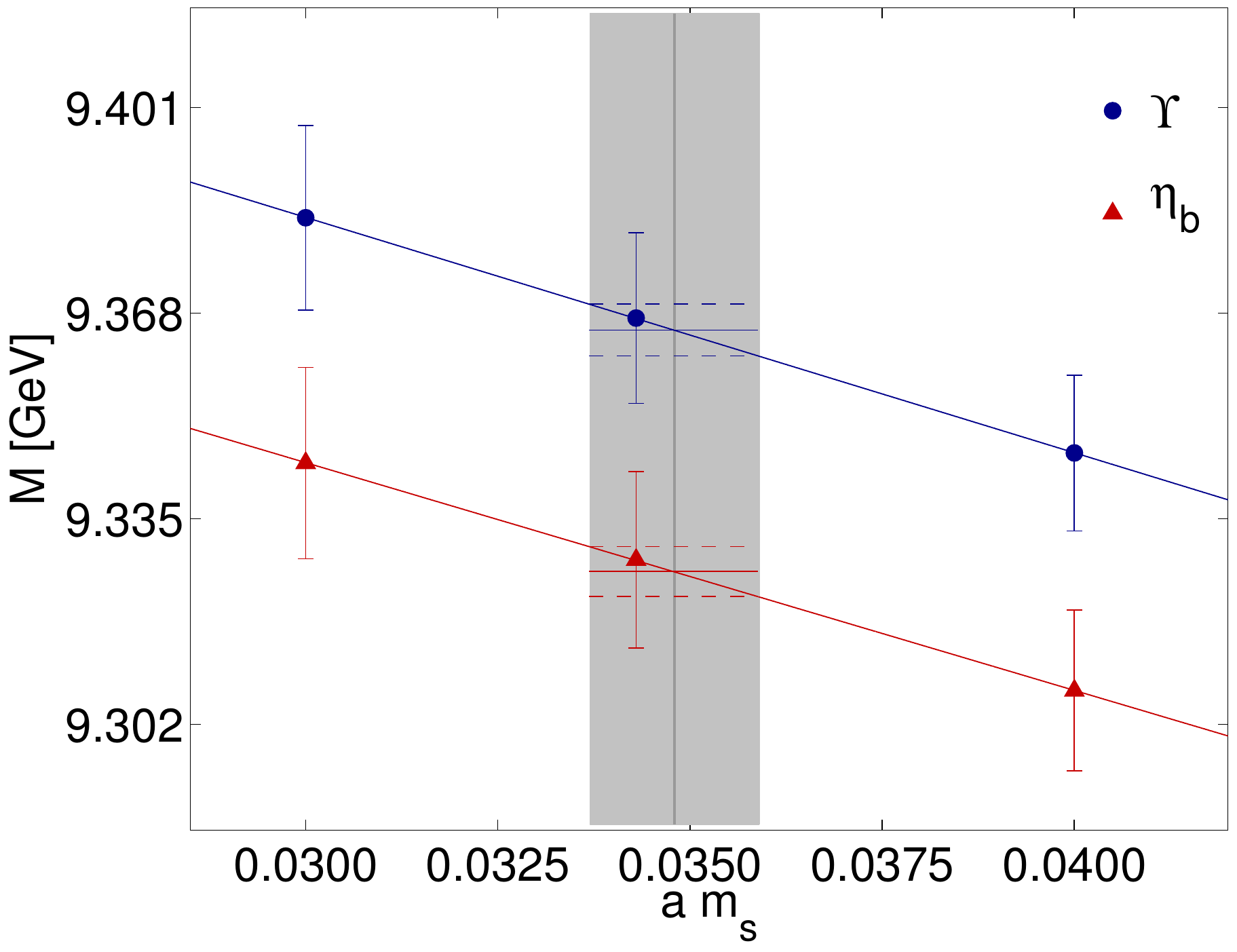}
\includegraphics[scale=0.43,clip]{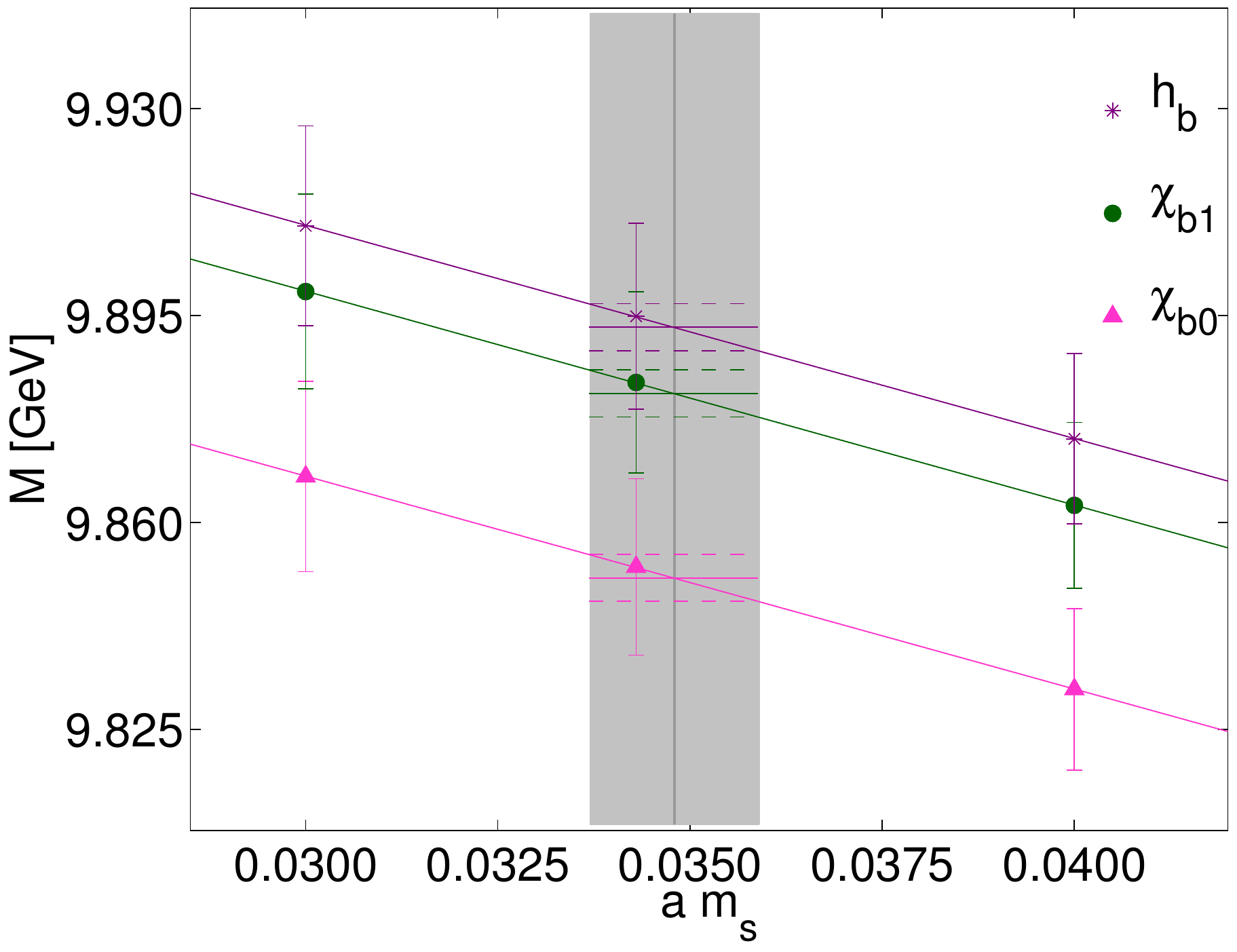}
\includegraphics[scale=0.43,clip]{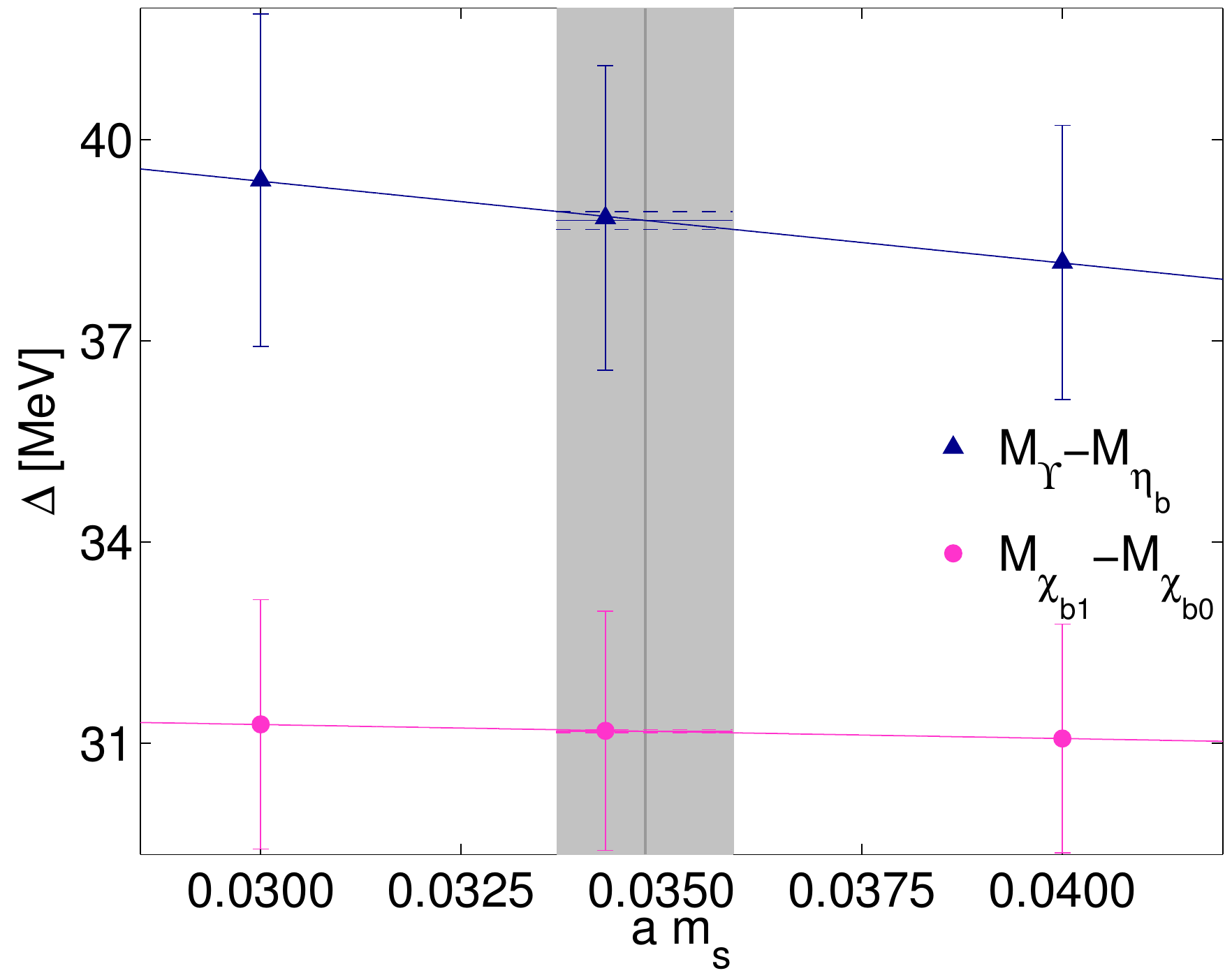}
\caption{Bottomonium masses and mass-splittings versus the valence strange-quark mass in the bottom-strange meson correlators used to tune the parameters of the RHQ action.  Results are shown for the $am_l = 0.005$ $24^3$ ensemble.  The meson states shown in each plot are specified in the legend.  For each quantity, the thicker line in the same color as the plotting symbol is an uncorrelated linear fit used to obtain the slope $\Delta M_{\bar{b}b} / \Delta m_s$.  The vertical solid line with gray error band denotes the value of the physical strange-quark mass obtained in Ref.~\cite{Aoki:2010dy}.  For each quantity, the two horizontal dashed lines show where the linear fit crosses the edges of the error band, thereby indicating the error due to the uncertainty in the strange-quark mass.}
\label{fig:Onium_vs_ms}
\end{figure}

Figure~\ref{fig:Onium_vs_ms} shows the dependence of the meson masses and mass-splittings on the valence strange-quark mass used to tune the parameters of the RHQ action on the $am_l = 0.005$ $24^3$ ensemble.  The results at the three strange-quark mass values are consistent within statistical error, and analogous plots on the $am_l = 0.004$ $32^3$ ensemble look similar.  Because the $\approx 1.2\%$ uncertainty in $m_s$ leads to a 0.1\% or less change in the bottomonium masses and a 0.3\% or less change in the mass-splittings, we can safely neglect its effect from our error budget.

\subsubsection{Input scale uncertainty}

At first glance, the value of the lattice spacing in physical units enters the computation of the bottomonium masses and mass-splittings in two ways.  It first enters indirectly through the parameters of the RHQ action, which we tune by matching the values of the $B_s$ and $B_s^*$ meson masses obtained on the lattice to the experimentally-measured values from the PDG~\cite{Nakamura:2010zzi}.  It then enters directly when we convert the lattice values of the bottomonium masses and mass splittings into GeV in order to compare with experiment.  In fact, however, the RHQ parameter tuning procedure allows us to avoid this second source of scale uncertainty.  This is because our lattice calculation of the mass $M_{\bar{b}b}$ of a $\bar{b}b$ meson gives directly the dimensionless ratio $M_{\bar{b}b}/M_{B_s}$ at the tuned values of the RHQ parameters $\{m_0a, c_P, \zeta \}$ without further reference to the lattice scale.  By construction, at the tuned point the $B_s$-meson mass is fixed to the experimentally-measured value; hence we can precisely obtain the bottomonium mass or mass-splitting in GeV by multiplying the ratio by $M_{B_s} = 5.3366$ GeV~\cite{Nakamura:2010zzi}.  We therefore need only consider the implicit dependence on the lattice spacing due to the RHQ parameters when estimating the scale uncertainty in the $\bar{b}{b}$-meson masses.

The absolute lattice scale ($a^{-1}$) has a quoted statistical error of $\sim 1\%$ ($1.5\%$) on the $32^3$ ($24^3$) lattices~\cite{Aoki:2010dy}, where the errors on the two lattice spacings are highly correlated because they come from a single fit to data at both lattice spacings.  We estimate the corresponding error in the bottomonium states by varying the lattice scale $a^{-1}$ used in the RHQ parameter tuning procedure by plus and minus a statistical sigma on each sea-quark ensemble.  For each bottomonium mass or mass-splitting, we then take the largest variation on any of the ensembles  to be the error due to the uncertainty in the lattice scale.  We find that the resulting uncertainty in the meson masses is 0.2\% or less, and in the mass-splittings is $\sim$1--3\%;  these errors are given in Table~\ref{tab:BottomErr}.

\subsubsection{Experimental inputs}

We tune the parameters of the RHQ action by using the experimental measurements of the spin-averaged $B_s$ meson mass and hyperfine splitting.  The $B_s$ and $B_s^*$ meson masses are both known to sub-percent precision~\cite{Nakamura:2010zzi}, so the experimental error in $\overline{M}_{B_s}$ contributes a negligible uncertainty to the tuned values of the RHQ parameters.  The experimental error in the hyperfine splitting $\Delta M_{B_s} = 49.0(1.5)$~MeV~\cite{Nakamura:2010zzi}, however, is $\sim 3.1\%$ and cannot be neglected.  We estimate the error in the bottomonium masses and mass-splittings due to the experimental uncertainty in the $B_s$ meson hyperfine splitting by varying the value of $\Delta M_{B_s}$ used in the RHQ tuning procedure by plus and minus 1.5~MeV.  For each bottomonium mass or mass-splitting, we then take the largest variation on any of the ensembles to be the corresponding error.  We find that the resulting uncertainty in the meson masses is 0.1\% or less, and in the mass-splittings is $\sim$4--6\%;  these errors are given in Table~\ref{tab:BottomErr}.

\subsubsection{Linear approximation}

We interpolate to the tuned values of the RHQ parameters assuming a linear dependence upon $\{m_0a, c_P, \zeta\}$.  Hence any deviation from linearity must be accounted for in the systematic error budget.  In practice, as shown in Figs.~\ref{fig:243_LinTest}, we do not see any statistically significant deviation from linearity for the heavy-strange states over a wide range of RHQ parameters.  Nor do we observe any statistically significant curvature for the $\chi$ states or the $h_b$ (see the right-hand plots in Fig.~\ref{fig:24c_Bott_LinTest1}).  Thus the systematic uncertainty in the $\chi$ states and the $h_b$ due to nonlinear dependence upon the RHQ parameters is negligible.  We can resolve nonlinear dependence of $\Upsilon$ and $\eta_b$ meson masses and the hyperfine splitting within the statistical errors in the measured effective masses, as shown in Figs.~\ref{fig:24c_Bott_LinTest1} and~\ref{fig:24c_Bott_LinTest2}.  The statistical errors in these data points, however, are almost two orders of magnitude smaller than the statistical errors in the $\Upsilon$ and $\eta_b$ meson masses and the hyperfine splitting interpolated to the tuned RHQ parameters given in Table~\ref{tab:BottomPred};  this is because the interpolated values include the uncertainty due to the statistical errors in $\{m_0a, c_P, \zeta\}$.  Hence we conclude that the systematic error due to deviations from linearity is negligible for all bottomonium quantities considered here.    

\begin{table*}
\caption{Error budget for bottomonium masses and mass-splittings.    The estimates of the size of each systematic uncertainty are given in the main text.  Each error is given as a percentage, and we obtain the total systematic by adding the individual systematic uncertainties in quadrature.  Errors that were considered but were found to be negligible ({\it i.e.} light-quark and gluon discretization errors,  strange-quark mass uncertainty, and linear approximation) are not shown.} 
\vspace{3mm}
  \begin{tabular}{lccr@{.}lccr@{.}lc} \hline\hline
     & $M_{\eta_b}$ &  $M_\Upsilon$ & \multicolumn{2}{c}{$M_\Upsilon$-$M_{\eta_b}$} & $M_{\chi_{b0}}$ & $M_{\chi_{b1}}$ & \multicolumn{2}{c}{$M_{\chi_{b1}}$-$M_{\chi_{b0}}$} & $M_{h_b}$ \\[0.5mm] \hline
    statistics & 0.4 & 0.3 & \quad3&3 & 0.4 & 0.4 & 3&7 & 0.4 \\ \hline
    heavy-quark discretization errors & 0.4 & 0.3 & 33&0 & 0.4 & 0.4 & \quad43&6 & 0.4 \\
    input scale uncertainty & 0.2 & 0.2 & 3&2 & 0.1 & 0.1 & 1&0 & 0.1 \\
    experimental inputs & 0.0 & 0.1 &6&2 & 0.0 & 0.0 & 4&3 & 0.0\\  \hline
    total systematic & 0.4 & 0.4 & 33&7 & 0.4 & 0.4 & 43&8 & 0.4 \\ 
    \hline\hline
  \end{tabular}
  \label{tab:BottomErr}
\end{table*}

\section{Results and conclusions}
\label{sec:Conc}

The relativistic heavy-quark formalism enables the description of systems involving $b$-quarks, such as $B$-mesons and bottomonium states, on currently available lattice spacings with lattice discretization errors from the heavy-quark sector of the same size as those from the light-quark sector.  We have determined the $b$-quark parameters for the RHQ action on the RBC/UKQCD 2+1 flavor domain-wall lattices with lattice spacings $a \approx 0.11$~fm and $a \approx 0.08$~fm.  This is a continuation of and improvement upon the work of Li and Peng, who each presented preliminary results for $B$-mesons and bottomonium at conferences~\cite{Li:2008kb,Peng:Lattice10}. 

In this work we tune the three parameters $\{m_0a, c_P, \zeta\}$ using the bottom-strange system, where discretization errors are expected to be of $\CO([\vec{p}a]^2)$ with $|\vec{p}| \approx \Lambda_\textrm{QCD}$.  We obtain the parameters nonperturbatively by imposing three simple conditions:  that the masses of the $B_s$ and $B_s^*$ mesons agree with the experimental measurements, and that the $B_s$ meson on the lattice obey the continuum relativistic dispersion relation $E^2 = \vec{p}^2 + M^2$.  We then test the reliability of the tuned  parameters and the validity of the relativistic heavy-quark approach by making predictions for the masses and mass splittings of several bottomonium states.  

\begin{figure}[t]
\centering
\includegraphics[scale=0.45,clip]{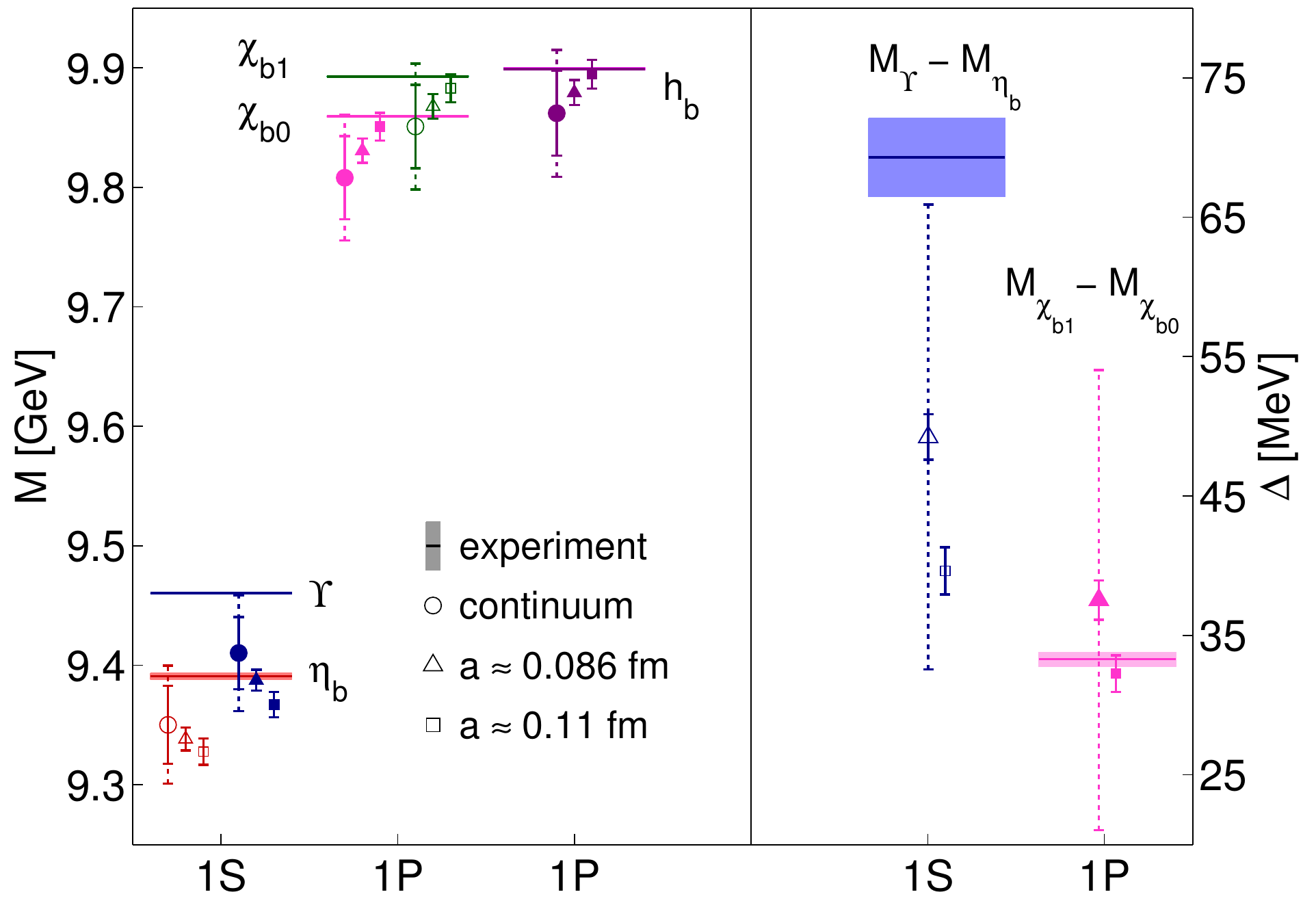}
\caption{Comparison of predicted bottomonium masses (left panel) and mass-splittings (right panel) with experiment.  For the bottomonium masses we extrapolate the results on the two lattice spacings to the continuum linearly in $a^2$, whereas for the fine-structure splittings we take the results on the finer $32^3$ ensembles as our central value.  The solid error bars on the data points show the statistical errors.  For our preferred results, we also show the systematic errors added in quadrature as dashed error bars.}
\label{fig:CompareExp}
\end{figure}

As shown in Fig.~\ref{fig:CompareExp} and Table~\ref{tab:CompareExp}, we obtain bottomonium masses with $\sim$0.5--0.6\% total uncertainties and mass-splittings with $\sim$35--45\% uncertainties, and find good agreement between our predicted values and experiment for all the quantities that we study.  In fact, the preliminary work of Li successfully predicted the mass of the $h_b$ meson~\cite{Li:2008kb} before it was first observed by the Belle collaboration~\cite{Adachi:2011ji}, thereby lending further credence to the relativistic heavy-quark formalism.  We also find agreement with calculations of $M_{\eta_b}$, the hyperfine splitting $M_{\Upsilon}-M_{\eta_b}$, and $M_{h_b}$ using the NRQCD formalism for the $b$-quark~\cite{Gray:2005ur,Meinel:2010pv} and with a calculation of the hyperfine splitting using the Fermilab formalism~\cite{Burch:2009az}.  Both the HPQCD and Fermilab/MILC works use the MILC collaboration's gauge configurations with 2+1 flavors of Asqtad-improved staggered sea quarks~\cite{Aubin:2004fs}; our study of $\bar{b}b$ meson spectroscopy using three flavors of dynamical domain-wall light quarks provides a fully independent check of these results.  Although the calculation by Meinel~\cite{Meinel:2010pv} uses the same RBC/UKQCD domain-wall + Iwasaki configurations as in this paper, our result is still largely independent of his work because statistical errors (which are somewhat correlated between the two results) are not the primary source of uncertainty.  

\begin{table*}
\caption{Comparison of predicted bottomonium masses and mass-splittings with experiment and, where possible, with other 2+1 flavor lattice calculations.  The HPQCD and Meinel calculations use the NRQCD action for the $b$-quarks~\cite{Lepage:1992tx}, while the Fermilab/MILC calculation uses the Fermilab action~\cite{ElKhadra:1996mp}.  For our results, the first error is statistical and the second is systematic; for the other results we add the errors in quadrature and quote the total.  All results are given in MeV.}
\vspace{3mm}
  \begin{tabular}{lr@{}lr@{.}lr@{}lcr@{.}l} \hline\hline
     &  \multicolumn{2}{c}{this work } & \multicolumn{2}{c}{Experiment} \quad &  \multicolumn{2}{c}{HPQCD~\cite{Dowdall:2011wh}}  \quad & Fermilab/MILC~\cite{Burch:2009az} \quad & \multicolumn{2}{c}{Meinel~\cite{Meinel:2010pv}} \\[0.5mm] \hline
    $M_{\eta_b}$  & 9350&(33)(37) & \quad9390&9(2.8)~\cite{Nakamura:2010zzi} & \quad9390&(9) & & 9400&0(7.7) \\
    $M_\Upsilon$ &  9410&(30)(38) & 9460&30(26)~\cite{Nakamura:2010zzi} &  &   \\
    $M_\Upsilon$-$M_{\eta_b}$ & 49&(02)(17) & 69&3(2.8)~\cite{Nakamura:2010zzi} & 70&(9) & 54.0$\left(^{+12.5}_{-12.4}\right)$ & 60&3(7.7) \\
    $M_{\chi_{b0}}$  & 9808&(35)(39) & 9859&44(52)~\cite{Nakamura:2010zzi} &  \\
    $M_{\chi_{b1}}$  & 9851&(35)(39) & 9892&78(40)~\cite{Nakamura:2010zzi} &  \\
    $M_{\chi_{b1}}$-$M_{\chi_{b0}}$ & 38&(01)(16) & 33&3(5)~\cite{Brambilla:2004wf} & \\ 
    $M_{h_b}$ & 9862&(36)(39) & 9899&1(1.1)~\cite{Bellehb} & 9905&(7)  & & 9899&8(1.0) \\ \hline\hline
  \end{tabular}
  \label{tab:CompareExp}
\end{table*}

Given the successful predictions of the bottomonium states, we now plan to use the nonperturbatively tuned parameters of the RHQ action to calculate $B$-meson weak matrix elements of interest to flavor physics phenomenology.  We are currently computing the leptonic decay constants $f_B$ and $f_{B_s}$ and the neutral $B^0$-$\bar{B^0}$ mixing parameters~\cite{VandeWater:2011gr}.  These calculations are particularly timely given the observed approximately $3\sigma$ tension in the CKM unitarity triangle~\cite{Bona:2009cj,Lenz:2010gu,Lunghi:2010gv,Laiho:2012ss} which currently favors the presence of new physics in $B_d$-mixing or $B\to\tau\nu$ decays.  Eventually we would also like to use the RHQ framework to calculate more challenging quantities such as $B\to\pi\ell\nu$ and $B\to D^{(*)}\ell\nu$ semileptonic form factors, which are needed to extract the CKM matrix elements $|V_{ub}|$ and $|V_{cb}|$, respectively, from exclusive channels.  Like the Fermilab interpretation, our relativistic heavy-quark formalism applies to any value of the quark mass, and allows for a continuum limit.  (This is in contrast to the NRQCD formalism, for which errors increase away from the infinite heavy-quark limit.)  Hence the same framework can be used for charm quarks, which are neither particularly heavy compared to $\Lambda_\textrm{QCD}$ nor light enough to be treated with a standard lattice light-quark formulation with $\CO(m_ca)^2$ errors that are well-controlled.  Treatment of both $b$- and $c$-quarks within the same framework allows for further tests of the methodology.  We therefore also plan to tune the parameters of the relativistic heavy-quark action for charm quarks, such that we can compute the leptonic decay constants $f_D$ and $f_{D_s}$, as well as other weak matrix elements such as the short-distance contribution to $D^0$-$\bar{D^0}$ mixing.

This work demonstrates the validity of the relativistic heavy quark action on bottom systems and opens a practical approach to obtain bottom and charm weak matrix elements with high precision given the computer resources currently available.  Lattice QCD calculations of heavy-light weak matrix elements provide critical inputs to the CKM unitarity triangle analysis.  Hence determinations with a variety of methods and independent sources of systematic uncertainty will be essential to definitively uncovering new physics in the flavor sector.  Use of the relativistic heavy-quark formalism for $b$-quarks on the RBC/UKQCD dynamical domain-wall lattices will provide phenomenologically-important, independent determinations of  key heavy-light weak-matrix elements with comparable errors to other methods.

\section*{Acknowledgments}

Computations for this work were carried out in part on facilities of the USQCD Collaboration, which are funded by the Office of Science of the U.S. Department of Energy.  We thank BNL, Columbia University, Fermilab, RIKEN, and the U.S. DOE for providing the facilities essential for the completion of this work.  

This work was supported in part by the U.S. Department of Energy under grant No. DE-FG02-92ER40699 and by the Grant-in-Aid of the Ministry of Education, Culture, Sports, Science and Technology, Japan (MEXT Grant), No.21540289, and No.22224003, No, 2254030, and No. 23105715.  JMF acknowledges support from STFC grant ST/J0003961.  This manuscript has been authored by employees of Brookhaven Science Associates, LLC under Contract No. DE-AC02-98CH10886 with the U.S. Department of Energy.  CL acknowledges support from the RIKEN FPR program.  

\appendix

\section{Heavy-quark mismatch functions}
\label{sec:Mismatch}

In this section we collect the forms of the mismatch functions used to estimate the size of heavy-quark discretization errors in heavy-heavy and heavy-light systems for the RHQ action.  

For each operator in the heavy-quark effective Lagrangian, the ``mismatch function" is defined as the difference between the short-distance coefficients in the lattice and continuum theories.  Hence the mismatch functions depend upon the parameters of the lattice action.  The mismatch functions have been calculated at tree-level for the anisotropic clover-improved Wilson action in Ref.~\cite{Oktay:2008ex}, but we present them here for completeness.  Although Oktay and Kronfeld derive general expressions for $c_E \neq c_B$ and $r_s \neq 1$ and include dimension 6 and higher order operators in the lattice action, here we show the mismatch functions specific to the RHQ action.  We obtain these expressions from those in Ref.~\cite{Oktay:2008ex} by setting  $c_E = c_B = c_P/\zeta$ and $r_s = 1$, and setting the coefficients of the dimension 6 and higher-order operators to zero.

There are five relevant tree-level mismatch functions that enter our estimates of heavy-quark discretization errors.  The first is 
\beq
	f_E (m_0a, c_P, \zeta) =  \frac{1}{8 m_E^2 a^2} - \frac{1}{8 m_2^2 a^2}, \label{eq:fE}
\eeq
where
\bea
	\frac{1}{m_2a} & = & \frac{2\zeta^2}{m_0a(2+m_0a)} 
		+ \frac{\zeta}{1+m_0a}, \label{eq:m2_tree} \\
	\frac{1}{4m_E^2a^2} &=&
		\frac{\zeta^2}{[m_0a(2+m_0a)]^2} +
		\frac{\zeta c_P}{m_0a(2+m_0a)} \,.
\eea
The function $f_E$ vanishes when the ``chromoelectric mass" $m_E$ equals the $b$-quark's kinetic mass $m_2$.
The second tree-level mismatch function is
\beq
	f_{w_4} (m_0a, c_P, \zeta) =  \frac{1}{6}w_4 \,, \label{eq:fw4}
\eeq
where
\bea
	w_4  & = & \frac{2\zeta^2}{m_0a(2+m_0a)}
		+ \frac{r_s\zeta}{4(1+m_0a)} \label{eq:w4_tree} \,.
\eea
The short-distance coefficient $w_4$ multiples the Lorentz-symmetry violating $p_i^4$ term in the lattice $b$-quark's energy-momentum dispersion relation; hence the mismatch function $f_{w_4}$ vanishes when $w_4 = 0$.  The third tree-level mismatch function is
\bea
	f_{m_4} (m_0a, c_P, \zeta) =  \frac{1}{8 m_4^3 a^3} - \frac{1}{8 m_2^3 a^3} \,, \label{eq:fm4}
\eea
where
\begin{eqnarray}
	\frac{1}{m_4^3a^3} & = & \frac{8\zeta^4}{[m_0a(2+m_0a)]^3} 
		+ \frac{4\zeta^4 + 8\zeta^3(1+m_0a)}{[m_0a(2+m_0a)]^2} \nonumber\\
		&& + \frac{\zeta^2}{(1+m_0a)^2} \,. \label{eq:m4_tree}
\end{eqnarray}
The short-distance coefficient $\frac{1}{m_4^3a^3}$ multiplies the $(\vec{p}^2)^2$ term in the $b$-quark's energy-momentum dispersion relation, so the mismatch function $f_{m_4}$ vanishes when $m_4 = m_2$.  The fourth tree-level mismatch function is
\beq
	f_{w'_B} (m_0a, c_P, \zeta) =  \frac{1}{12}w'_B \,, \label{eq:f_wBprime}
\eeq
where
\bea
	w'_B & = & \frac{c_P}{1+m_0a} \,.
\eea
The coefficient $w'_B$ leads to a spin-dependent contribution to the lattice quark-gluon vertex, so the mismatch function $f_{w'_B}$ vanishes when $w'_B = 0$.    The fifth tree-level mismatch function is
\beq
	f_{m_{B'}} (m_0a, c_P, \zeta) =  \frac{1}{4 m^3_{B'} a^3} - \frac{1}{4 m^3_{2} a^3}  \,, \label{eq:f_mBprime}
\eeq
where
\bea
		\frac{1}{m_{B'}^3a^3} & = & \frac{1}{m_4^3a^3} - \frac{\zeta^2 - \zeta c_P}{(1+m_0a)^2}  \,.  
\eea
The function $f_{m_{B'}}$ vanishes when $m_4 = m_2$ (as above) and $c_P = \zeta$.

To estimate the size of heavy-quark discretization errors in our numerical simulations, we evaluate the mismatch functions in Eqs.~(\ref{eq:fE}), (\ref{eq:fw4}), (\ref{eq:fm4}), (\ref{eq:f_wBprime}), and~(\ref{eq:f_mBprime}) at the tuned values of the RHQ parameters given in Tables~\ref{tab:243_RHQParams} and~\ref{tab:323_RHQParams}.  For the $24^3$ ensembles we use $\{m_0a, c_P, \zeta \} = \{ 8.45, 5.8, 3.10 \}$ and for the $32^3$ ensembles we use $\{m_0a, c_P, \zeta \} = \{ 3.99,  3.57,  1.93 \}$.  The results are presented in Table~\ref{tab:Mismatch_Fcns}.  Because the size of the heavy-quark discretization errors is sensitive to the numerical values of the tree-level mismatch functions, we have also tried evaluating Eqs.~(\ref{eq:fE}), (\ref{eq:fw4}), (\ref{eq:fm4}), (\ref{eq:f_wBprime}), and~(\ref{eq:f_mBprime}) at the tree-level values of the RHQ parameters $\{m_0a, c_P, \zeta \}$.  We find that the results are similar to those in Table~\ref{tab:Mismatch_Fcns}.  We therefore conclude that the mismatch functions given in Table~\ref{tab:Mismatch_Fcns} reflect the typical size of such coefficients for our simulations, and use them for estimating the heavy-quark discretization errors in the following appendices. 

\begin{table}[tb]
	\centering 
	\caption{Tree-level mismatch functions for the nonperturbatively-tuned parameters of the RHQ action on the $24^3$ and $32^3$ ensembles.} \vspace{2mm}
	\label{tab:Mismatch_Fcns}
\begin{tabular}{lccccc}
	\hline\hline
	& $f_E$ & $f_{w_4}$ & $f_{m_4}$ & $f_{w'_B}$ & $f_{m_{B'}}$ \\\hline
	$a \approx$~0.11 fm & 0.0640  & 0.0499  &  0.0353 &  0.0505  &  0.0934 \\
	$a \approx $~0.086 fm & 0.0864  & 0.0681 & 0.0521 & 0.0596 & 0.1359 \\
	\hline\hline
\end{tabular}
\end{table}

\section{Discretization errors in heavy-heavy meson masses and fine-structure splittings}
\label{sec:HHErrs}

In this section we estimate the size of heavy-quark discretization errors in heavy-heavy mesons and fine-structure mass-splittings using the framework described in Sec.~\ref{sec:HQDiscErr}.  To estimate the numerical size of the operator matrix elements, we use the NRQCD power-counting given in Eq.~(\ref{eq:NRQCD_PC}), and for the size of the coefficients we use the mismatch functions on the $32^3$ ensembles given in Table~\ref{tab:Mismatch_Fcns}. 

\subsection{Masses}
\label{sec:HHMass}

Here we consider operators of $\CO(v^4)$, which produce the dominant discretization errors in bottomonium masses.
Oktay and Kronfeld enumerate all dimension 6 and 7 bilinear operators in the heavy-quark effective Lagrangian consistent with symmetries in Table III of Ref.~\cite{Oktay:2008ex}.  We do not need to consider contributions from dimension 8 bilinears because they will be of $\CO(v^6)$ or higher.

\subsubsection{$\CO(a^2)$ errors}
\label{sec:BBbar_Oa2}

There are two dimension six bilinears that are of $\CO(v^4)$ in the NRQCD power-counting:
\bea
& \bar{h}\{\bm{\gamma}\cdot\bm{D}, \bm{\alpha}\cdot\bm{E}\}h \,, \\
& \bar{h}\gamma_4(\bm{D}\cdot\bm{E}-\bm{E}\cdot\bm{D})h \,.
\eea
The expected size of these operators is
\beq
	\langle \CO_{E} \rangle^{\rm NRQCD} \sim a^2 m_b^3 v^4 \,.
\eeq
At tree level the coefficients of these operators are both equal to $f_E$, Eq.~(\ref{eq:fE}).  We therefore estimate the contribution to the error from each of these operators to be
\beq
	\textrm{error}_{E} = f_E \langle \CO_{E} \rangle^{\rm NRQCD} / 2m_b \sim 0.15 \% \,,
\eeq
where we obtain the relative error in the $b\bar{b}$ meson masses by dividing by $2m_b$, the size of the meson masses in the NRQCD power counting.  

\subsubsection{$\CO(a^3)$ errors}
\label{sec:BBbar_Oa3}

There are two dimension seven bilinears that are also of $\CO(v^4)$ in the NRQCD power-counting:
\bea
& \bar{h}D_i^4h \,, \\
& \bar{h}(\bm{D}^2)^2h \,
\eea
and the expected size of these operators is
\beq
	\langle \CO_4 \rangle^{\rm NRQCD} \sim a^3 m_b^4 v^4 \,.
\eeq
At tree-level the mismatch function for the first operator is given by $f_{w_4}$, Eq.~(\ref{eq:fw4}), so we estimate its contribution to the error in $\bar{b}b$ meson masses to be
\beq
	\textrm{error}_{w_4} = f_{w_4} \langle \CO_4 \rangle^{\rm NRQCD}  / 2m_b \sim 0.21 \% \,.
\eeq
The tree-level mismatch function for the second operator is given by $f_{m_4}$, Eq.~(\ref{eq:fm4}), so we estimate its contribution to the error in $\bar{b}b$ meson masses to be
\beq
	\textrm{error}_{m_4} = f_{m_4} \langle \CO_4 \rangle^{\rm NRQCD}  / 2m_b \sim 0.16 \% \,.
\eeq
%

\subsubsection{Total error}
\label{sec:BBbar_Error}

We obtain the total heavy-quark discretization error in the $\bar{b}b$ meson masses by adding the errors from the different operators in quadrature, including $\CO_E$ twice because there are two such operators:
\bea
\textrm{error}^{M_{b\bar{b}}}_\textrm{total} &=& \left( 2 \times \textrm{error}_{E}^2  + \textrm{error}_{m_4}^2 + \textrm{error}_{w_4}^2\right)^{1/2} \nonumber\\ &\sim& 0.34\% \,.
\eea
%

\subsection{Hyperfine splittings}
\label{sec:HFErrs}

Only spin-dependent operators containing the term $\vec{\sigma} \cdot \vec{B}$ where $\vec{B}$ is the chromomagnetic field (and permutations thereof), contribute to hyperfine splittings such as the mass difference $M_{\Upsilon} - M_{\eta_b}$~\cite{Eichten:1980mw,Peskin:1983up}.  There are five dimension 7 bilinear operators of this form in the heavy-quark effective action at $\CO(v^6)$:
\bea
	&\sum_{i\neq j}\bar{h}\{D_j^2,i\Sigma_iB_i\}h\,, \label{eq:HFOp1}\\
	& \bar{h}\{\bm{D}^2,i\bm{\Sigma}\cdot\bm{B}\}h \,, \label{eq:HFOp2}\\
	&\sum_{i\neq j}\bar{h}i\Sigma_iD_jB_iD_jh\,, \label{eq:HFOp3}\\
	& \bar{h}\bm{\gamma}\cdot\bm{D}i\bm{\Sigma}\cdot\bm{B} \bm{\gamma}\cdot\bm{D}h\,, \label{eq:HFOp4}\\
	& \bar{h}D_ii\bm{\Sigma}\cdot\bm{B}D_ih\,.  & \label{eq:HFOp5}
\eea
Only the first two operators in Eqs.~(\ref{eq:HFOp1}) and~(\ref{eq:HFOp2}) have nonzero matching coefficients at tree-level~\cite{Oktay:2008ex}.  The matching coefficients of the remaining three operators in Eqs.~(\ref{eq:HFOp3})--(\ref{eq:HFOp5}) are zero at tree-level~\cite{Oktay:2008ex}, and have not been computed to one-loop.  Higher-dimension operators in the heavy-quark effective Lagrangian such as $\bar{h} \{ \bm{D}^2, \bm{\sigma} \cdot (\bm{D} \times \bm{E} - \bm{E} \times \bm{D}) \}  h$ also contribute to hyperfine splittings at $\CO(v^6)$, but the full set of dimension 8 heavy-heavy bilinears has not been worked-out in the literature.  

Given our incomplete knowledge of the $\CO(v^6)$ bilinear operators and corresponding mismatch functions, we use a more naive error estimation procedure for the bottomonium hyperfine splittings.  The leading contribution to the hyperfine splittings is $\sim m v^4$, so contributions of $\CO(v^6)$ are suppressed by by a factor of $v^2 \sim 0.1$.  Hence we expect that neglected $\CO(v^6)$ operators lead to 10\% errors in hyperfine splittings.  We can check this estimate for the two cases in which the mismatch functions are known, as shown below.

\subsubsection{$\CO(a^3)$ errors}

The expected size of the operators in Eqs.~(\ref{eq:HFOp1}) and~(\ref{eq:HFOp2}) is
\beq
	\langle \CO_{\mathbf{\sigma} \cdot \mathbf{B}} \rangle^{\rm NRQCD} \sim a^3 m_b^4 v^6 \,.
\eeq
The tree-level mismatch function for the first operator is given by $f_{w'_B}$, Eq.~(\ref{eq:f_wBprime}), so we estimate its contribution to the error to be
\beq
	\textrm{error}_{w'_B} = f_{w'_B} \langle \CO_{\mathbf{\sigma} \cdot \mathbf{B}} \rangle^{\rm NRQCD}  / m_b v^4 \sim 3.72 \% \,,
\eeq
where we obtain the relative error in $\bar{b}b$ meson hyperfine splittings by dividing by $m_b v^4$, the size of the hyperfine splittings in the NRQCD power counting.  The tree-level mismatch function for the second operator is given by $f_{m_{B'}}$, Eq.~(\ref{eq:f_mBprime}), so we estimate its contribution to the error in bottomonium hyperfine splittings to be
\beq
	\textrm{error}_{m_{B'}}  = f_{m_{B'}} \langle \CO_{\mathbf{\sigma} \cdot \mathbf{B}} \rangle^{\rm NRQCD}  / m_b v^4 \sim 8.48 \% \,.
\eeq
Both of these estimates are consistent with the naive power-counting expectation of 10\% based on the order in the $b$-quark velocity $v$.

\subsubsection{Total error}

There are five dimension 7 and an unknown number of dimension 8 operators in the heavy-quark effective action that contribute to the hyperfine splittings at $\CO(v^6)$ in the NRQCD power-counting.  If we assume that there are the same number of $\CO(v^6)$ operators at dimensions 7 and  8, we arrive at the estimate
\beq
	\textrm{error}^{\Delta M_{HF}}_\textrm{total} = \Big( 10 \times ({v^2})^2 \Big)^{1/2} = 31.62\% \,.
\eeq

\subsection{$\chi$-state splittings}
\label{sec:ChiErrs}

The fine-structure splitting between $\chi$ mesons $(M_{\chi_{b1}} - M_{\chi_{b0}})$ is a linear combination of the spin-orbit and tensor splittings:
\begin{align}
	\Delta_M^\textrm{spin-orbit} &= \frac{1}{9} \left( 5 M_{\chi_b2} - 2 M_{\chi_b0} - 3M_{\chi_b1} \right) , \\
	\Delta_M^\textrm{tensor} &= \frac{1}{9} \left( 3 M_{\chi_b1} - M_{\chi_b2} - 2 M_{\chi_b0} \right) .
\end{align}
Hence it receives contributions from both the spin-dependent operators containing $\sigma \cdot \vec{B}$ considered above (which lead to the tensor splitting~\cite{Eichten:1980mw}) and from spin-dependent operators containing $\vec{D} \times \vec{E}$ where $\vec{E}$ is the chromoelectric field (which lead to the spin-orbit splitting~\cite{Peskin:1983up}).  

\subsubsection{$\CO(v^4)$ errors}

There is one relevant bilinear at dimension 6  which is of $\CO(v^4)$ in the NRQCD power-counting:
\beq
	 \bar{h}\{\bm{\gamma}\cdot\bm{D}, \bm{\alpha}\cdot\bm{E}\}h \,.
\eeq
We estimate the size of its contribution to the error in the $\chi$-state splittings to be
\beq
	\textrm{error}_{v^4} = f_E \langle \CO_{E} \rangle^{\rm NRQCD} / m_b v^4 \sim 29.30 \% \,.
\eeq
Note that the contribution of this operator to the $\chi$-state splittings is not as large as the order in the $b$-quark velocity $v$ would suggest because of the small numerical size of $f_E$.

\subsubsection{$\CO(v^6)$ errors}

The same $\CO(v^6)$ operators that contribute to the hyperfine splittings also contribute to the splitting between the $\chi$ states.  We therefore estimate their contributions to be the same size as for the hyperfine splittings:
\beq
	\textrm{error}_{v^6} = 31.62\% \,.
\eeq
%

\subsubsection{Total error}

We obtain the total heavy-quark discretization error in the $\chi$ state splittings by adding the $\CO(v^4)$ and $\CO(v^6)$ errors in quadrature, yielding
\beq
	\textrm{error}^{\Delta M_{\chi}}_\textrm{total} = \left( \textrm{error}_{v^4}^2 + \textrm{error}_{v^6}^2  \right)^{1/2} = 43.11 \% \,.
\eeq

\section{Discretization errors in heavy-strange meson masses and hyperfine splitting}
\label{sec:HSErrs}

In this section we estimate the size of heavy-quark discretization errors in the heavy-strange meson quantities -- the spin-averaged mass, hyperfine splitting, and ratio of rest-to-kinetic masses -- used in the RHQ parameter tuning procedure.  Again we use the framework described in Sec.~\ref{sec:HQDiscErr}.  To estimate the numerical size of the operators, we use the HQET power-counting given in Eq.~(\ref{eq:HQET_PC}), and for the size of the coefficients we use the mismatch functions on the $32^3$ ensembles given in Table~\ref{tab:Mismatch_Fcns}. 

\subsection{Rest mass}
\label{sec:Disc_M1}

Because we tune the coefficients of the dimension 5 operators in the RHQ action nonperturbatively, the leading discretization errors come from operators of dimension 6 and higher in the effective theory.  There are two dimension 6 bilinears of $\CO(\lambda^2)$ in the HQET power-counting:
\bea
& \bar{h}\{\bm{\gamma}\cdot\bm{D}, \bm{\alpha}\cdot\bm{E}\}h\,, \label{eq:Op1} \\
& \bar{h}\gamma_4(\bm{D}\cdot\bm{E}-\bm{E}\cdot\bm{D})h \label{eq:Op2} \,.
\eea
The estimated size of these operators is
\beq
	\langle \CO_{E} \rangle^{\rm HQET} \sim a^2 \Lambda_{\rm QCD}^3 \,.
\eeq
We do not consider operators of dimension 7 and higher because they are all at least of  $\CO(\lambda^3)$.  At tree-level the coefficients of the operators in Eqs.~(\ref{eq:Op1}) and~(\ref{eq:Op2}) are both given by Eq.~(\ref{eq:fE}), so we estimate their contributions to the error in the spin-averaged $B_s$ meson rest mass to be
\beq
	\textrm{error}_{E} = f_E \langle \CO_{E} \rangle^{\rm HQET} / \bar{M}_{B_s} \sim 0.04 \% \,.
\eeq
By construction, we tune the RHQ parameters such that the spin-averaged rest mass equals the experimental value of $\frac{1}{4}(M_{B_s} + 3M_{B_s}^*)$, so we obtain the relative error in $M_1$ by dividing by $\bar{M}_{B_s} = 5.4028 $~GeV.   We obtain the total heavy-quark discretization error in the spin-averaged $B_s$ meson rest mass by adding the contributions from the two operators in quadrature, which yields:
\beq
	\textrm{error}^{M_{1,B_s}}_\textrm{total} = \left( 2 \times \textrm{error}_{E}^2 \right)^{1/2} = 0.05 \% \,,
\eeq
or $\sim$ 3~MeV. 

\subsection{Kinetic mass}
\label{sec:Disc_M2}

Discretization errors in the kinetic meson mass $M_2$ arise from both the constituent quarks' kinetic energies and from the binding energy.  The Appendix of Ref.~\cite{Bernard:2010fr} provides a semi-quantitative estimate of the discretization error in $M_2$ (see also Ref.~\cite{Kronfeld:1996uy}).  Although this estimate is made assuming that both quarks in the meson are nonrelativistic, the result is interpreted {\it a posteriori} under the assumption that the strange quark is light and relativistic.  We follow the same approach here.

The tree-level discretization error in $M_2$ through $\CO(v^4)$ in the nonrelativistic expansion is given by~\cite{Bernard:2010fr}
\beq
	\delta M_2 = \frac{1}{3m_2} \frac{\langle \vec{p}^2 \rangle}{2}  \left[  5 \left( \frac{m_2^3}{m_4^3} -1 \right) + 4 w_4 (m_2 a)^3 \right] \,, \label{eq:Err_M2}
\eeq
where this result applies to $S$-wave states.  Note that the $\delta M_2$ is zero if the masses $m_4 = m_2$ and the Lorentz-symmetry violating coefficient $w_4 = 0$.  To estimate the numerical size of the discretization error in $M_2$ we replace $\langle \vec{p}^2 \rangle$ with $\Lambda_{\rm QCD}^2$ following the HQET power-counting prescription and use the expressions for $m_2$, $m_4$, and $w_4$ given in Eqs.~(\ref{eq:m2_tree}), (\ref{eq:w4_tree}), and~(\ref{eq:m4_tree}).  By construction, we tune the RHQ parameters such that the kinetic meson mass equals the experimental value of the $B_s$ meson mass, so we obtain the relative error in $M_2$ by dividing by $M_{B_s} = 5.366$~GeV.  We obtain
\beq
	\textrm{error}^{M_{2,B_s}}_\textrm{total} = 2.59 \%\,,
\eeq
or $\sim$139~MeV.

\subsection{Hyperfine splitting}
\label{sec:Disc_HF}

The bottom-strange hyperfine splitting receives contributions from spin-dependent operators containing the term $\vec{\sigma} \cdot \vec{B}$ where $\vec{B}$ is the chromomagnetic field (and permutations thereof)~\cite{Eichten:1980mw,Peskin:1983up}.  The leading contribution is from the dimension 5 operator $\bar{h} i \Sigma \cdot \bm{B} h$ and is of $\CO(\lambda)$ in the HQET power-counting.  Because we tune the coefficient of this operator nonperturbatively, there are no associated discretization errors.  Thus we consider discretization errors from operators of $\CO(\lambda^2,\lambda^3)$.  There are five dimension 7 bilinear operators of the type $\vec{\sigma} \cdot \vec{B}$ in the heavy-quark effective action at $\CO(\lambda^3)$; these are given in Eqs.~(\ref{eq:HFOp1})--(\ref{eq:HFOp5}).  Operators of dimension 8 and higher in the heavy-quark effective Lagrangian are all of $\CO(\lambda^4)$ or higher in the HQET power-counting.

\subsubsection{$\CO(a^3)$ errors}
\label{sec:Tree_Err}

The expected size of the operators in Eqs.~(\ref{eq:HFOp1}) and~(\ref{eq:HFOp2}) is 
\beq
	\langle \CO_{\sigma \cdot B} \rangle^{\rm HQET} \sim a^3 \Lambda_{\rm QCD}^4 \,.
\eeq
By construction, we tune the RHQ parameters such that we reproduce the experimental value of the bottom-strange hyperfine splitting $M_{B_s}^*$-$M_{B_s}$.  Hence we divide the contributions of these operators by $\Delta M_{B_s} = 49$~MeV to obtain the relative error in the $B_s$ hyperfine splitting.  The tree-level mismatch functions for the two operators are $f_{w'_B}$ [Eq.~(\ref{eq:f_wBprime})]  and $f_{m_{B'}}$ [Eq.~(\ref{eq:f_mBprime})], so we estimate their contribution to the error in the bottom-strange hyperfine splitting to be
\bea
	\textrm{error}_{w'_B} & = & f_{w'_B} \langle \CO_{\sigma \cdot B} \rangle^{\rm HQET} / \Delta M_{B_s} \nonumber \\ &\sim& 0.64 \% \,, \\
	\textrm{error}_{m_{B'}} & = & f_{m_{B'}} \langle \CO_{\sigma \cdot B} \rangle^{\rm HQET} / \Delta M_{B_s} \nonumber \\ &\sim& 1.46 \% \,.
\eea
%

\subsubsection{$\CO(\alpha_s a^3)$ errors}
\label{sec:Loop_Err}

The expected size of the operators in Eqs.~(\ref{eq:HFOp3})--(\ref{eq:HFOp5}) is also
\beq
	\langle \CO_{\sigma \cdot B} \rangle^{\rm HQET} \sim a^3 \Lambda_{\rm QCD}^4 \,.
\eeq
The mismatch functions of these operators, however, vanish at tree-level~\cite{Oktay:2008ex}.  Because they have not been computed to one-loop, we simply estimate their size to be $\alpha_s^{\bar{{\rm MS}}} (1/a_{32^3}) = 0.22$.  Under this assumption, we estimate that the contribution of each of these operators to the bottom-strange hyperfine splitting is
\bea
	\textrm{error}_{\alpha_s} & = & \alpha_s \langle \CO_{\sigma \cdot B} \rangle^{\rm HQET} / \Delta M_{B_s} \sim 2.36\% \,.
\eea
This estimate is likely conservative, given that we would naively expect $\CO(\alpha_s a^3)$ errors to be smaller than $\CO(a^3)$ errors, due to the fact that we have not considered any possible suppression from the 1-loop mismatch functions.

\subsubsection{Total error}

We obtain the total heavy-quark discretization error in the bottom-strange hyperfine splitting by adding the errors from the different operators in quadrature, including $\textrm{error}_{\alpha_s}$ three times because there are three 1-loop operators:
%
\bea
	\textrm{error}^{\Delta M_{B_s}}_\textrm{total} &=& \left( \textrm{error}_{w'_B}^2 + \textrm{error}_{m_{B'}}^2 + 3 \times \textrm{error}_{\alpha_s}^2  \right)^{1/2} \nonumber\\ & \sim & 4.40 \% \,,
\eea
or $\sim$ 2~MeV. 


\bibliography{B_meson}
\bibliographystyle{apsrev4-1} 
\end{document}